\documentclass[aps,prd,showpacs,twocolumn,
superscriptaddress]{revtex4-1}

\usepackage{graphicx}
\usepackage{dcolumn}

\usepackage{amsmath}
\usepackage{amssymb}
\usepackage{bm}
\usepackage{color}
\usepackage[normalem]{ulem}
\usepackage[dvipsnames]{xcolor}
\usepackage{hyperref}
\hypersetup{
colorlinks=true, % false: boxed links; true: colored links
linkcolor=blue,  % color of internal links
citecolor=cyan,  % color of links to bibliography
}

\newcommand{\bea}{\begin{eqnarray}}
\newcommand{\eea}{\end{eqnarray}}
\newcommand{\be}{\begin{equation}}
\newcommand{\ee}{\end{equation}}

\begin{document}
\title{
Charged particle motion and electromagnetic field in $\gamma$ spacetime 
}

\author{Carlos A. Benavides-Gallego}
\email{abgcarlos17@fudan.edu.cn}
\affiliation{Center for Field Theory and Particle Physics and Department of Physics, Fudan University, 200438 Shanghai, China }

\author{Ahmadjon Abdujabbarov}
\email{ahmadjon@astrin.uz}
\affiliation{Center for Field Theory and Particle Physics and Department of Physics, Fudan University, 200438 Shanghai, China }
\affiliation{Ulugh Beg Astronomical Institute, Astronomicheskaya 33,
	Tashkent 100052, Uzbekistan }

\author{Daniele~Malafarina}
\email{daniele.malafarina@nu.edu.kz}
\affiliation{Department of Physics, Nazarbayev University, 53 Kabanbay Batyr avenue, 010000 Astana, Kazakhstan }

\author{Bobomurat Ahmedov}
\email{ahmedov@astrin.uz}
\affiliation{Ulugh Beg Astronomical Institute, Astronomicheskaya 33,
	Tashkent 100052, Uzbekistan }

\author{Cosimo Bambi}
\email{bambi@fudan.edu.cn}
\affiliation{Center for Field Theory and Particle Physics and Department of Physics, Fudan University, 200438 Shanghai, China }
%\affiliation{Theoretical Astrophysics, Eberhard-Karls Universit\"{a}t T\"{u}bingen, Auf der Morgenstelle 10, 72076 T\"{u}bingen, Germany}

%
\date{\today}
\begin{abstract}

We consider the electromagnetic field occurring in the background of a static, axially symmetric vacuum solution of Einstein's field equations immersed in an external magnetic field. The solution, known as the $\gamma$ metric (or Zipoy-Voorhees), is related to the Schwarzschild spacetime through a real positive parameter $\gamma$ that describes its departure from spherical symmetry. We study the motion of charged and uncharged particles in this spacetime and particle collision in the vicinity of the singular surface and compare with the corresponding result for Schwarzschild. We show that there is a sharp contrast with the black hole case; in particular, in the prolate case ($\gamma<1$) particle collision can occur with an arbitrarily high center of mass energy. This mechanism could in principle allow one to distinguish such a source from a black hole.

\end{abstract}


\maketitle

\section{Introduction}

The electromagnetic field occurring around a rotating Kerr black hole immersed in an external asymptotically uniform magnetic field was first considered by Wald~\cite{Wald74}. Since then, the electromagnetic field structure and charged particle motion around axially-symmetric black holes have been studied by several authors~\cite{Aliev02,Aliev2004,Aliev05,Aliev06,Frolov10,Abdujabbarov10,Abdujabbarov11a,Frolov12,Karas12a,Hakimov13,Stuchlik14a,Stuchlik16}. 

The authors of~\cite{Banados09} have shown that a Kerr black hole can accelerate particles near the event horizon. The role of the magnetic fields in the charged particle acceleration mechanism has been considered in~\cite{Frolov12,Abdujabbarov13a}, acceleration of neutral and charged particles has also been studied by different authors~\cite{Tursunov13,Tursunov16,Kolos15,Kolos17,Tursunov18a,Tursunov18,Shaymatov18,Kimura11,Zaslavskii12a,Banados11,Oteev16,Hakimov14a,Igata12,Toshmatov15d,Gao11,Liu11,Patil11b,Patil12,Patil10,
Zaslavskii11,Zaslavskii10b,Zaslavskii11c,Ghosh14,Abdujabbarov14,Shaymatov13}, whereas energetic processes around black holes immersed in magnetic fields have been studied in~\cite{Dadhich18,Zaslavskii12b,dadhich12b,wagh85,Dhurandhar83,Dhurandhar84b,
Dhurandhar84,Abdujabbarov11b}.

In this paper we study a similar situation which does not involve a black hole with the aim of investigating the possible role played by deformations. Therefore, neglecting rotation, we consider the static and axially symmetric field outside a non spherical compact object, as described by the so-called $\gamma$ metric (also known as Zipoy-Voorhees metric)
\cite{Zipoy66,Voorhees70}
and study the effects of the presence of an external asymptotically uniform magnetic field. 

The main interest for studying such scenarios is to understand how robust are the current predictions of observable features of black holes~\cite{Bambi17c}. The main question is: 'Is there some other mathematical solution that can mimic the appearance of a black hole?'
\cite{Mazur2004,Chirenti07,Chirenti16,Cardoso16,Cardoso16erratum,Carballo-Rubio18}
If such black hole mimickers exist, they could hint at the existence of some kind of exotic object in astrophysics, or indicate a regime where modifications to classical GR are not negligible. 

Typically when studying departures from the black hole scenarios, people have considered either solutions that are perturbations of the usual Kerr black hole (see for example 
\cite{Johannsen11}), 
or exact solutions in modified theories of gravity
\cite{Yagi16}.
In the first case, one obtains a perturbed geometry that, strictly speaking, is not a solution of Einstein's equations in vacuum. In the second case, one needs to justify the rationale behind the modifications to classical GR.

On the other hand there are many exact solutions of Einstein's equations in vacuum that can serve the purpose of investigating the observable behavior when departing from the Kerr or Schwarzschild geometry.
The main advantage of considering the $\gamma$ metric with respect to other approaches is that the $\gamma$ metric is an exact solution of Einstein's equations in vacuum that is continuously linked to the Schwarzschild solution through the value of one parameter. Therefore by varying the parameter $\gamma$ one can investigate how the behavior of a determined observable feature (in our case the motion and collision of test particles in the presence of an external magnetic field) is affected by the change in geometry. This matter is not trivial, in light of the fact that for all values of $\gamma$ for which the solution is not spherically symmetric the surface $r=2m$ (which in the Schwarzschild case corresponds to the event horizon) becomes a singular Cauchy horizon.

One can expect that, similarly to the black hole case, the presence of magnetic fields, induces an electromagnetic field in the spacetime, which affects the motion of charged test particles.
In the following, we derive particles' motion in the $\gamma$ spacetime immersed in an external magnetic field with particular attention to energetic processes, such as the particle's acceleration and collisions, in the vicinity of the Cauchy horizon.

The  paper is organized as follows:
In Sect~\ref{gamma} a brief overview of the properties of the $\gamma$ metric is outlined. Sect.~\ref{emfield} is devoted to the study of 
electromagnetic fields in $\gamma$ spacetime in the presence of external asymptotically uniform magnetic field. In Sect~\ref{chparmot},
we consider the motion of charged particles in the $\gamma$ spacetime in the presence of a magnetic field. In Sect.~\ref{energetics}, we study the center of mass energy for collisions of test particles in the $\gamma$ spacetime.
Finally, in Sect.~\ref{Summary} we summarize
our obtained results and outline some possible connections with observable quantities for astrophysical black hole candidates.
Throughout the paper we use a 
signature $(-,+,+,+)$, a system of units
in which $G = c = 1$. 
Greek indices run from $0$ to $3$, Latin indices from $1$ to $3$.

\section{The $\gamma$ metric}\label{gamma}

The $\gamma$ metric can be described in Erez-Rosen coordinates \cite{Erez59} by the line element 
\begin{eqnarray}\label{metric}
ds^2&=&- F dt^2 \\ \nonumber
&&+ F^{-1}[G dr^2+H d\theta^2 + (r^2-2 mr)\sin^2\theta d\phi^2] ,
\end{eqnarray}
with
\begin{eqnarray}
F&=&\left(1-\frac{2m}{r}\right)^\gamma\ , \\
G&=&\left(\frac{r^2-2m r }{r^2-2mr +m^2\sin^2\theta}\right)^{\gamma^2-1}\ ,\\ 
H&=&\frac{(r^2-2mr)^{\gamma^2}}{(r^2-2mr +m^2\sin^2\theta)^{\gamma^2-1}}\ ,
\end{eqnarray}
\\
where $\gamma$ is the mass density parameter which is related to the axially symmetric deformations. The Schwarzschild solution is immediately obtained when $\gamma=1$. However, one needs to be careful not to consider the coordinates as spherical when $\gamma \neq 1$. Performing the multipole expansion of the line element (1) and considering up to monopole moment of the spacetime one might conclude that the spacetime in relatively good approximation is  vacuum Schwarzschild one with a total mass (monopole moment of the spacetime)  
\be 
M=\gamma m \; ,
\ee 
while the non vanishing quadrupole moment is given by 
\be 
Q=(1-\gamma^2)\frac{\gamma m^3}{3} \; .
\ee
Notice that in the case $\gamma=1$ all higher multipole moments vanish identically
\cite{Herrera00,Hernandez-Pastora94}.
Interior solutions for the $\gamma$ metric have been studied in 
\cite{Hernandez67,Stewart82,Herrera05}, showing that there exist viable interiors describing oblate and prolate spheroids that match the vacuum exterior.

The most peculiar property of the $\gamma$ metric is the fact that the curvature invariants such as the Kretschmann scalar diverge (at least directionally) for $r\rightarrow 2m$
\cite{Virbhadra96}.
This shows that, with the exception of the Schwarzschild case when $\gamma=1$, the surface $r=2m$ is not an event horizon, in accordance with what expected from the no hair theorem, but it is a genuine singularity. 

The behavior of geodesics in the $\gamma$ spacetime was studied in \cite{Herrera99} and the application to accretion disks around astrophysical compact objects was studied in \cite{Liu18,Chowdhury12}.
As expected, the trajectory of test particles differs from the Schwarzschild case. In particular the deviations become more pronounced when the departure from spherical symmetry is larger and also in the vicinity of the Cauchy horizon. This last fact is particularly important as, even for values of $\gamma$ very close to one, one can obtain a significantly different trajectory for the motion of the test particle when the particle approaches $r=2m$.

In the following we will implement a similar analysis for the case where the source of the $\gamma$ metric is immersed in an external magnetic field.

\section{Electromagnetic field produced by an external magnetic field 
%in the $\gamma$ spacetime 
\label{emfield}}

In the following we shall assume that the $\gamma$ spacetime is immersed in an external magnetic field. 
For simplicity, we will consider the case where the source of the $\gamma$ metric is placed in a magnetic field chosen to be asymptotically uniform and aligned in the direction of the axis of symmetry of the metric.
Due to the curvature of the spacetime the structure of the electromagnetic field induced near the source will change. Using the Lorentz gauge ($A^{\alpha}_{;\alpha} = 0$) one can rewrite the Maxwell equations in vacuum as 
\begin{eqnarray}\label{maxwa}
A^{\alpha;\beta}_{\ \ \ ;\beta} = \Box A^\alpha =0\ ,
\end{eqnarray}
where $A^\alpha$ is the four vector potential of the electromagnetic field.
On the other hand one can rewrite the Killing equation for this spacetime
\begin{eqnarray}
\xi_{\alpha;\beta}+\xi_{\beta;\alpha}=0
\end{eqnarray}
as
\begin{eqnarray}
\xi^{\alpha;\beta}_{\ \ \ ;\beta}=R^{\alpha}_{\ \beta}\xi^{\alpha}\ . 
\end{eqnarray}
For the vacuum spacetime the Ricci tensor vanishes $R^\alpha_{\ \beta}=0$ and for Killing vectors one gets
\begin{eqnarray}\label{killbox}
\Box\xi^\alpha=0. 
\end{eqnarray}
Note that here we are interested in the particular case of the external magnetic field configuration: asymptotically uniform magnetic field. From an astrophysical point of view this is most relevant configuration (see e.g.~\cite{Wald74,Aliev02,Aliev2004,Aliev05,Aliev06,Frolov10,Abdujabbarov10,Abdujabbarov11a,Frolov12,Karas12a,Hakimov13,Stuchlik14a,Stuchlik16}). 
Using the similarity of Eqs.~(\ref{maxwa}) and (\ref{killbox}) and the existence of the time-like $\xi^\alpha_{(t)} = (-1,0,0,0)$ and space-like $\xi^\alpha_{(\phi)} = (0,0,0,1)$ Killing vectors for the metric (\ref{metric}) one can rewrite the solution of Eq.~(\ref{maxwa}) as
\begin{eqnarray}
A^{\alpha} = C_1\xi^{\alpha}_{(t)} + C_2 \xi^{\alpha}_{(\phi)}\ ,
\end{eqnarray}
where the constants $C_1 = 0$ and $C_2=B/2$ can be found from the asymptotic properties of the spacetime and the electromagnetic field. The constant $B$ is the asymptotic value of the uniform magnetic field aligned along the axis of symmetry. Finally the covariant components of the four potential of the electromagnetic field are 
\begin{eqnarray}
&&A_0=A_1=A_2=0\ ,\nonumber\\ 
&&A_3= \frac{B}{2} (r-2m)^{1-\gamma}r^{1+\gamma} \sin^2\theta\ .
\end{eqnarray}

As the next step we calculate the components of the electric and magnetic field, namely $E^\alpha$ and $B^\alpha$ respectively, by using the expressions
\begin{eqnarray}
E_\alpha&=&F_{\alpha\beta}u^\beta\ ,\\
B^\alpha&=&\frac12 \eta^{\alpha\beta\mu\nu}F_{\beta\mu} u_\nu\ ,
\end{eqnarray}
where $F_{\alpha\beta}=A_{\beta;\alpha} -A_{\alpha;\beta} $ is the electromagnetic field tensor and  $\eta^{\alpha\beta\mu\nu}$ is the Levi-Civita tensor. The four velocity of the observer $u^\alpha$ can be chosen as 
\begin{eqnarray}
u^{\alpha}&=& \left(1-\frac{2m}{r}\right)^{-\gamma/2}\left(-1,0,0,0\right)\ , \\
u_{\alpha} &=& \left(1-\frac{2m}{r}\right)^{\gamma/2}\left(1,0,0,0\right)\ , 
\end{eqnarray}
and non-zero orthonormal components of the electromagnetic field will take the form
\begin{eqnarray}
B^{\hat{r}}&=&\frac{B\cos\theta}{\sqrt{G}}\  ,\\
B^{\hat{\theta}}&=&\frac{B\sin\theta}{\sqrt{H}}(r-m-m\gamma)\   .\end{eqnarray}

\begin{figure*}
	\includegraphics[width=0.48 \textwidth]{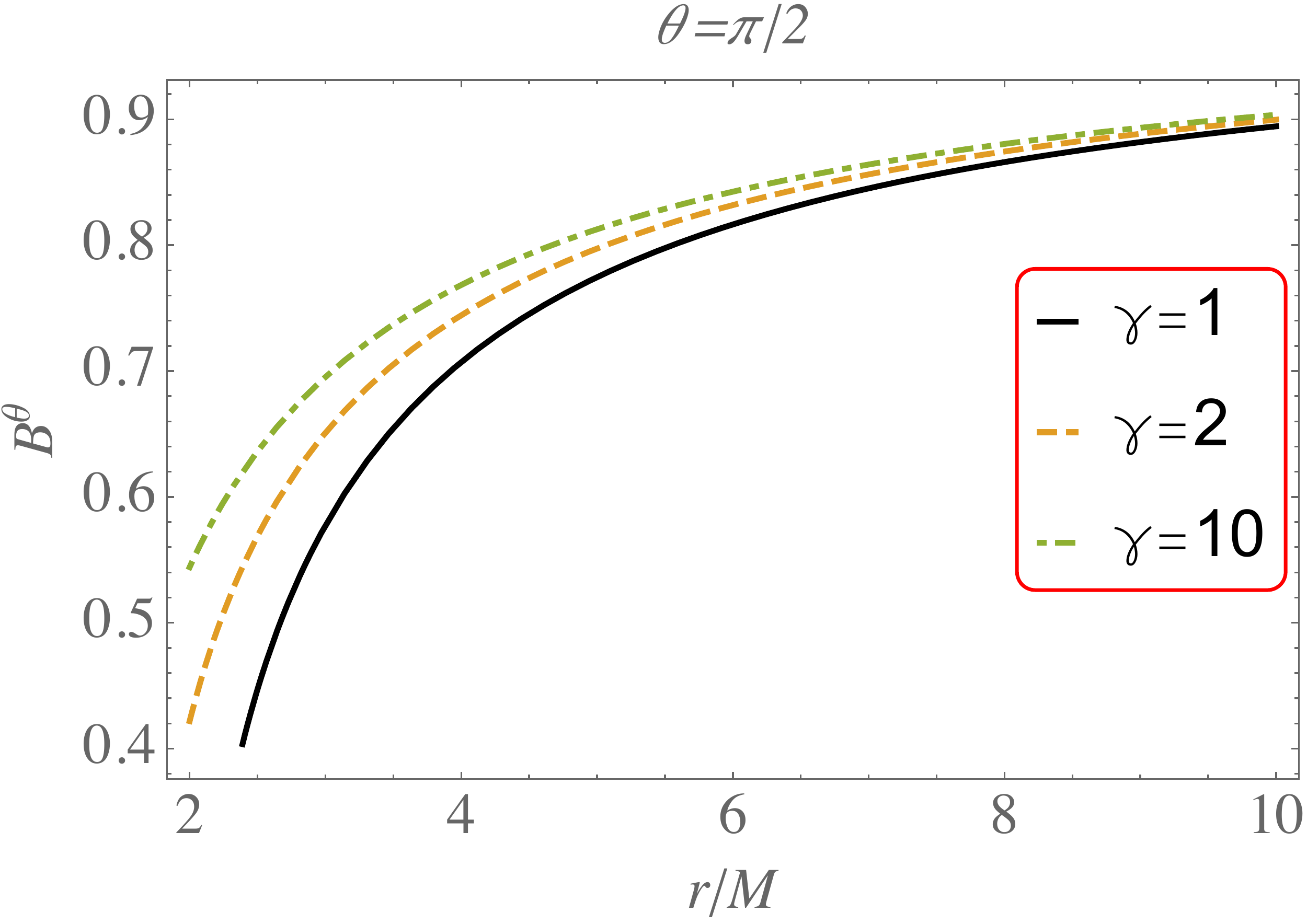}
	\includegraphics[width=0.48 \textwidth]{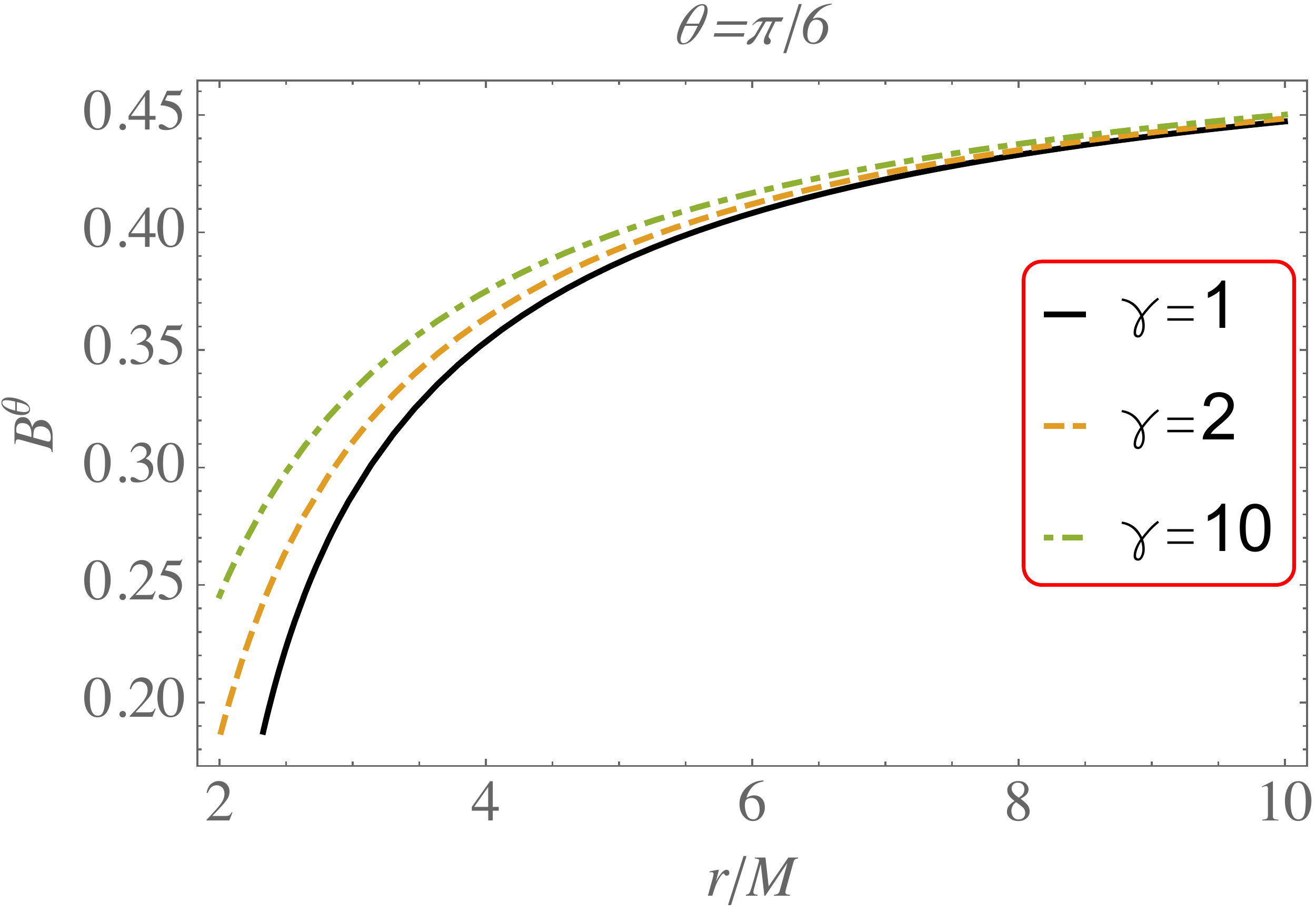}
	
	\includegraphics[width=0.48 \textwidth]{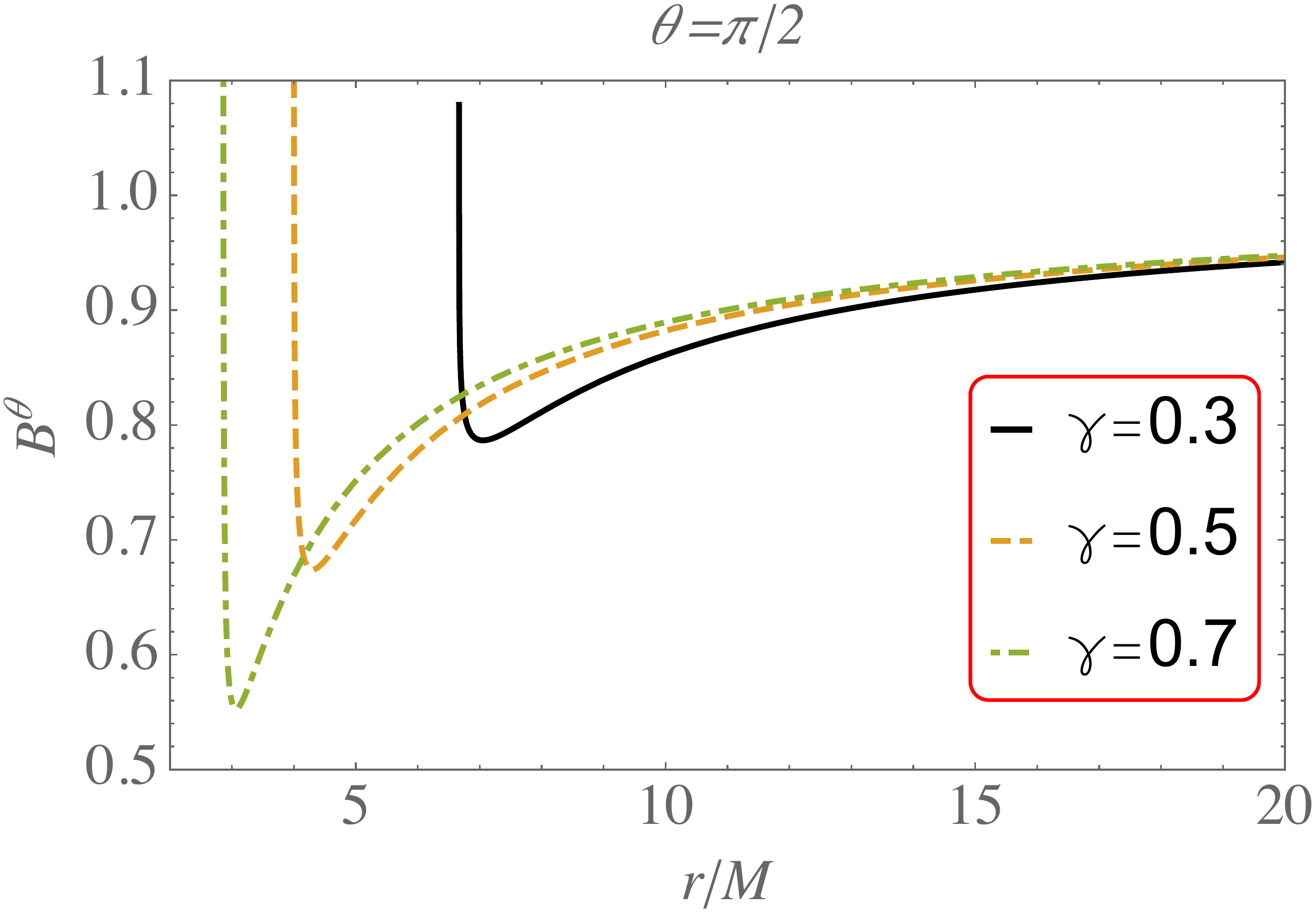}
	\includegraphics[width=0.48 \textwidth]{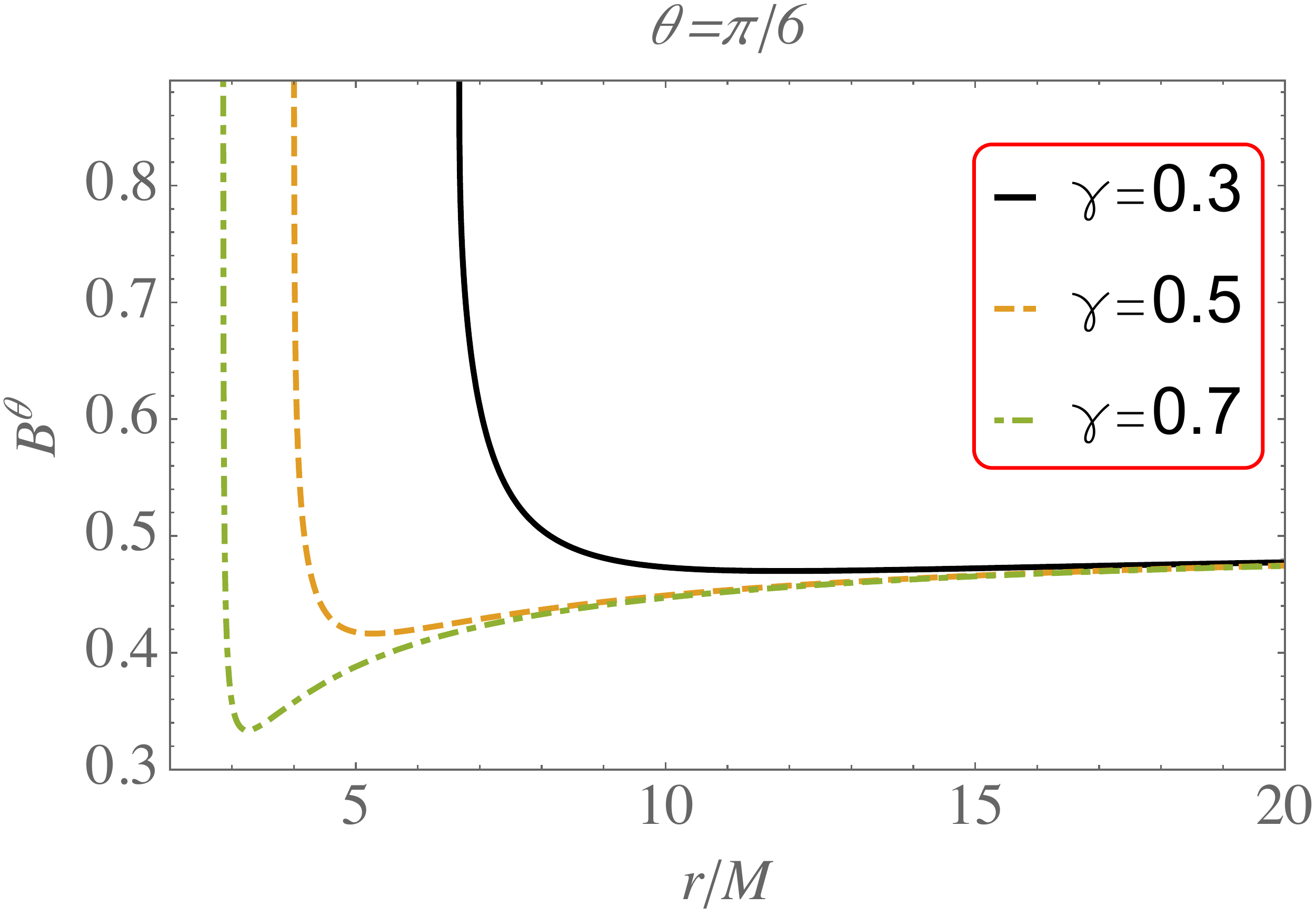}
	
	\caption{Radial dependence of the azimuthal component of the magnetic field in $\gamma$ spacetime for the different values of $\gamma$ and the azimuthal angle $\theta$. \label{figem}}
\end{figure*}

\begin{figure*}
	\includegraphics[width=0.48 \textwidth]{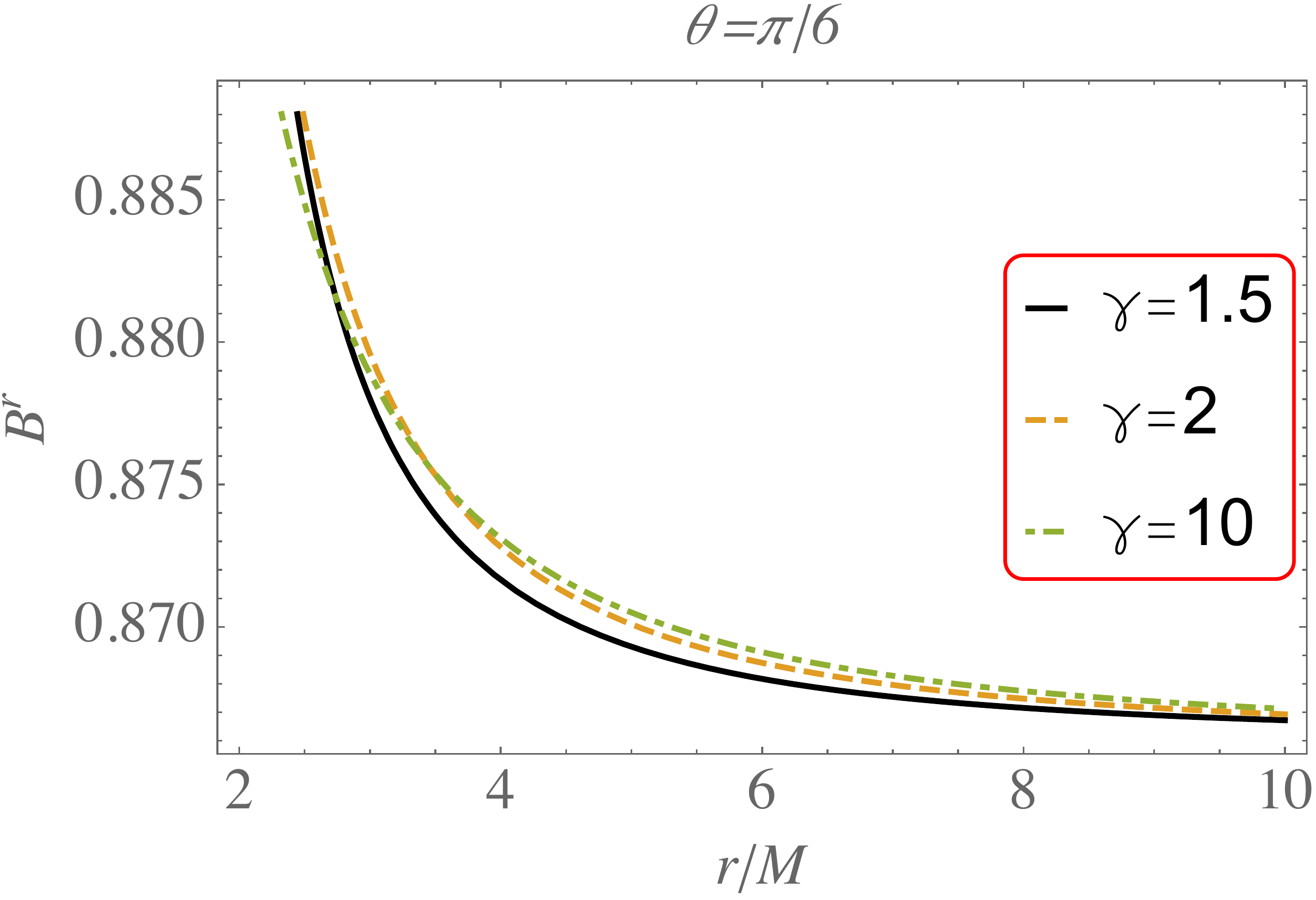}
	\includegraphics[width=0.48 \textwidth]{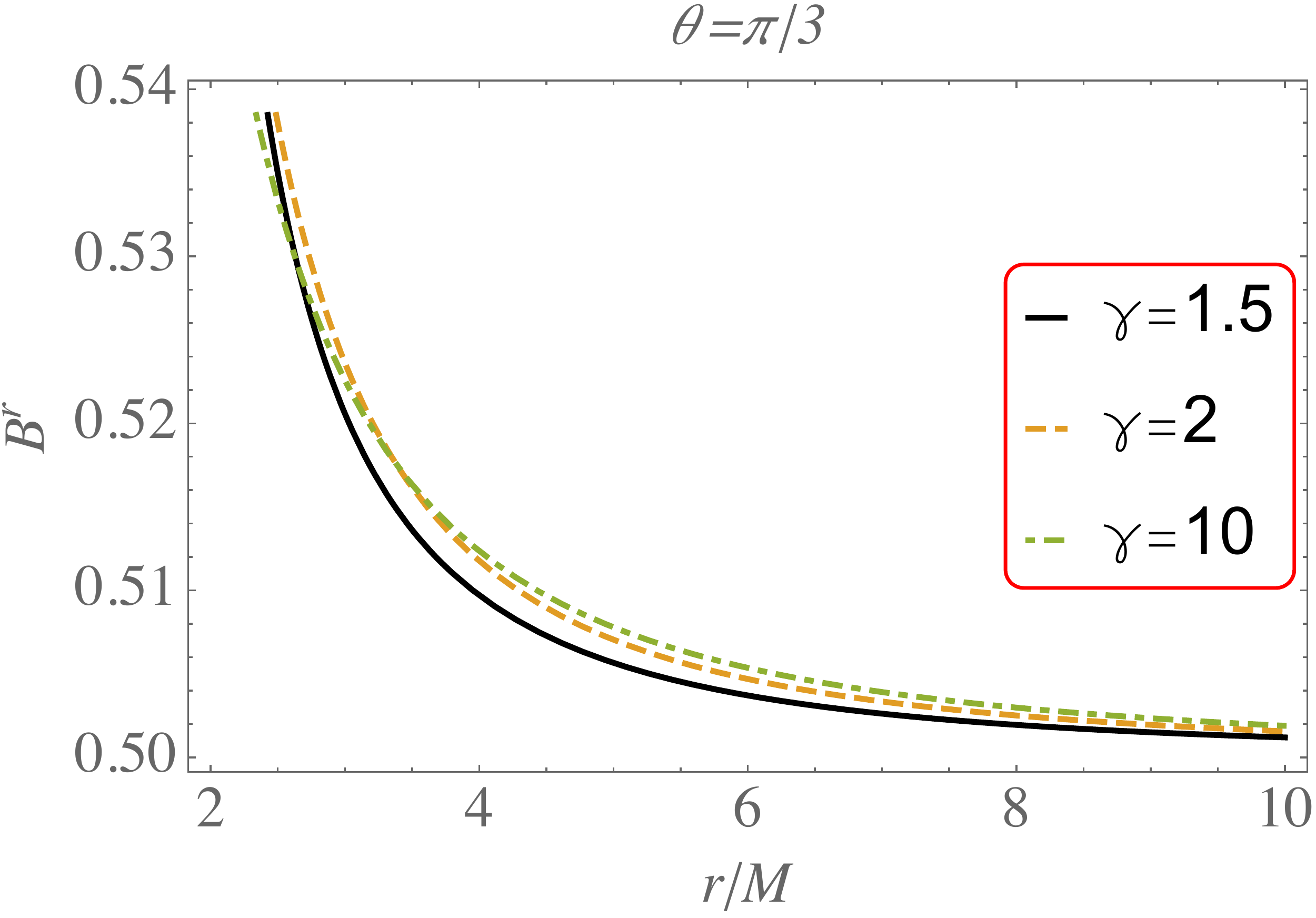}
	
	\includegraphics[width=0.48 \textwidth]{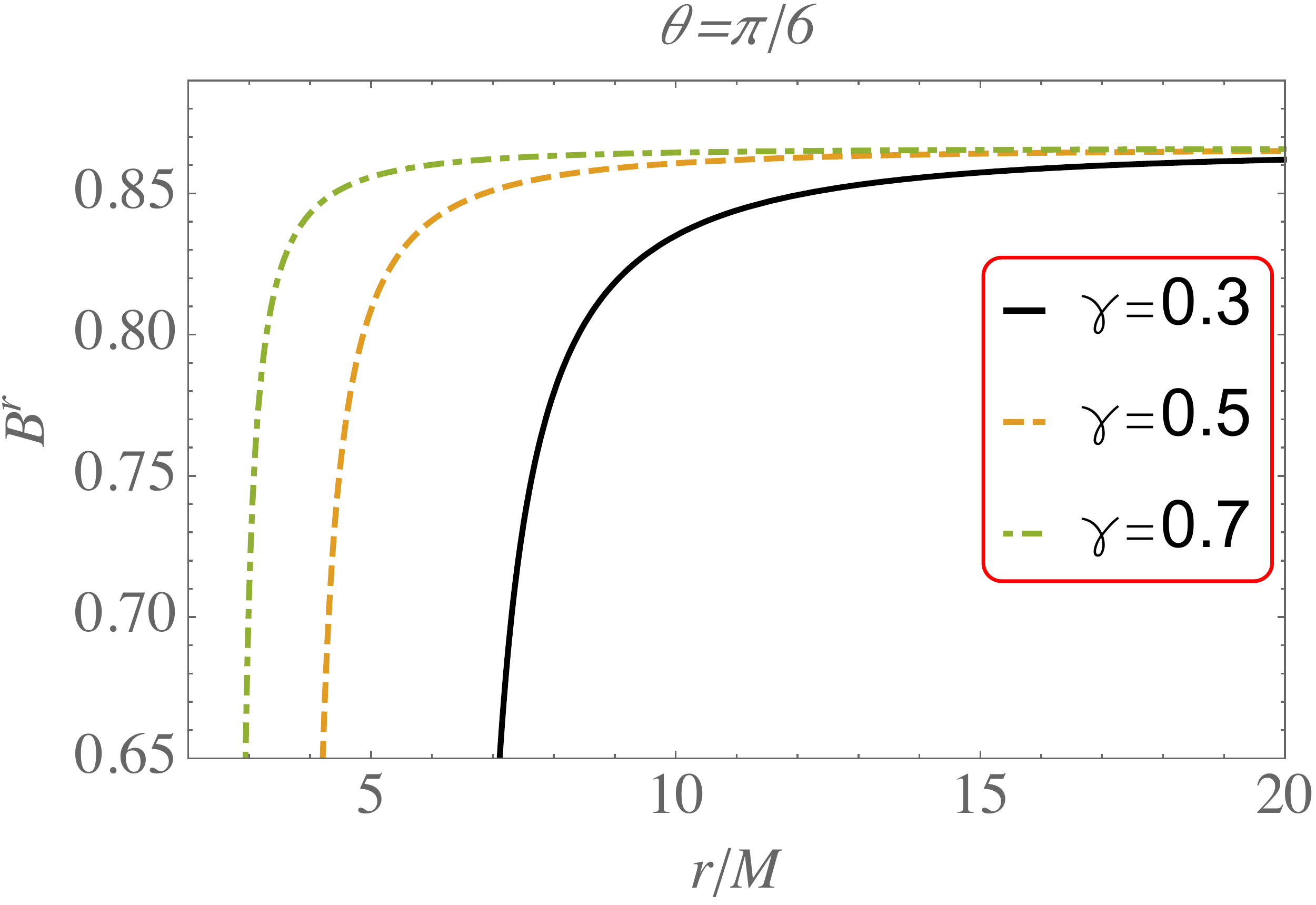}
	\includegraphics[width=0.48 \textwidth]{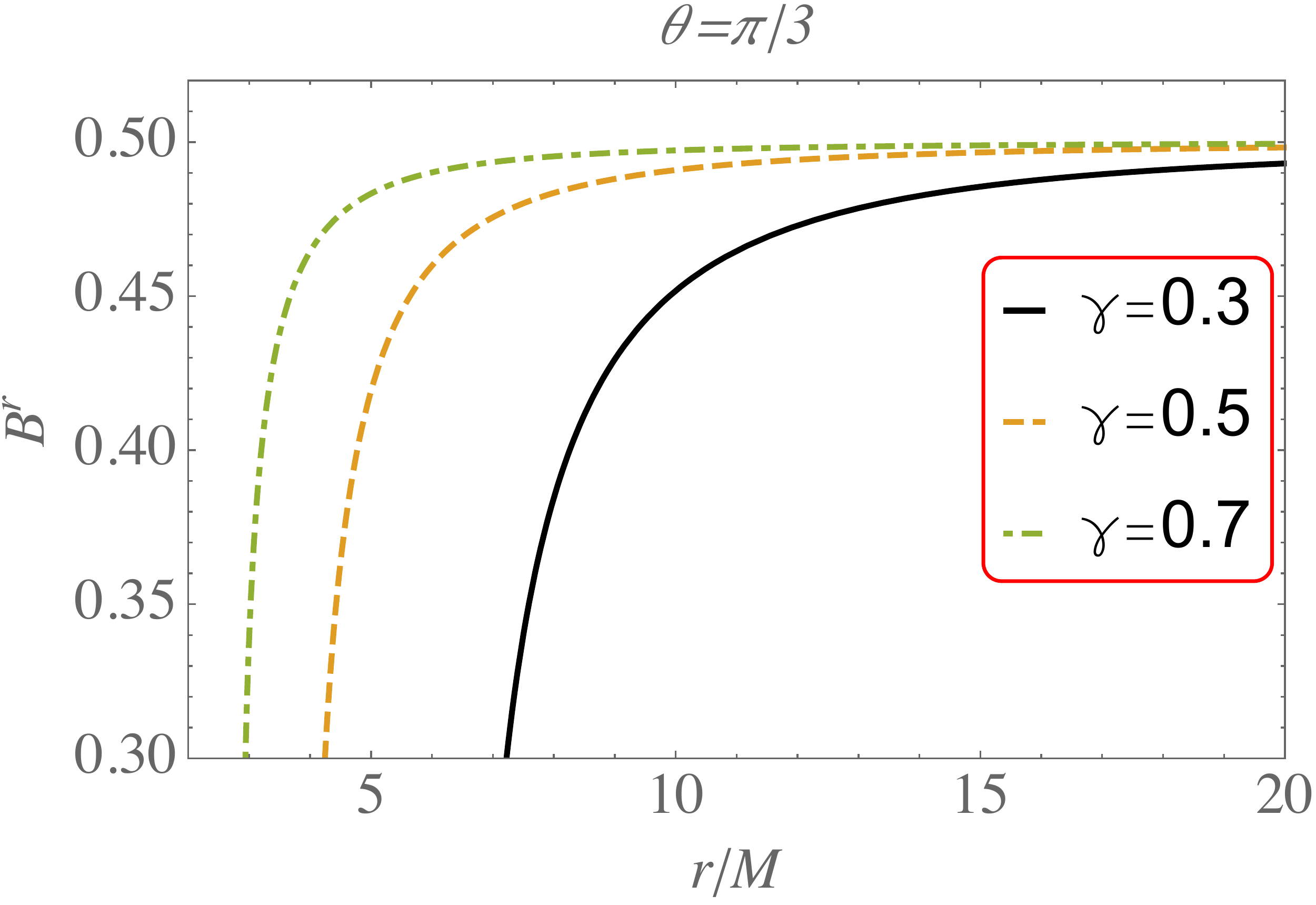}

	\caption{\label{figem2} Radial dependence of the radial component of the magnetic field in $\gamma$ spacetime for  different values of $\gamma$ and the azimuthal angle $\theta$. }
\end{figure*}

\begin{figure*}

	\includegraphics[width=0.48 \textwidth]{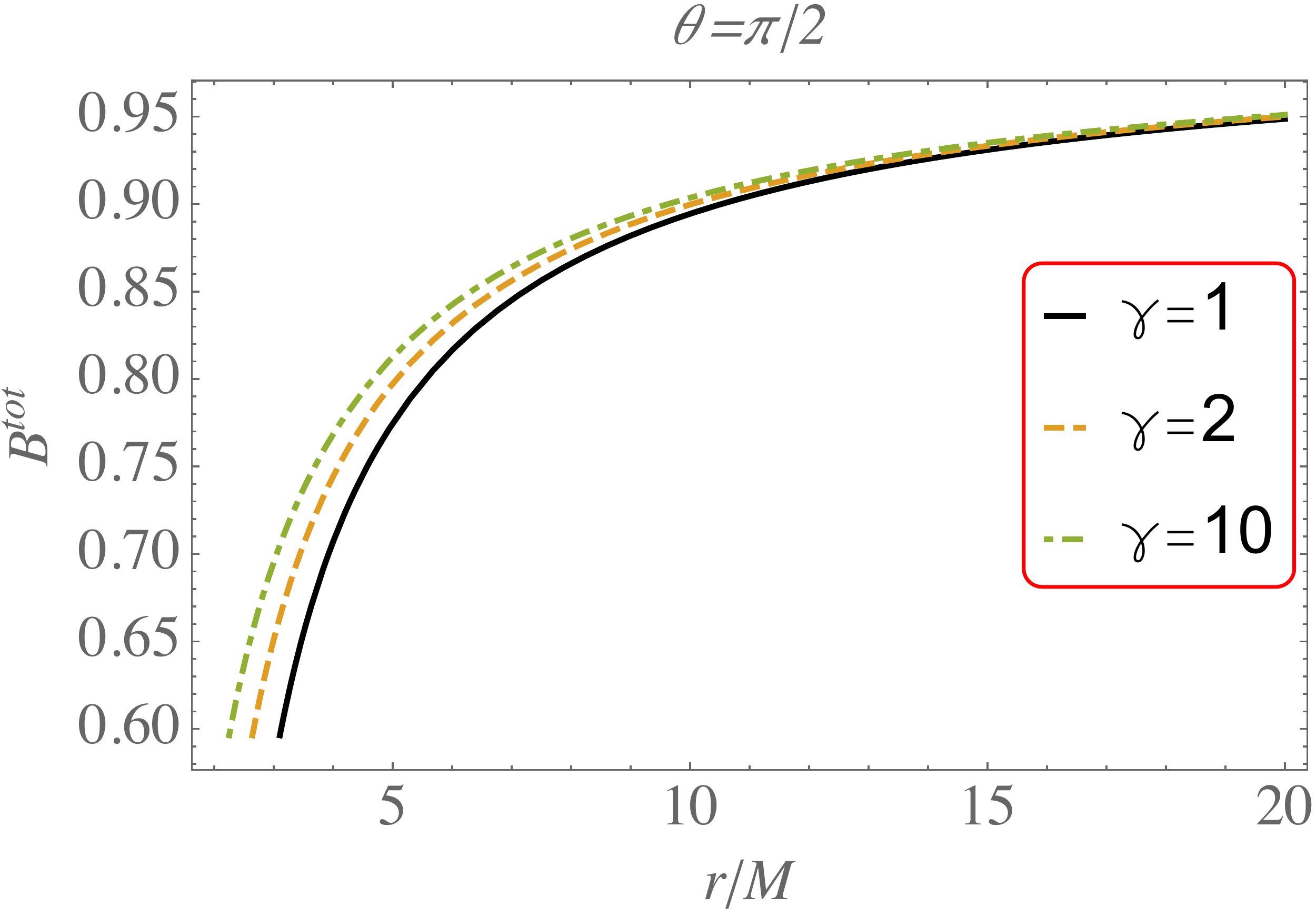}
	\includegraphics[width=0.48 \textwidth]{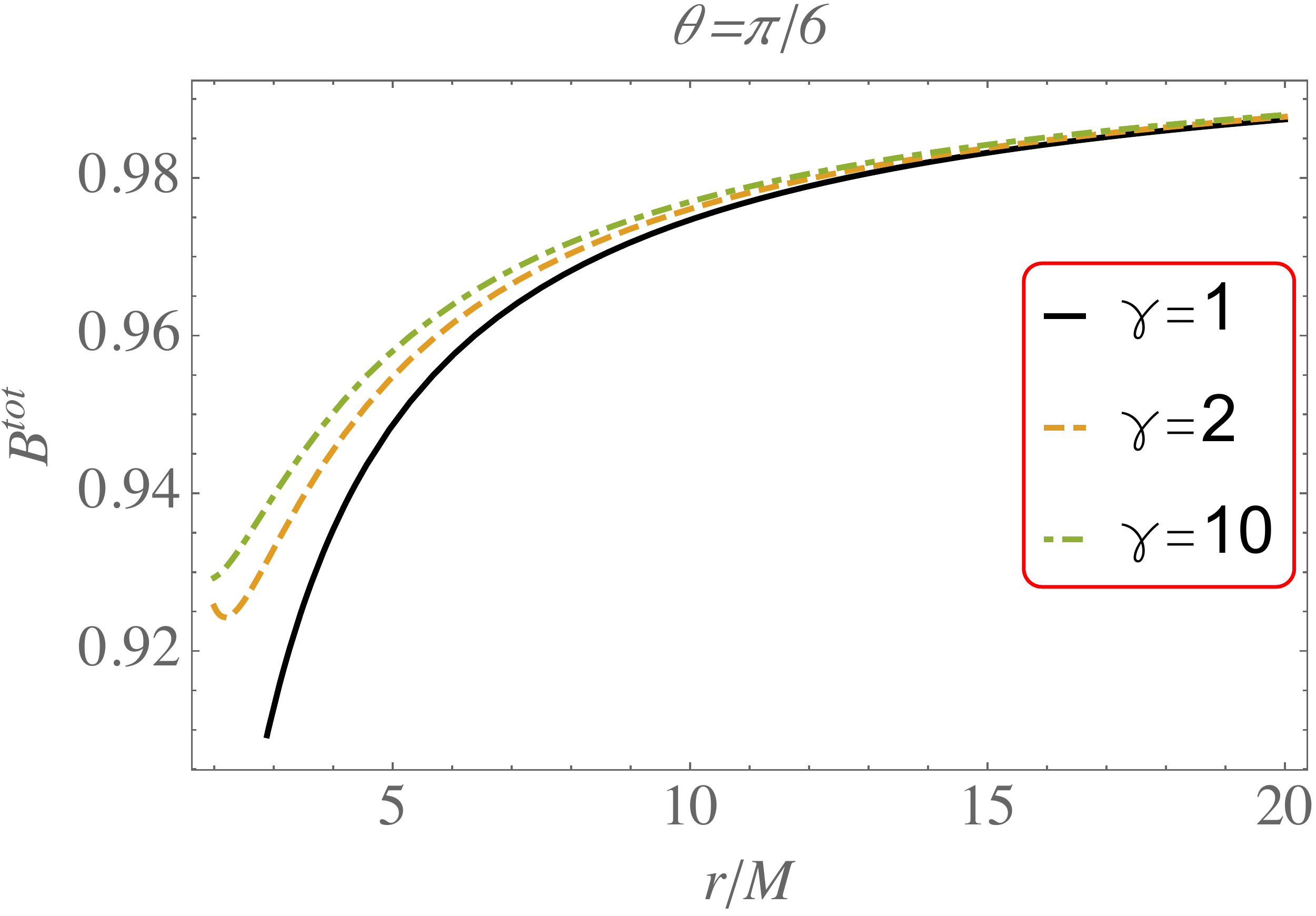}
	
	\includegraphics[width=0.48 \textwidth]{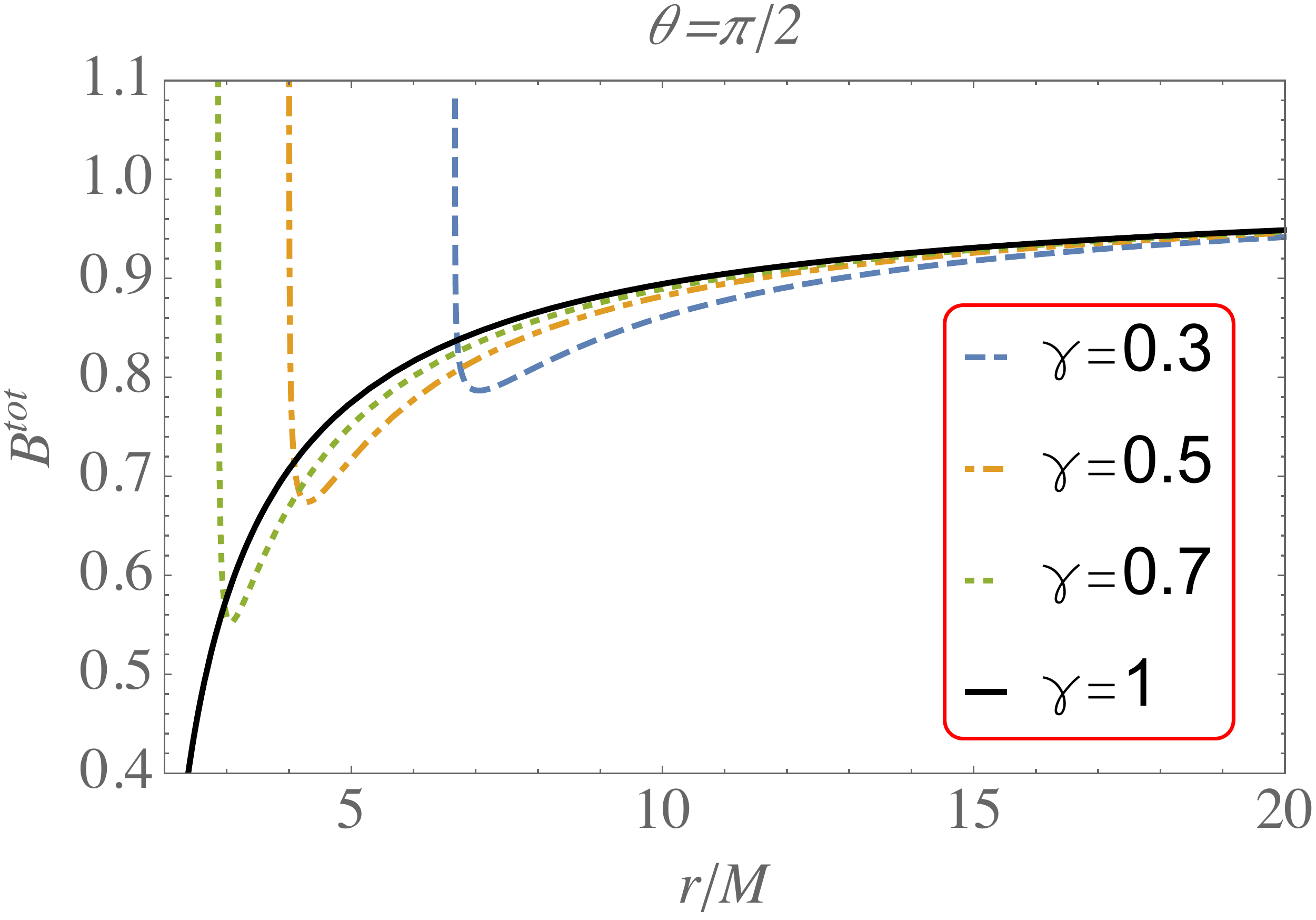}
	\includegraphics[width=0.48 \textwidth]{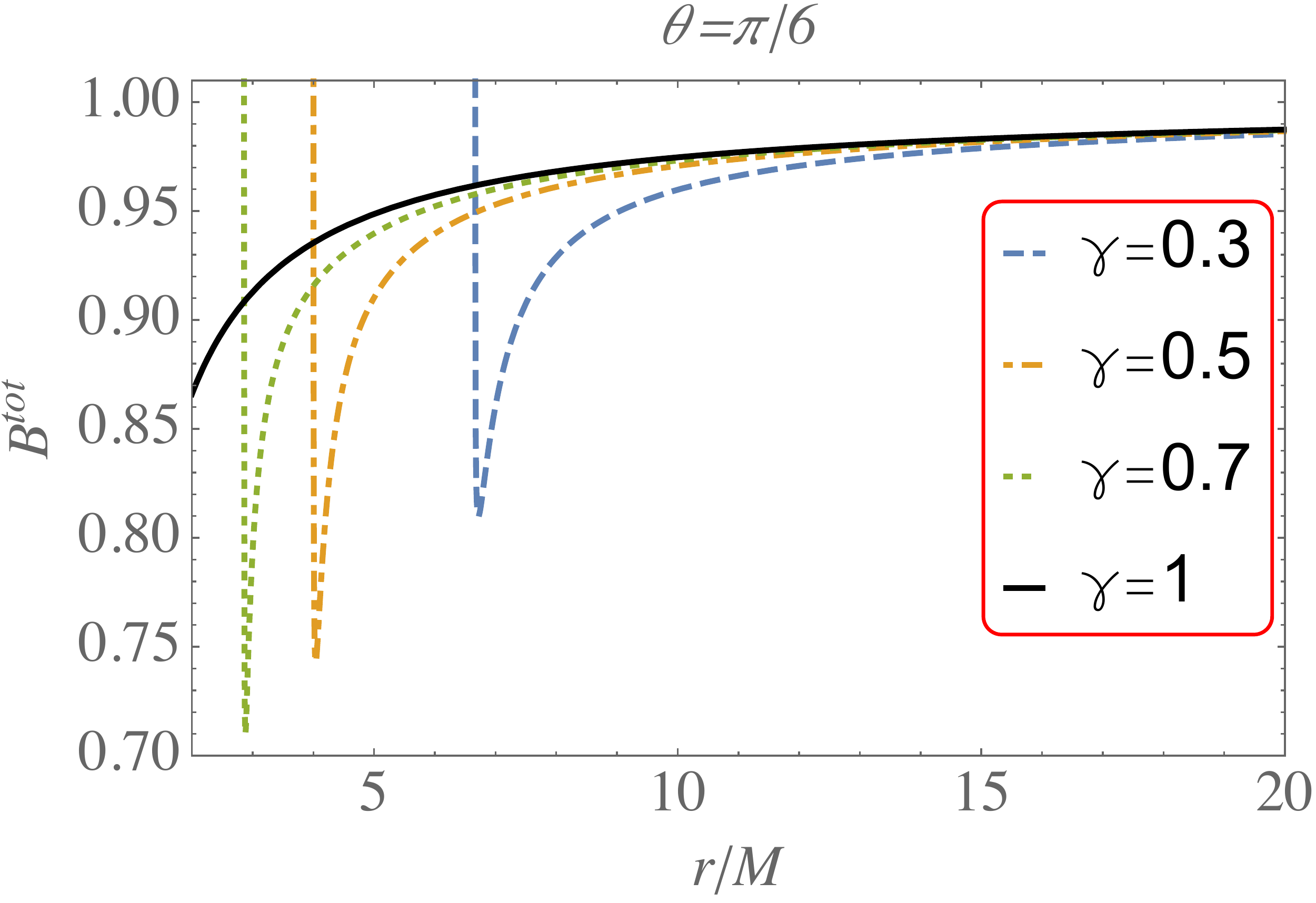}

	\caption{\label{figem2} Radial dependence of the total magnetic field $B^{\rm tot} = \sqrt{(B^{r})^2+(B^{\theta})^2}$ in $\gamma$ spacetime for different values of $\gamma$ and the azimuthal angle $\theta$. }
\end{figure*}

One can easily check that in the limiting case when $m/r\rightarrow 0,$ the nonzero components of the magnetic field coincide with their Newtonian values: $B^{\hat{r}} = B\cos\theta, \ \ B^{\hat{\theta}}=B\sin\theta$. In Figs.~\ref{figem}-\ref{figem2} the radial dependence of the components of the magnetic field in the $\gamma $ spacetime have been shown. From the figures one can easily see that the magnetic field components are not affected by the value of the parameter $\gamma $ towards asymptotic infinity, and they tend to the Newtonian limits. 
 The difference in the magnetic field strength becomes more pronounced as the radius decreases. With increasing the value of $\gamma$ one can observe an increase of the absolute value of the magnetic field components. In the case when $\gamma<1$ one can see that the azimuthal component of the magnetic field does not cross the surface of infinite redshift $r=2m$, which is similar to the Meissner effect, when the magnetic field does not penetrate the surface of a superconductor.

\section{Charged particle motion \label{chparmot} }

In this section, we will consider the motion of charged particles in the $\gamma$ spacetime in the presence of the external magnetic field described above. For this purpose we will use the Hamilton-Jacobi equation 
which in our case can be written as
\begin{eqnarray}
g^{\alpha\beta} \left(\frac{\partial S}{\partial x^{\alpha}} +e A_\alpha\right) \left(\frac{\partial S}{\partial x^{\beta}} +e A_\beta\right) = -m_0^2\ , \label{hamjam}
\end{eqnarray}
where $m_0$ and $e$ are the mass and the electric charge of the test particle, respectively. Due to existence of the time-like and space-like Killing vectors, the action for the test particle in Eq.~(\ref{hamjam}) can be written as 
\begin{eqnarray}
S=-{\cal E} t + {\cal L} \phi + S_{\rm r \theta} (r, \theta)\ , \label{action}
\end{eqnarray}
where the conserved quantities $\cal E$ and $\cal L$ are understood as representing the energy and angular momentum of the test particle, respectively, and $ S_{\rm r \theta} (r, \theta)$ is the part of the action depending on $r$ and $\theta$ coordinates. 

For the line element (\ref{metric}) the Hamilton-Jacobi equation (\ref{hamjam}) with the action (\ref{action}) is, in general, not separable. However, one can easily find the equation of motion for the charged particle for the particular case where the motion is restricted to the equatorial plane $\theta= \pi/2 $. In this case we obtain
\begin{eqnarray}
\frac{dt}{d\tau}&=&\frac{{\cal E}}{F} \ ,\\
\left(\frac{dr}{d\tau}\right)^2&=&F \left(\frac{r^2 -2 m r }{r^2-2m r+m^2}\right)^{1-\gamma ^2} \nonumber\\
&&\times\bigg[\frac{{\cal E}^2}{F} -\frac{F}{r^2-2 mr } \left({\cal L}+ \frac{r^2-2 mr }{2F}\beta \right)^2-1\bigg] ,\ \ \ \ \ \\
\frac{d\phi}{d\tau}&=&\frac{F}{r^2-2 m r} {\cal L}+\frac{1}{2} \beta \ , 
\end{eqnarray}
where ${\cal E}$ and ${\cal L}$ are now normalized to the mass of the particle and $\beta= eBm/m_0$ is the magnetic parameter related to the interaction of the charged particle with the magnetic field. One may define the effective potential $V$ for the radial motion of the charged particle from
\begin{eqnarray}
 \left(\frac{r^2 -2 m r }{r^2-2m r+m^2}\right)^{\gamma ^2-1}\left(\frac{dr}{d\tau}\right)^2+V^2={\cal E}^2\ .
\end{eqnarray}
and the explicit form of $V$ becomes
\begin{eqnarray}
V^2&=&F+\frac{F^2}{r^2-2 mr} \left({\cal L}+ \frac{r^2-2 mr }{2F}\beta \right)^2\ .
\end{eqnarray}

\begin{figure*}
	\includegraphics[width=0.45 \textwidth]{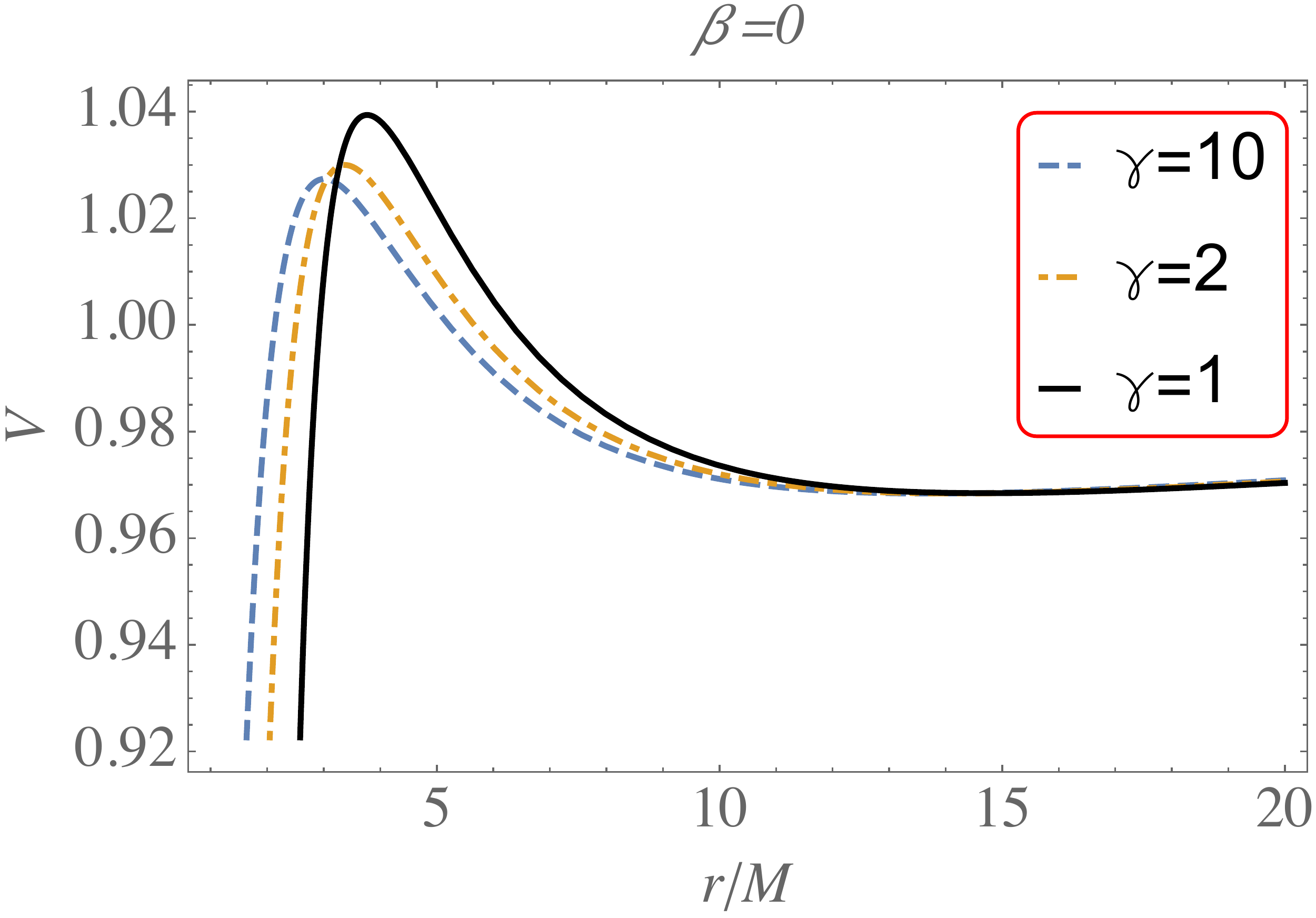}
	\includegraphics[width=0.45 \textwidth]{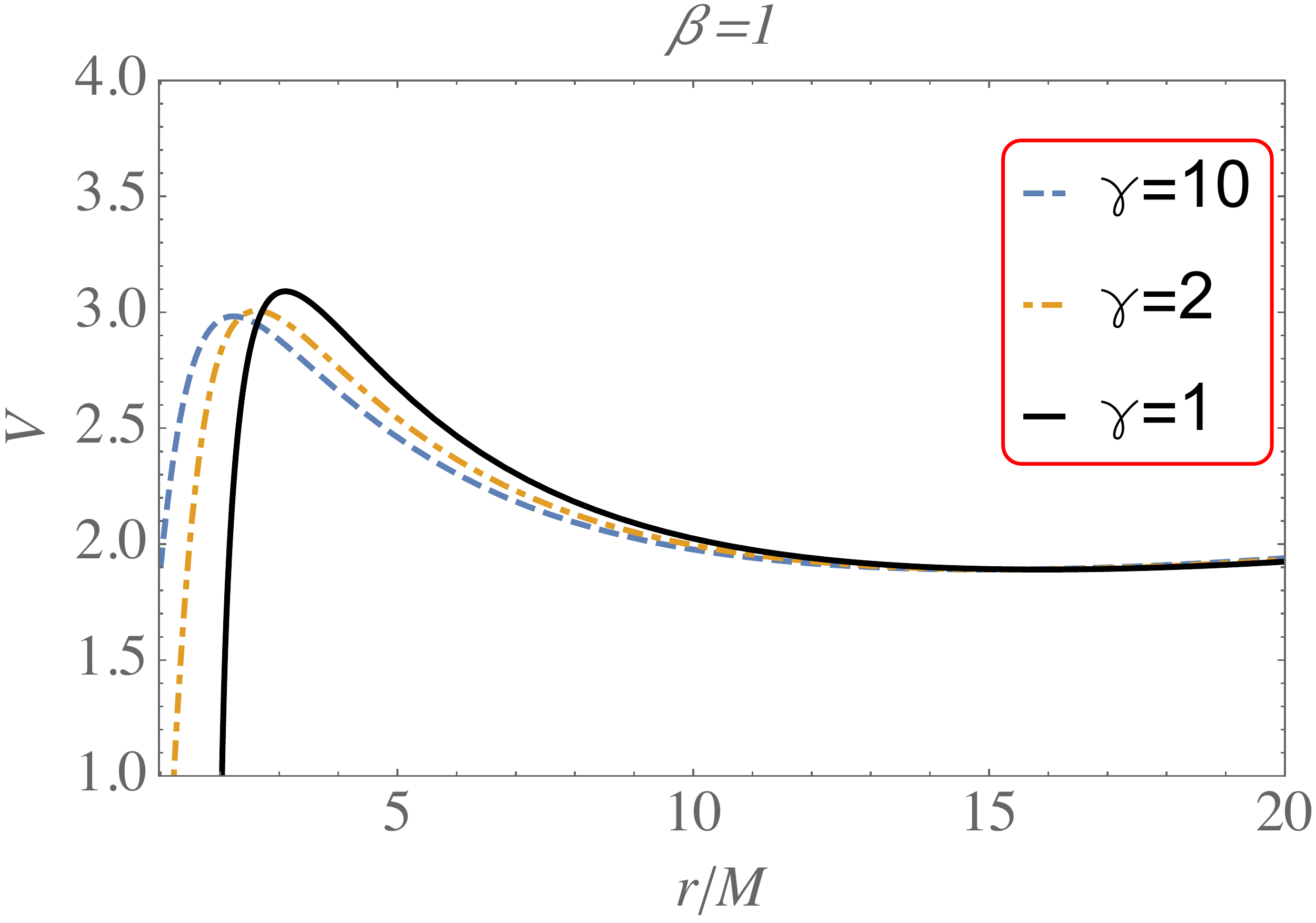}

	\includegraphics[width=0.45 \textwidth]{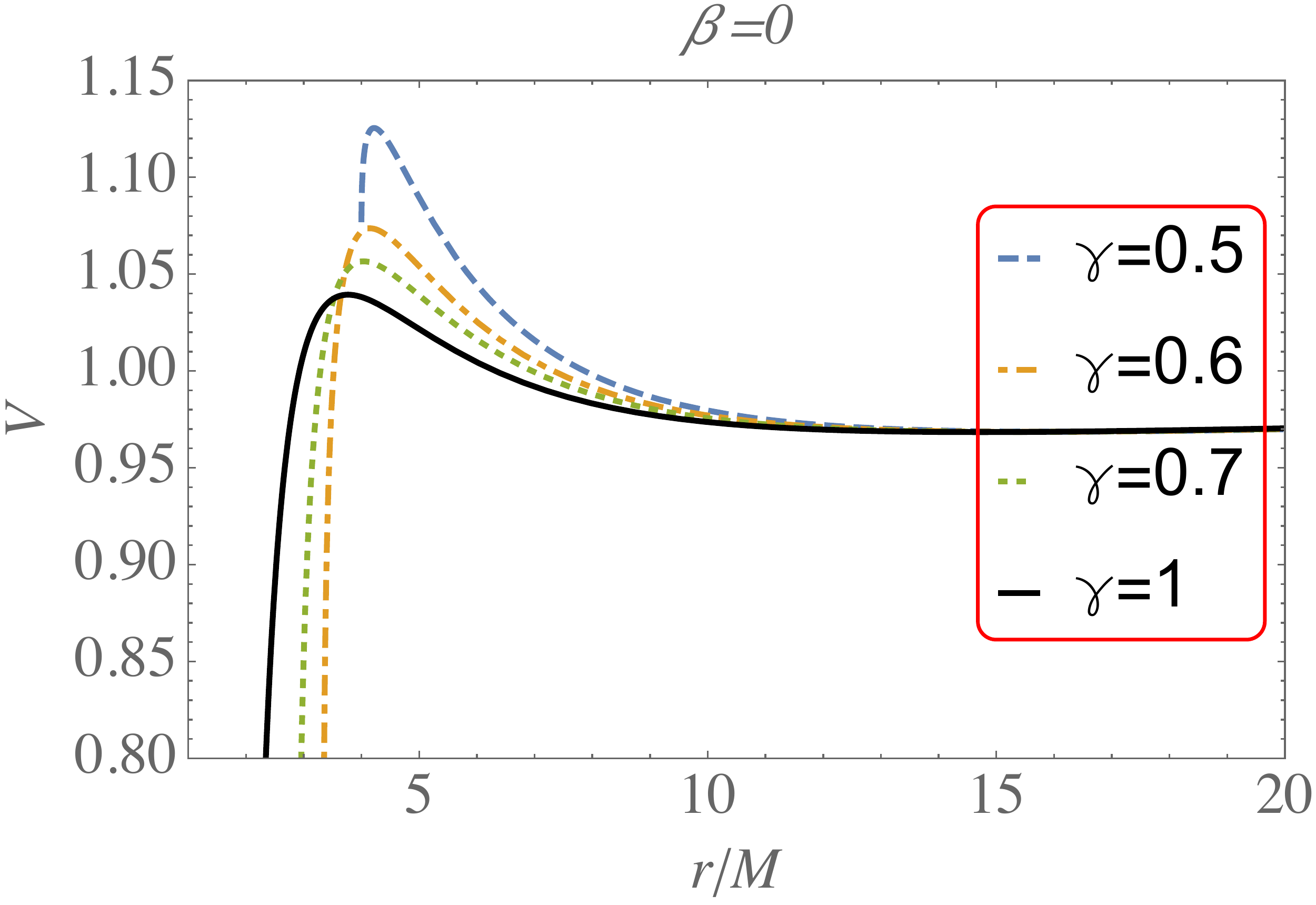}
	\includegraphics[width=0.45 \textwidth]{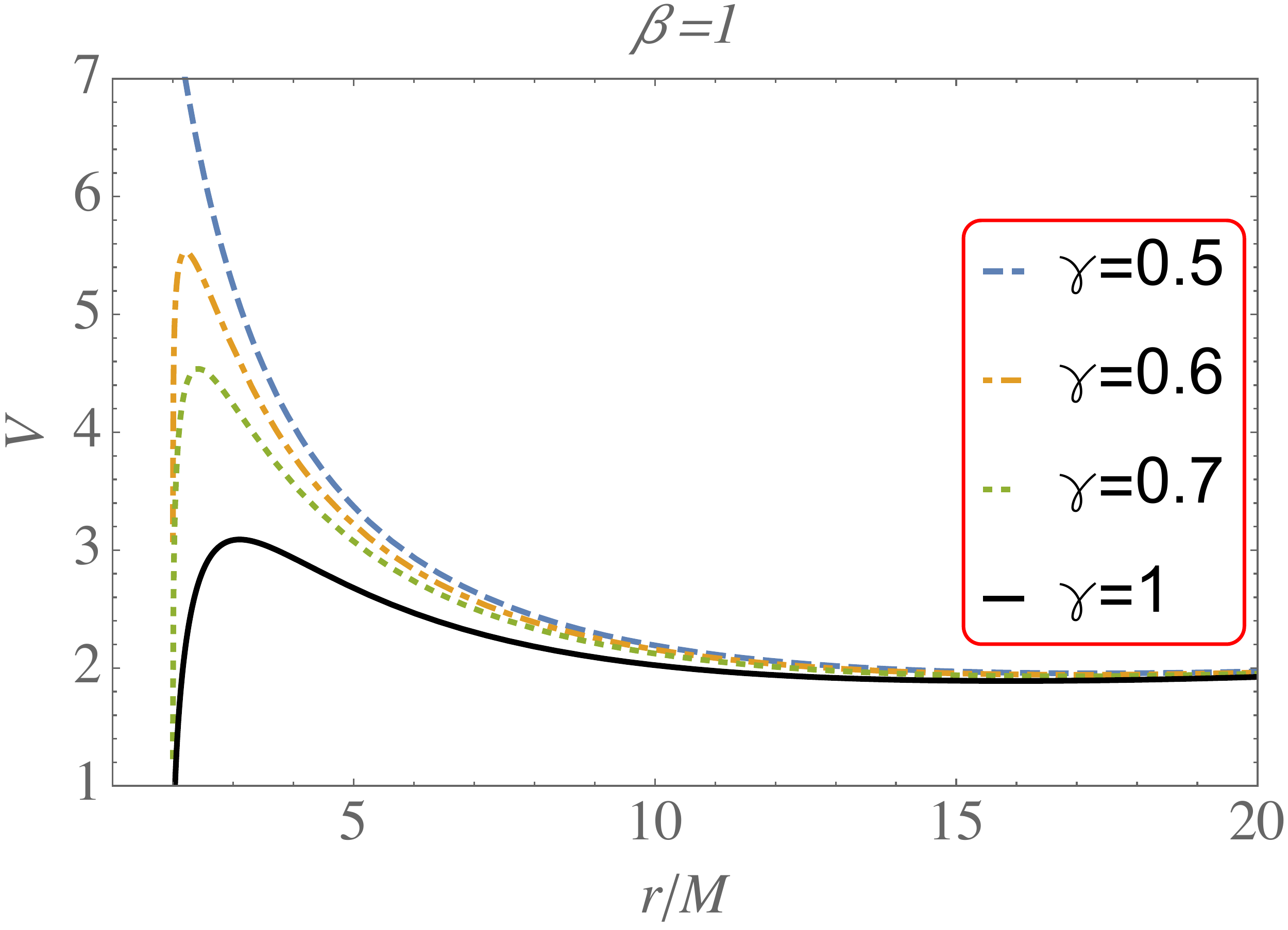}
	
\includegraphics[width=0.45 \textwidth]{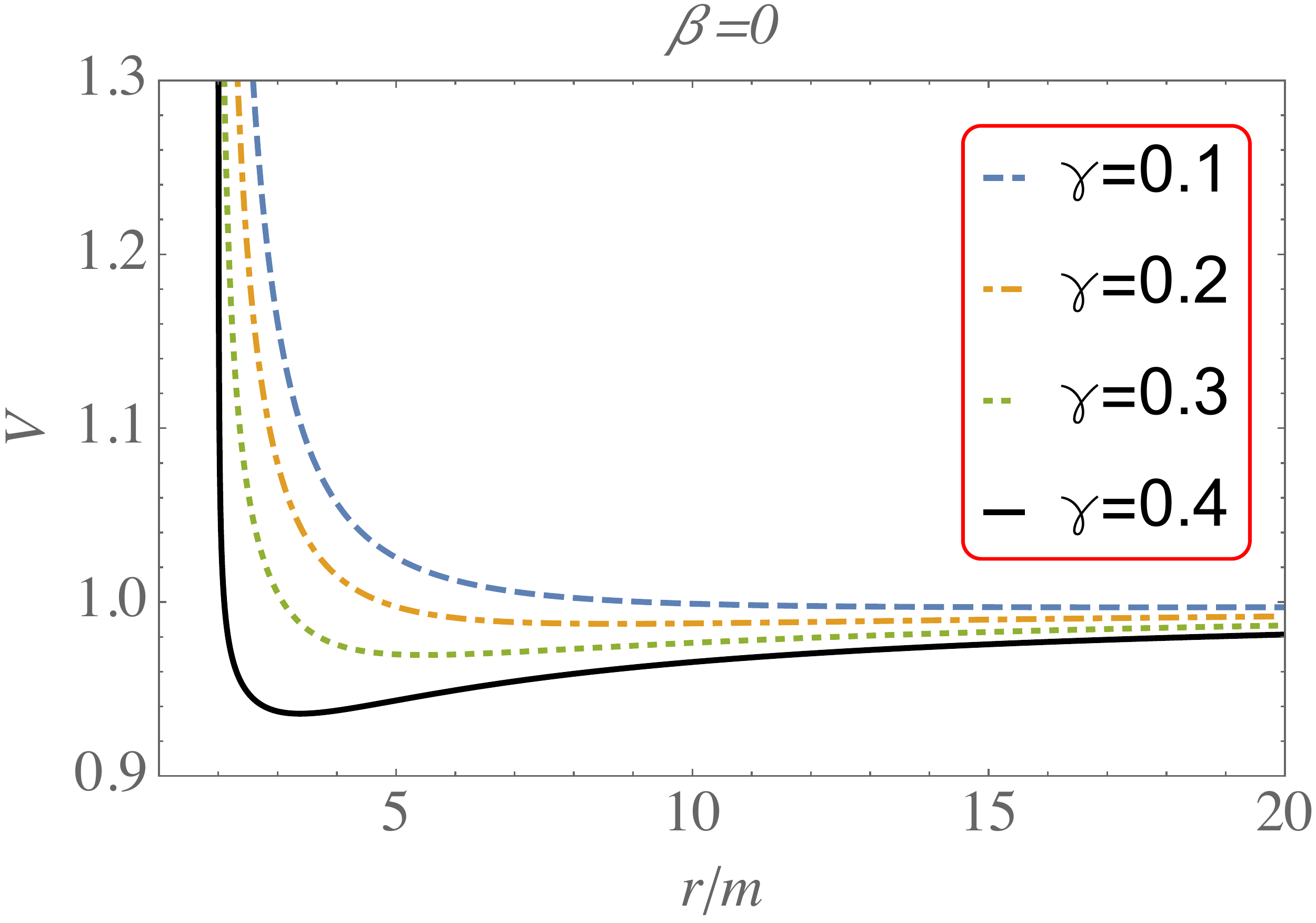}
\includegraphics[width=0.45 \textwidth]{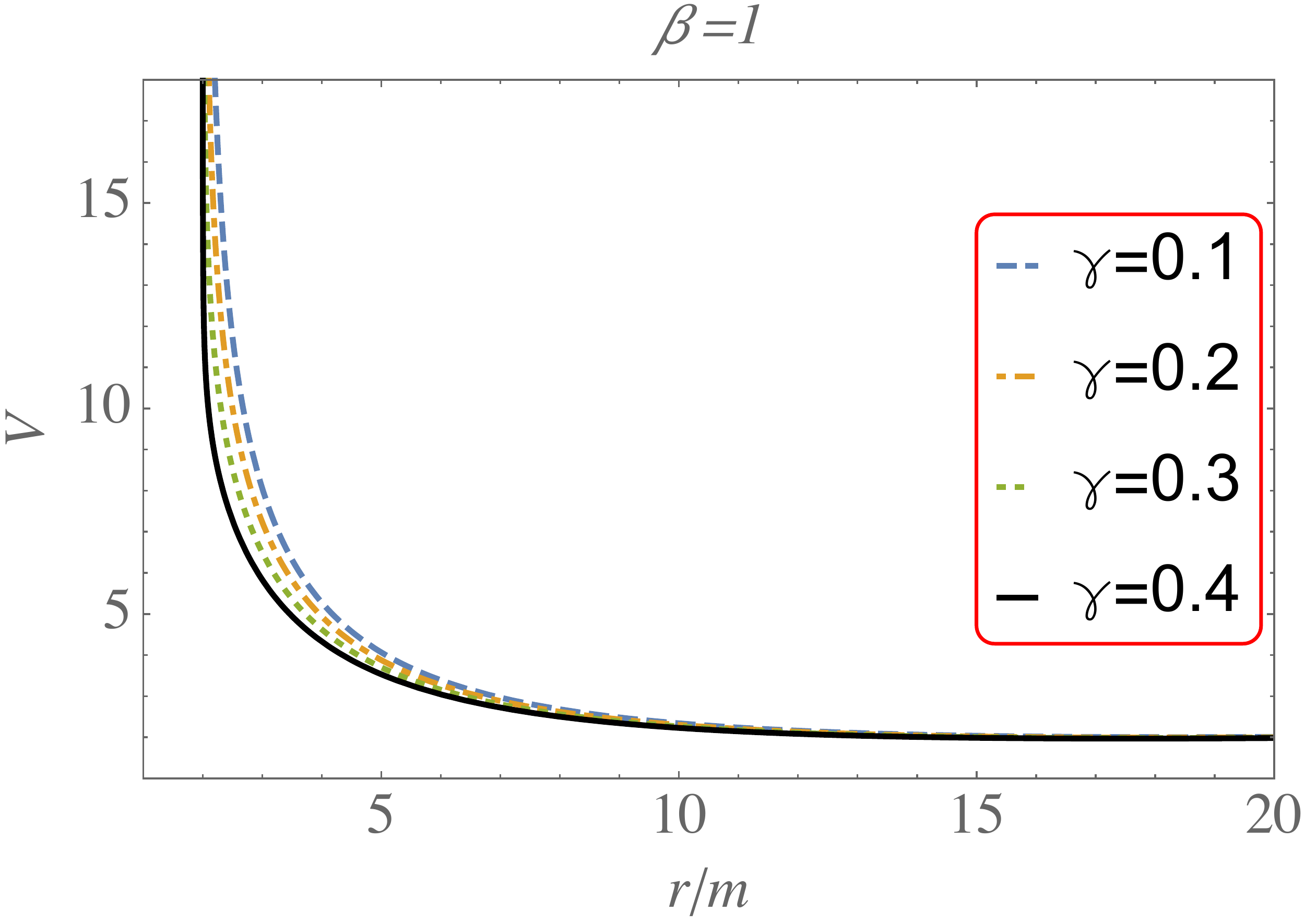}

	\caption{Radial dependence of the effective potential for a charged particle on the equatorial plane for different values of $\gamma$ and magnetic parameter $\beta$. \label{effpot}}
\end{figure*}

The radial dependence of the effective potential of the charged particle's motion is shown in Fig.~\ref{effpot}. As expected at large distance from the Cauchy horizon the effect of the magnetic field becomes dominant and the latter starts to play the role of barrier preventing the charged particles from falling towards smaller radii from infinity. Near the surface $r=2m$ the effects of deformation become more significant, and with increasing value of the $\gamma$ parameter the effective potential decreases and stable bound orbits starts to become unstable.

\begin{figure*}
	\includegraphics[width=0.45 \textwidth]{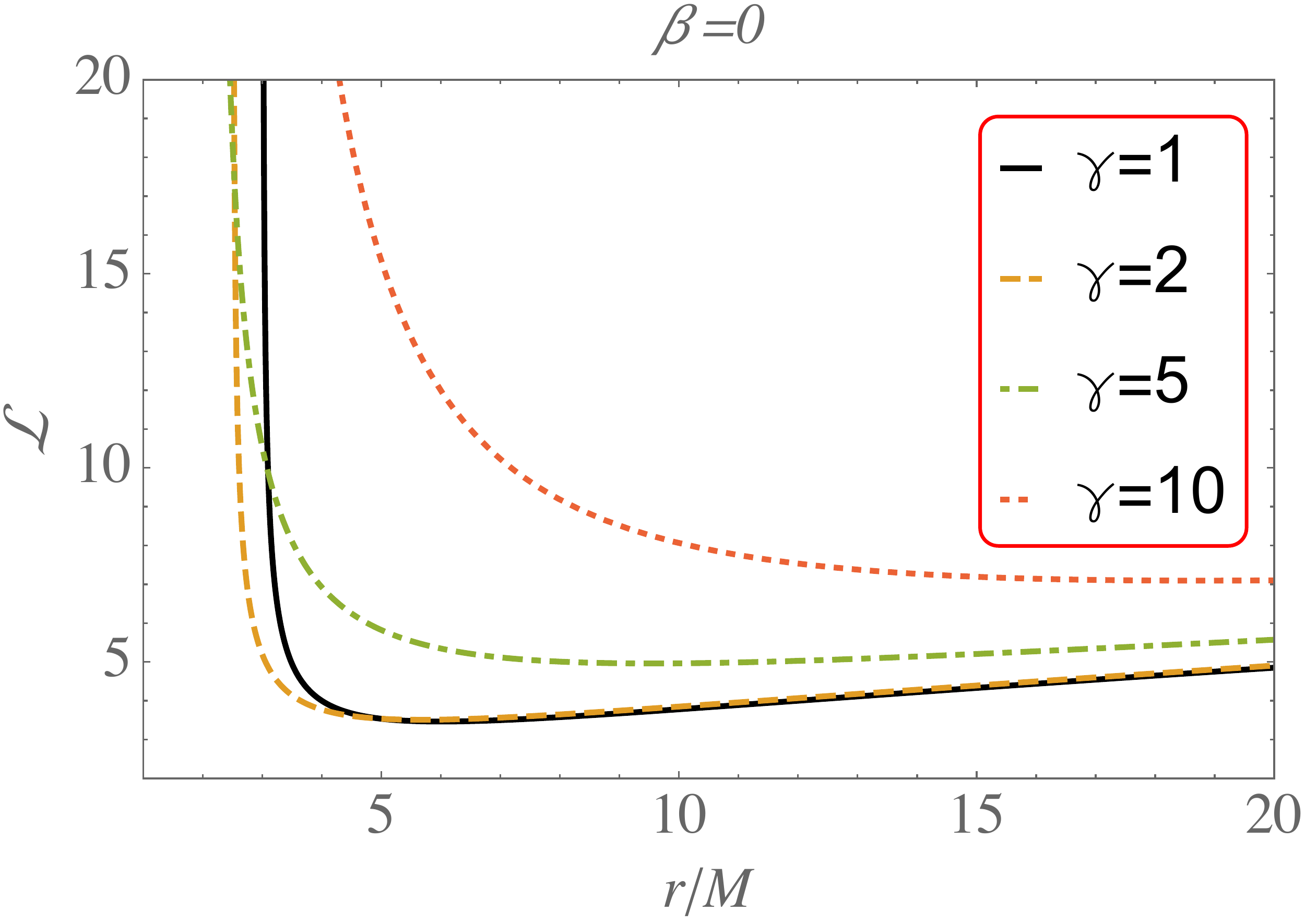}
	\includegraphics[width=0.45 \textwidth]{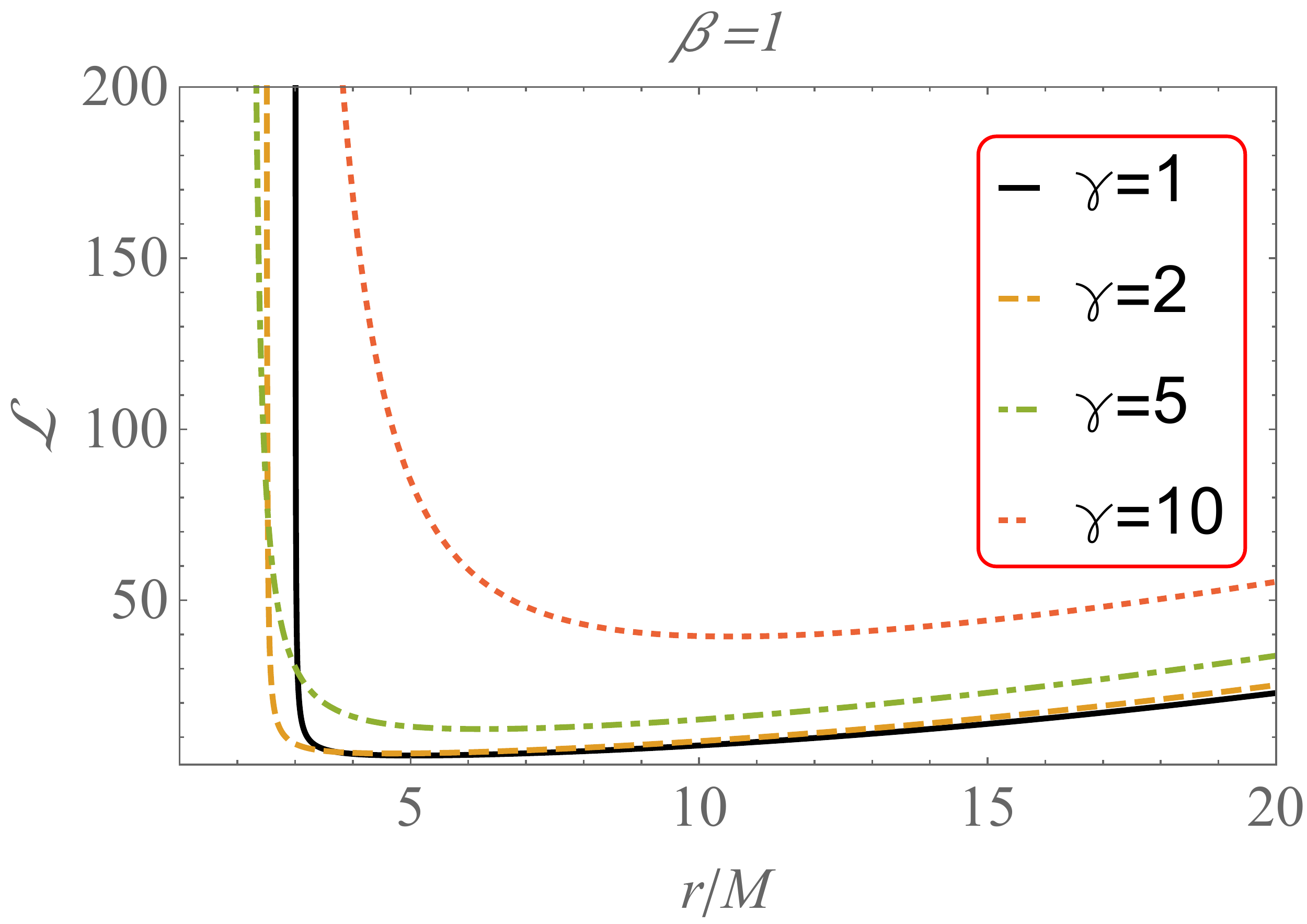}
	
	\includegraphics[width=0.45 \textwidth]{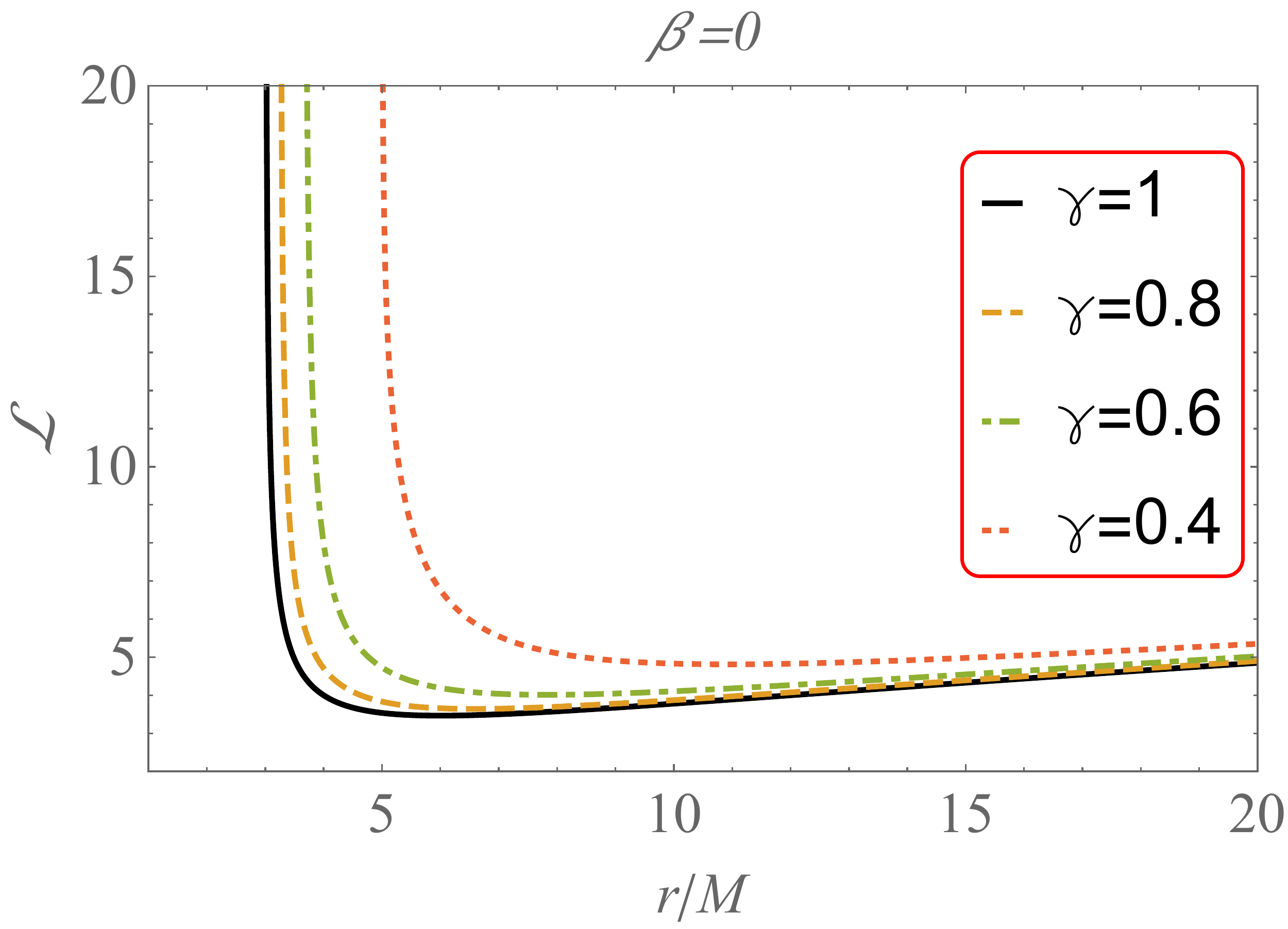}
	\includegraphics[width=0.45 \textwidth]{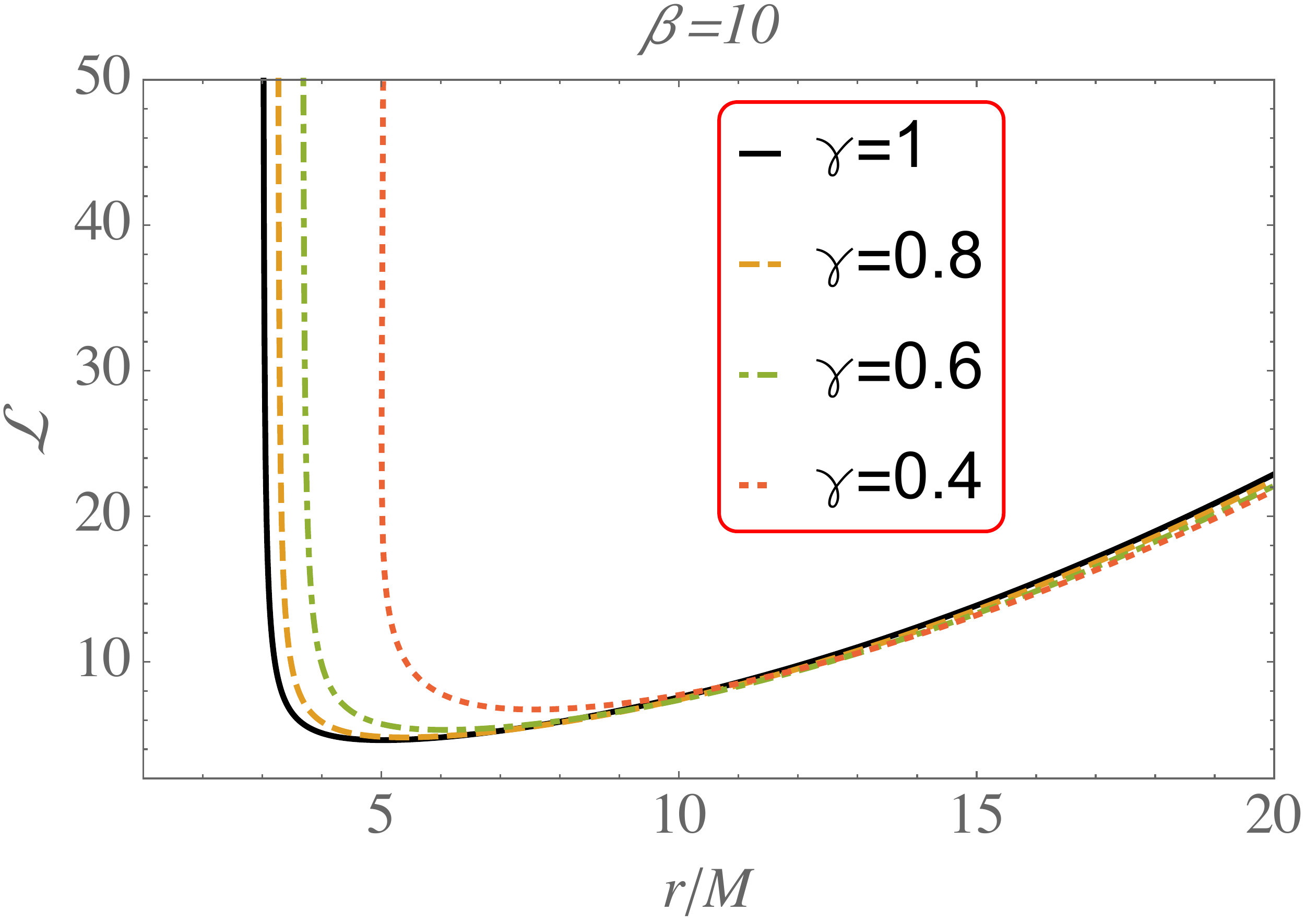}
	
	\caption{Radial dependence of the angular momentum of the charged particle moving along circular orbit on equatorial plane for different values of $\gamma$ and magnetic parameter $\beta$. \label{angmom}}
\end{figure*}

Now we shall consider particles moving on circular orbits. In order for circular orbits to exist two conditions are required, namely ${\cal E}^2-V^2=0$ and $V'(r)=0$, where $'$ denotes the derivative with respect to the radial coordinate. Using these conditions one can find the values of the energy ${\cal E}$ and angular momentum ${\cal L}$ for charged particles in circular orbits. These are 
\begin{eqnarray}\label{lcc}
{\cal L}_{c \pm}&=	& \frac{r^2 F^{1-\gamma }}{2 (r-2 \gamma  m-m)} \bigg(\beta  \gamma  m\pm\\&& \sqrt{\beta ^2 (r-\gamma  m-m)^2+\frac{4 \gamma  m F^{\gamma } (r-2 \gamma  m-m)}{r^2-2 m r}}\bigg)\nonumber,\\
{\cal E}_{c \pm}&=&V (r,\ {\cal L}_{c\pm} )\ .\label{ecc}
\end{eqnarray}
In (\ref{lcc})-(\ref{ecc}) the sign $\pm$ corresponds to the different orientation of magnetic field or/and different sign of the electric charge of test particle. This means that on the same circular orbits a charged particle, in principle, may have two different sets of energy and angular momentum, depending on the sign of the particle's charge. In Fig.~\ref{angmom} we show the radial dependence of the angular momentum of charged particles in circular orbits for different values of the $\gamma$ parameter. One can see that with increasing value of $\gamma$ the angular momentum of a test particle at a fixed circular orbit  increases while the radius of minimal circular orbits decreases.

Note that the left and right panels of Figs. \ref{effpot} and~\ref{angmom} correspond to vanishing and nonvanishing external magnetic field, respectively. In the presence of magnetic field the value of effective potential increases (see Fig.~\ref{effpot}.). This corresponds to the case when magnetic field starts to play a role of a potential barrier for charged particles and Lorentz force changes the direction of radially infalling charged particles. Similarly the presence of a magnetic field increases the value of angular momentum of charged particles on circular orbits (see Fig.~\ref{angmom}.). Charged particles need higher angular momentum in order to stay in circular orbits in the presence of a magnetic field.

\begin{figure}
	\includegraphics[width=0.49 \textwidth]{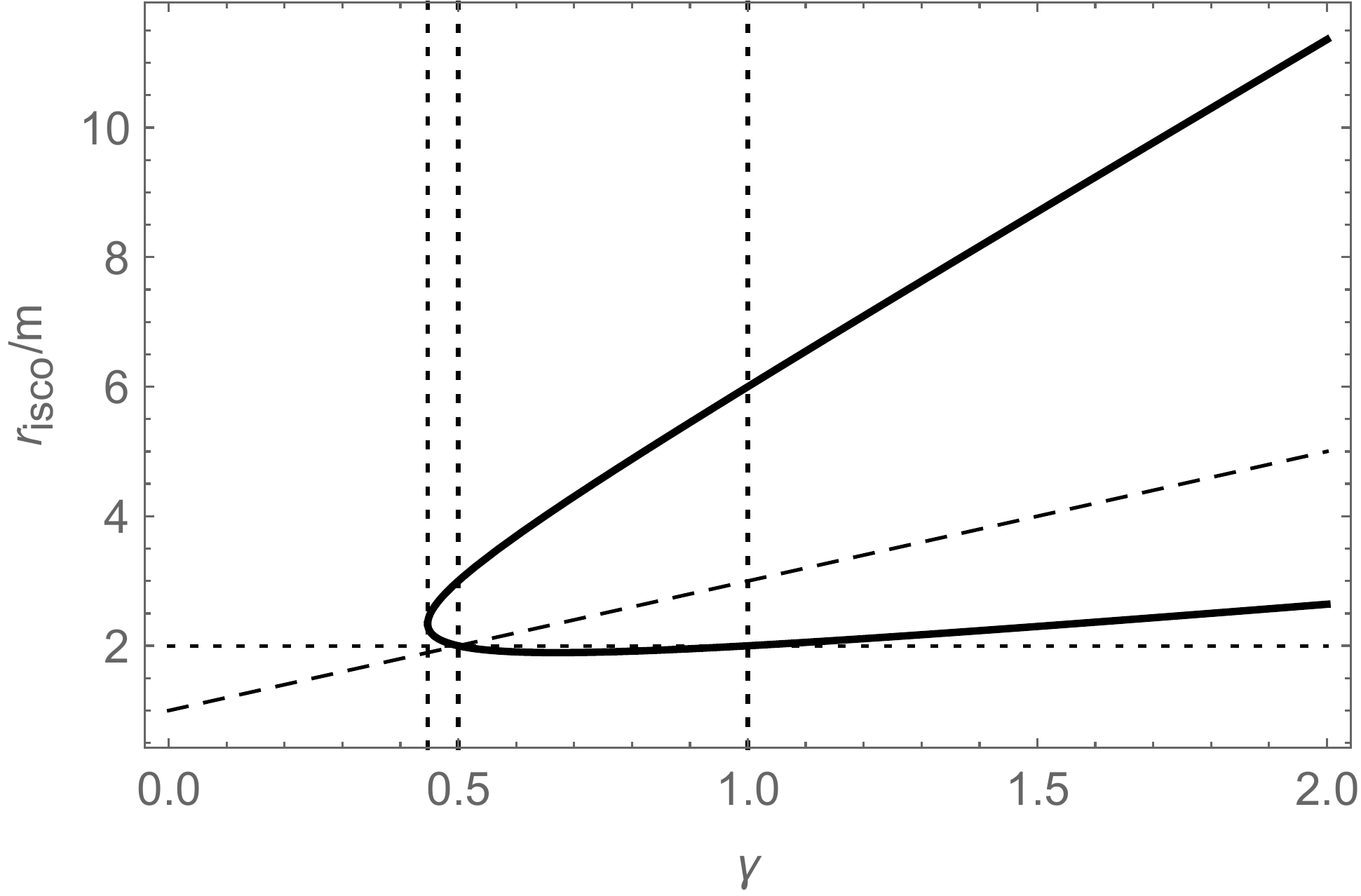}
	
	\caption{Value of the ISCO for neutral particles on circular orbits in the equatorial plane for different values of $\gamma$. Vertical dotted lines define the borders of regions with different ISCO structure (see text for details). The corresponding values of $\gamma$ are: $\gamma= 1/\sqrt{5};\ 1/2;\ {\rm and}\ 1 $. The horizontal dotted line corresponds to the radius of singular surface $r=2m$. The dashed line corresponds to the radius of photon sphere. \label{isconull}}
\end{figure}

\begin{figure*}

	\includegraphics[width=0.45 \textwidth]{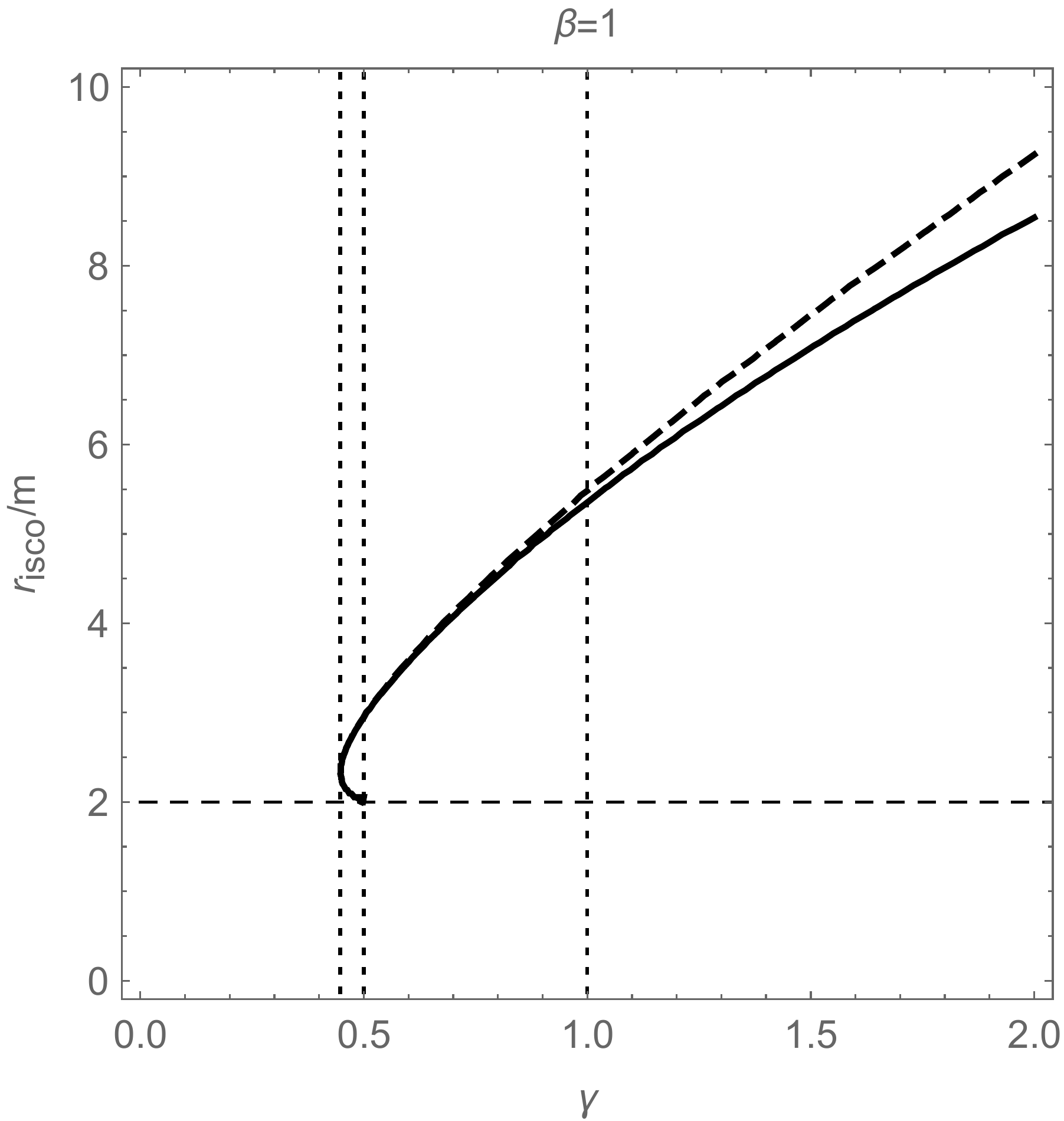}
	\includegraphics[width=0.45 \textwidth]{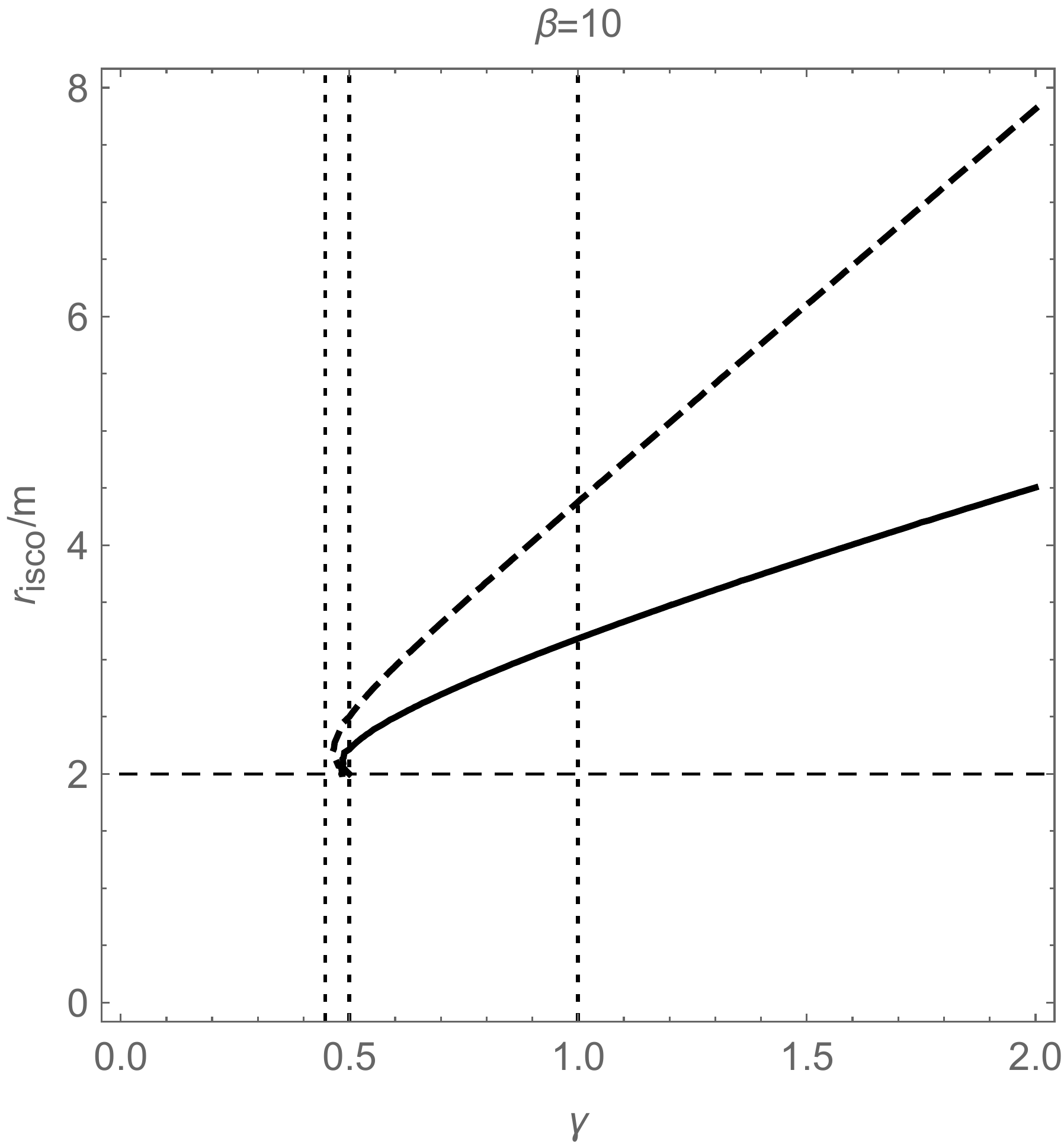}

	\caption{Value of the ISCO radius for charged particles on circular orbits in the equatorial plane for different values of $\gamma$ and different values of the magnetic parameter $\beta$. Vertical dotted lines are the borders of the regions, significant in the neutral case, corresponding the following values: $\gamma= 1/\sqrt{5};\ 1/2;\ {\rm and}\ 1 $. It can be seen that the values of $\gamma$ defining region (ii) change with $\beta$. The horizontal dotted line corresponds to the radius of singular surface $r=2m$. In the case of the motion of charged particles we have two separate curves for the location of the ISCO, depending on the sign of the charge of the particle (solid and dashed lines correspond to positive and negative charges, respectively). \label{isco}}
\end{figure*}

Circular orbits on the equatorial plane are used to describe particle's motion in accretion disks around astrophysical objects. In particular, concerning astrophysical black holes, the spectrum of light coming from accretion disks is often used to suggest the presence of a black hole candidate. One of the most important features of such accretion disks around black holes is the possible existence of an innermost stable circular orbit (ISCO), which determines the inner boundary of the disk. The study of the properties of the ISCO around exotic sources can then be used to determine if such sources can in principle be distinguished from black holes (see for example, \cite{Joshi11,Joshi14,Bambi13d}).
Therefore, as the next step we will consider the ISCO of the $\gamma$ metric, which can be defined as a inner edge of the solution of inequality $V''(r)>0$. In the case without an external magnetic field the ISCO was studied in \cite{Chowdhury12}. 
The dependence of the ISCO radius upon the $\gamma$ parameter for neutral particles is shown in Fig.~\ref{isconull}. Here we can distinguish four different regions depending on the values of $\gamma$: i) $\gamma<1/\sqrt{5}$, ii) $1/\sqrt{5}<\gamma<1/2$, iii) $1/2<\gamma<1$, and iv) $\gamma>1$. In Fig.~\ref{isconull} these regions are separated by vertical dotted lines. The dashed line corresponds to the border for the real and imaginary values of angular momentum. Above that dashed line the angular momentum of the particles remains real, whereas below the line the value of the latter is imaginary. The infinite redshift (singular)  surface is noted by the dotted line. In region (i) we have no ISCO, which means above the singular surface there exist stable circular orbits at every radius. In region (ii) we have two regions for the existence of stable circular orbits. Notice that the existence of a second region close to the singular surface seems to suggest the presence of repulsive effects that are capable of holding a particle on the circular orbit. In regions (iii) and (iv) we have only one region (above the solid line) with stable circular orbits, similarly to the black hole case. 
The numerical results for the ISCO radius depending on the value of $\gamma$ are shown in Fig.~\ref{isco} for different values of the magnetic parameter. From Fig.~\ref{isco} one can easily see that the radius of the ISCO decreases with increasing values of $\beta$. The dashed and solid curves in Fig.~\ref{isco} correspond to the '-' and '+' types of circular orbits, respectively (see Eq.~(\ref{lcc})).

\section{Particle collisions in the presence of magnetic field \label{energetics}}

In this section we consider collisions of charged and neutral particles in the vicinity of the Cauchy horizon of the $\gamma$ spacetime in the presence of a magnetic field. We shall consider several scenarios with particles collision: i) two neutral particles  -- when there is no effect due to the external magnetic field, ii) two charged particles and iii) one charged particle and one neutral particle, where there is no effect due to the magnetic field on the neutral particle's motion.  
In general, we calculate the center of mass energy of two colliding particles with four momenta $p_1^\alpha$ and $p_2^\alpha$. The energy of the center of mass of this system can be found from the relation
\begin{eqnarray}
E_{\rm cm}^2=-g_{\alpha \beta}\ p_{\rm tot}^\alpha\ p_{\rm tot}^{\beta} = 2 m_0^2 \left(1- g_{\alpha \beta} u_1^\alpha u_2^\beta\right)\ , 
\end{eqnarray}
where $u_{i} ^\alpha$ is the four velocity of the particle ($i=1, 2$) and the total momentum of the system is defined as 
\begin{eqnarray}
p_{\rm tot }^\alpha = p_1^\alpha + p_2^\alpha\ . 
\end{eqnarray}

Within this section for simplicity we consider particles with ${\cal E}=1$, corresponding to the situation where the particle is at rest at spatial infinity. For simplicity we also consider particles with the same rest mass
and the motion of particles restricted to the equatorial plane. 

\subsection{Collision of neutral particles with opposite angular momentum}

First we consider the case of two neutral particles, where both particles fall from infinity with opposite angular momentum. The four velocities of the particles take the form
\begin{eqnarray}\label{npvel}
u_{1,2}^\alpha &=& \bigg(\frac{1}{F},  \left(\frac{r^2 -2 m r }{r^2-2m r+m^2}\right)^{\frac{1-\gamma ^2}{2}}  \\ &&\times \sqrt{1 -F \left(1+\frac{F {\cal L}^2}{r^2-2 mr } \right)}, 0,\pm\frac{F}{r^2-2 m r} {\cal L} \bigg). \nonumber  
\end{eqnarray} 
The center of mass energy for this case becomes
\begin{eqnarray}\label{ecm1}
\frac{E_{\rm cm}^2}{4m_0^2} = 1+\frac{F}{r^2-2mr} {\cal L}^2\ .
\end{eqnarray}
Using the expression for the metric function $F$, we can rewrite Eq.~(\ref{ecm1}) in the form:
\begin{eqnarray}\label{ecm2}
\frac{E_{\rm cm}^2}{4m_0^2} = 1+\frac{(r-2m)^{\gamma-1}}{r^{\gamma +1}} {\cal L}^2\ .
\end{eqnarray}

From Eq.~(\ref{ecm2}) one can see that if we take the Schwarzschild limit when $\gamma=1$ we get the standard expression for $E_{\rm cm}$~\cite{Banados09} 
\begin{eqnarray}
\frac{E_{\rm cm}^2}{4m_0^2} = 1+\frac{{\cal L}^2}{r^2} \ ,
\end{eqnarray}
and in the limit of $r\rightarrow 2m$ we get a finite value for the center of mass energy. This is given by $E_{\rm cm} =2\sqrt{5} m_0$, when we take the critical value of the angular momentum for the particle to be captured by the central object ${\cal L}=4 m$. On the other hand, when $\gamma>1$ we observe that the second term on the right hand side of Eq.~(\ref{ecm2}) tends to zero when  $r\rightarrow 2m$ and we get $E_{\rm cm} =2 m_0$ irrespectively of the value of ${\cal L}$. Finally, in the case of $\gamma<1$ the center of mass energy of two colliding particles tends to infinity $E_{\rm cm} \rightarrow \infty$ when $r\rightarrow 2m$.

\subsection{Head on collision of charged particles at~circular orbits}

Now we consider the head-on collision of charged particles in circular orbits. The four velocity of a charged particle moving along a circular orbit has the following general form
\begin{eqnarray}\label{cpvel}
u^{\alpha}_{c} = \left(\Gamma F^{-1/2},0,0, \Gamma {v}/{r}\right)\ , 
\end{eqnarray}
where $v$ is the velocity of the charged particle at the radius $r$ and $\Gamma$ is the Lorentz factor. Using the normalization condition $u^\alpha u_\alpha=-1$ and $d\phi/d\tau=\Gamma v/r$, one can easily find the relations 
\begin{eqnarray}
\Gamma^2 &=& 1+ \frac{r^2-2 mr }{F} {\cal B}^2\ ,\\
v&=&\frac{\cal B}{\sqrt{1+{\cal B} (r^2-2mr)/F}}\ ,
\end{eqnarray}
where we have used the notation 
\begin{eqnarray}
{\cal B} = \frac{F}{r^-2mr}{\cal L}_c +\frac12\beta\ . 
\end{eqnarray}

Consider two charged particles in equatorial circular obits moving in opposite directions with  four velocities  
\begin{eqnarray}
u^{\alpha}_{1,2} = \left(\Gamma F^{-1/2},0,0, \pm \Gamma {v}/{r}\right)\ . 
\end{eqnarray}
The center of mass energy for this system will be 
\begin{eqnarray}
\frac{E_{\rm cm}^2}{2m_0^2} = 2 \Gamma^2\ ,
\end{eqnarray}
or 
\begin{eqnarray}
E_{\rm cm} = 2 m_0 \Gamma\ .
\end{eqnarray}

\begin{figure*}
	\includegraphics[width=0.45 \textwidth]{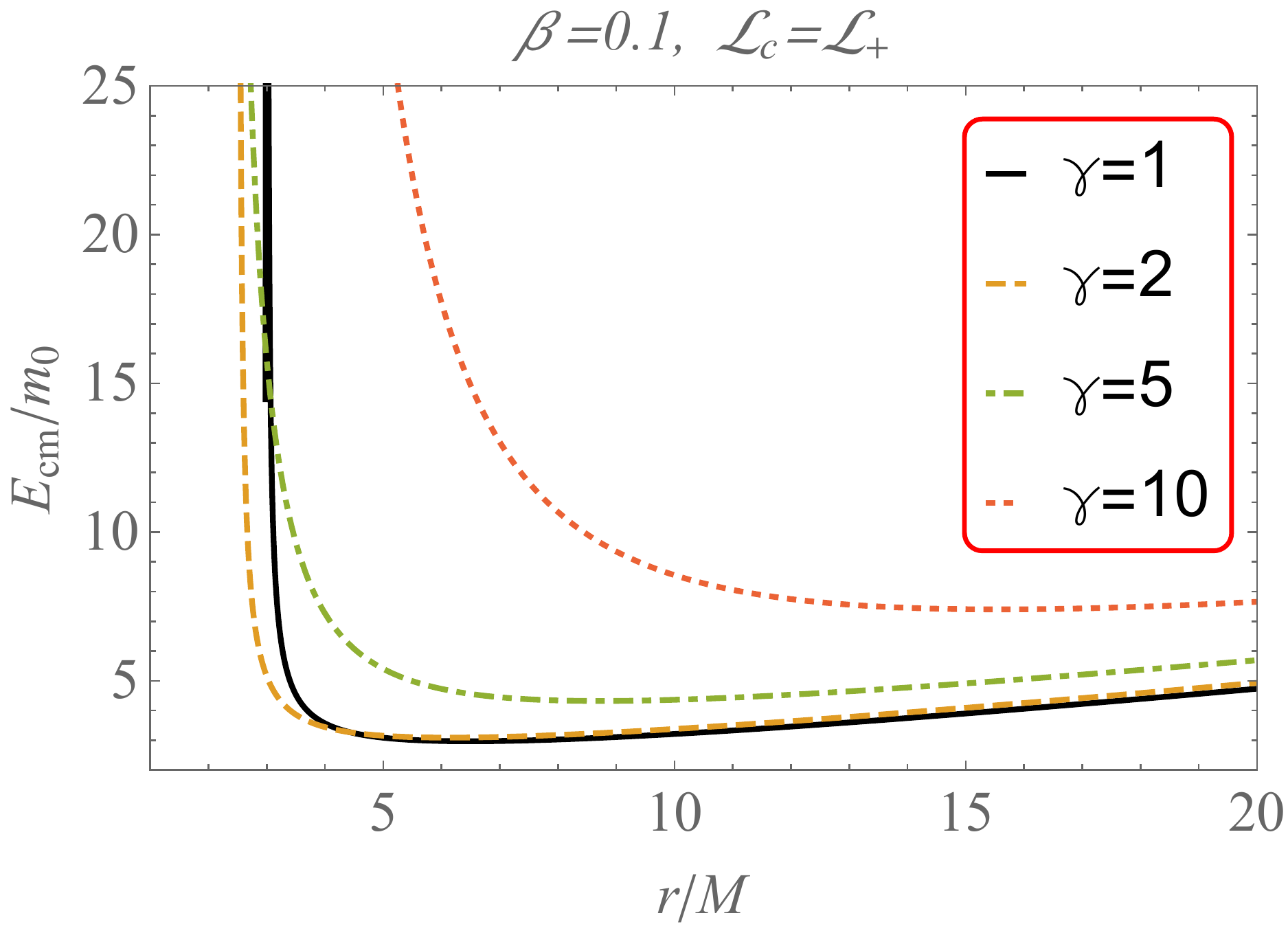}
	\includegraphics[width=0.45 \textwidth]{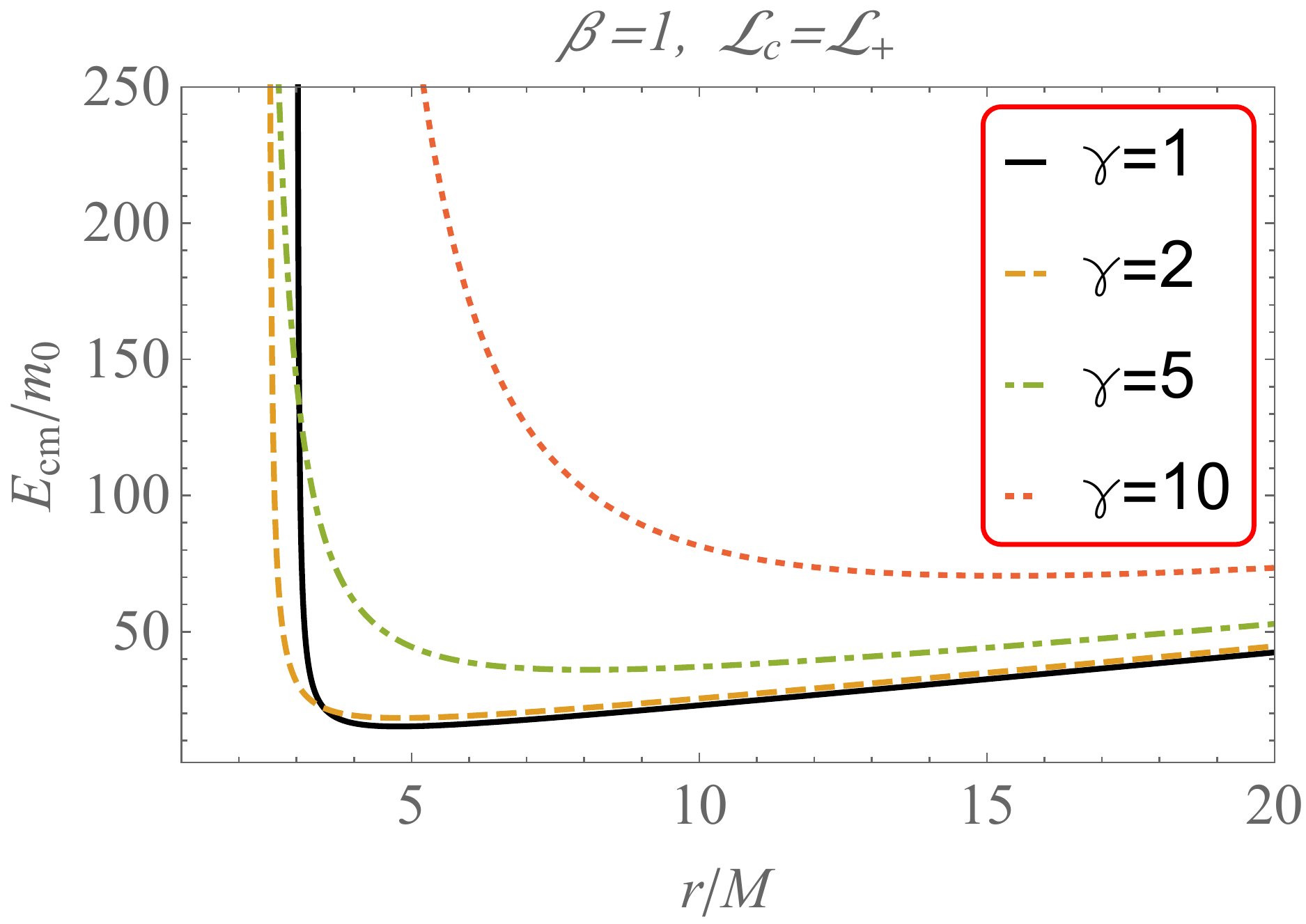}

	\includegraphics[width=0.45 \textwidth]{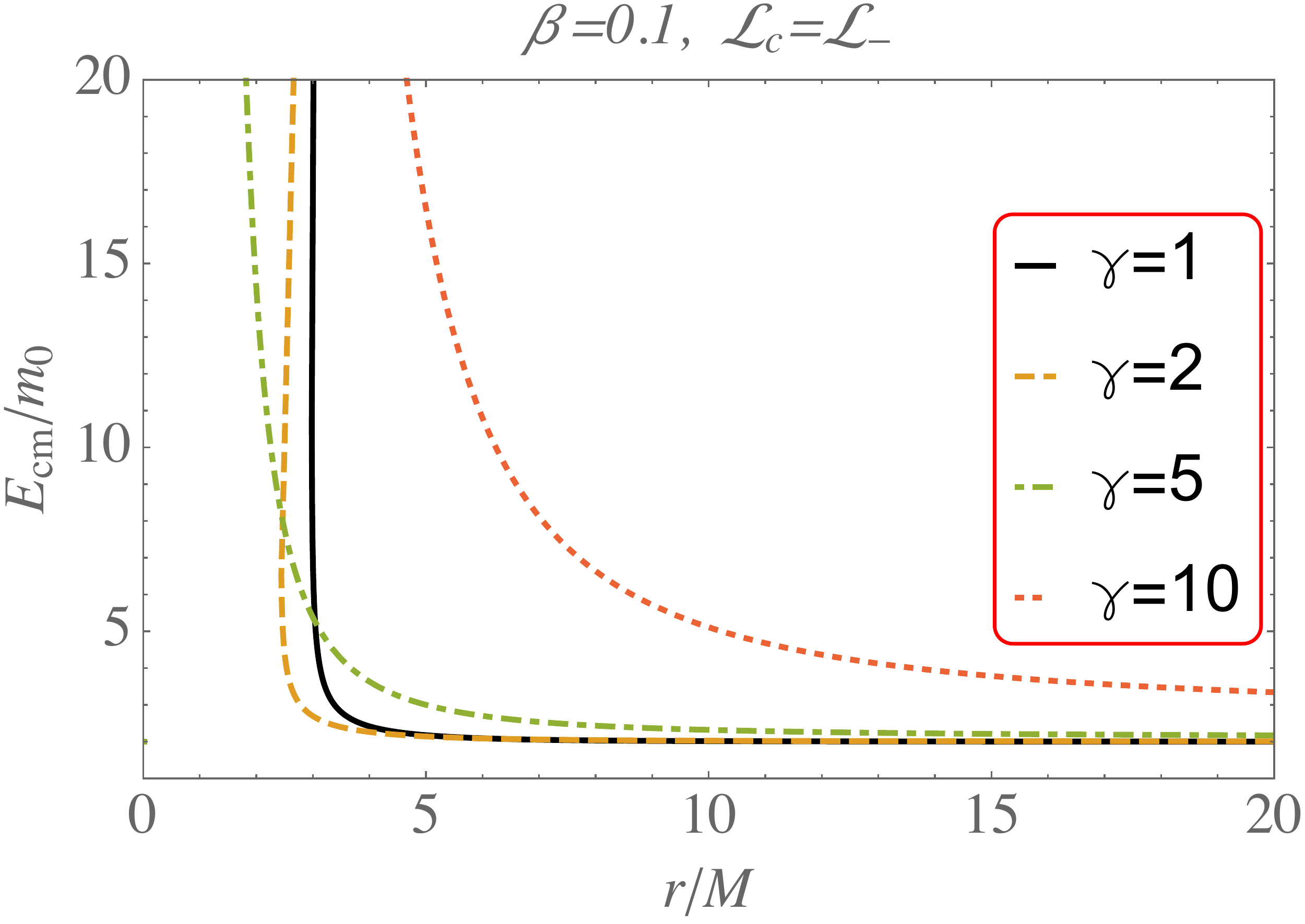}
	\includegraphics[width=0.45 \textwidth]{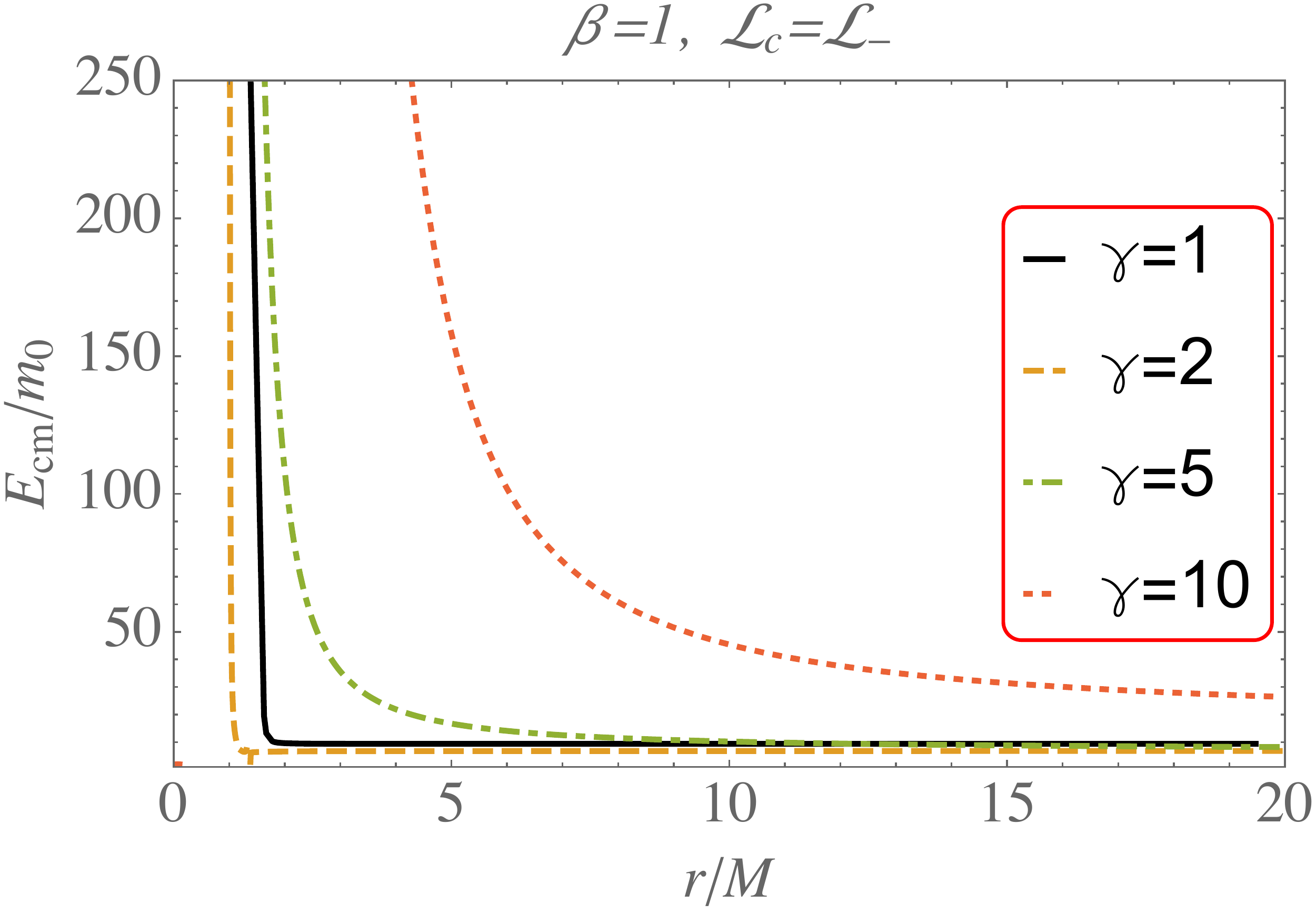}
		
	\includegraphics[width=0.45 \textwidth]{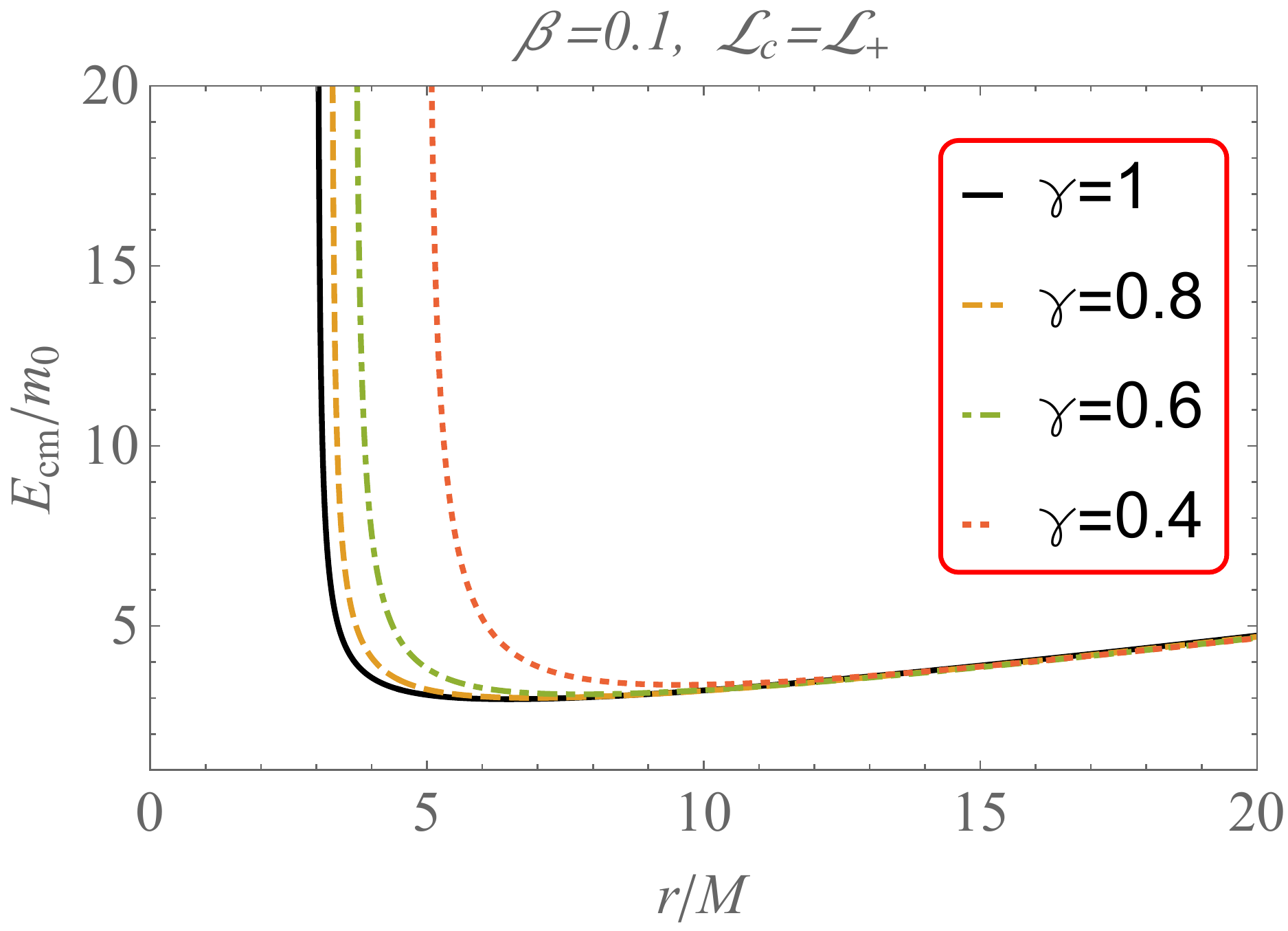}
	\includegraphics[width=0.45 \textwidth]{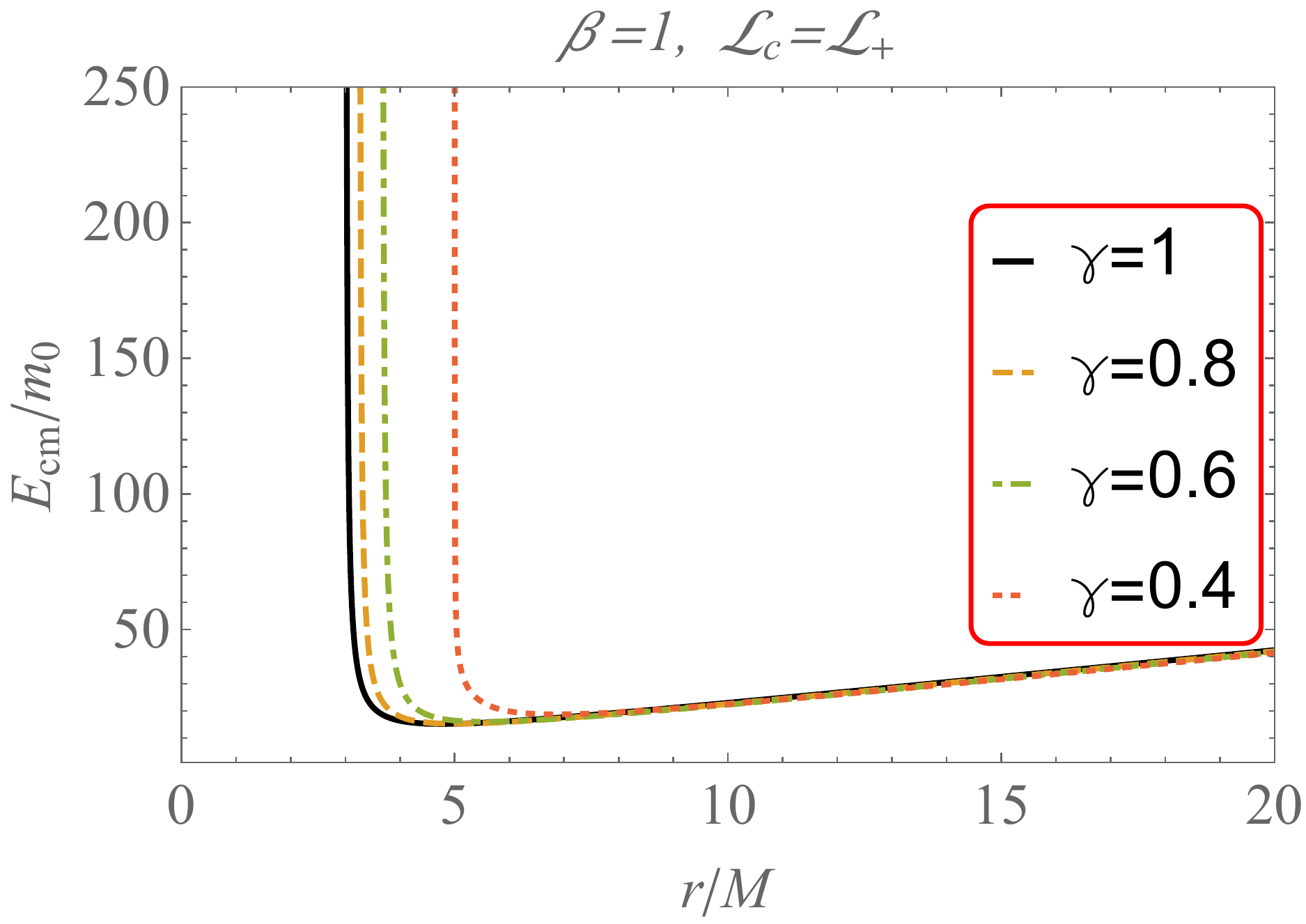}
	
	\includegraphics[width=0.45 \textwidth]{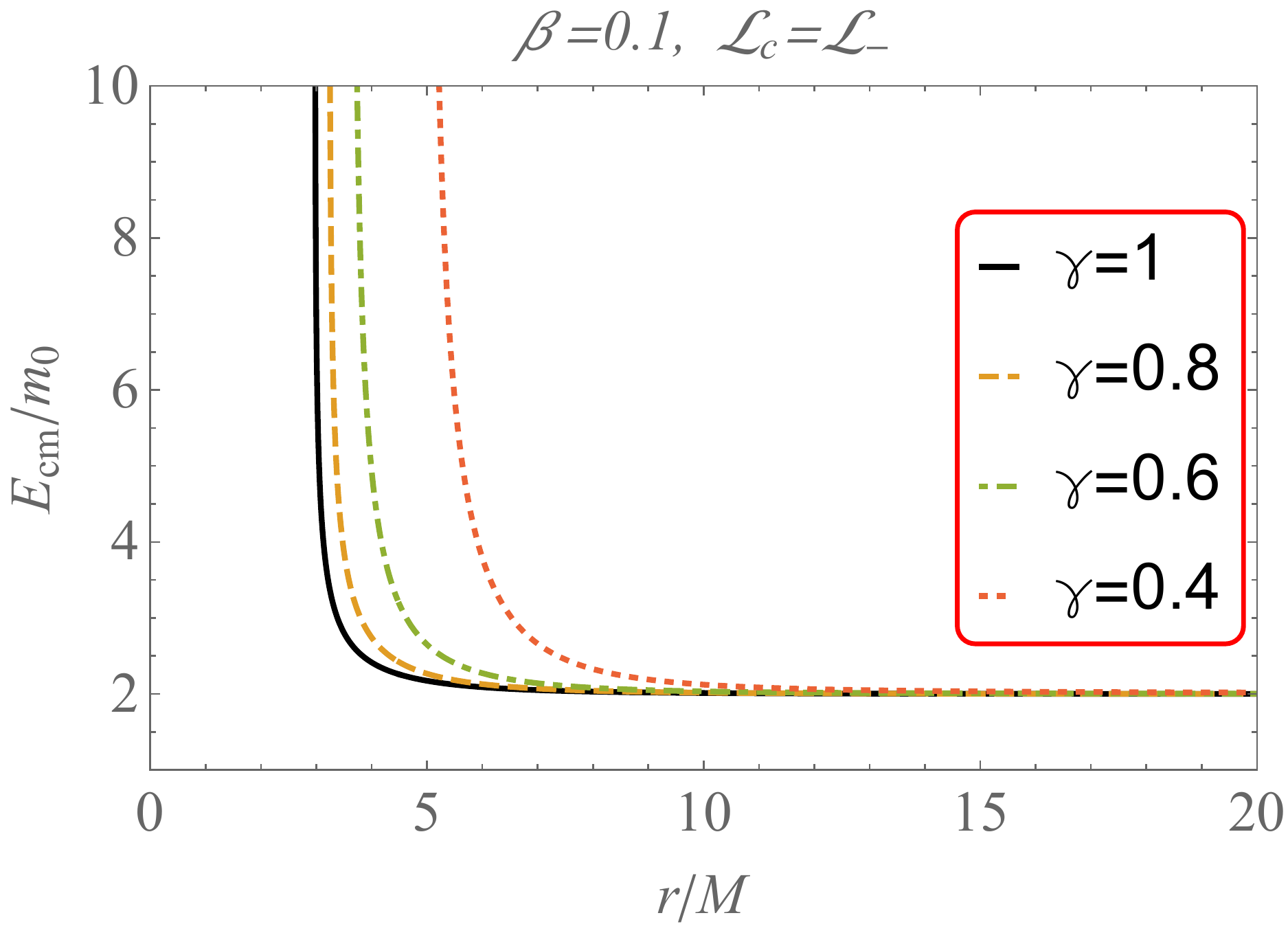}
	\includegraphics[width=0.45 \textwidth]{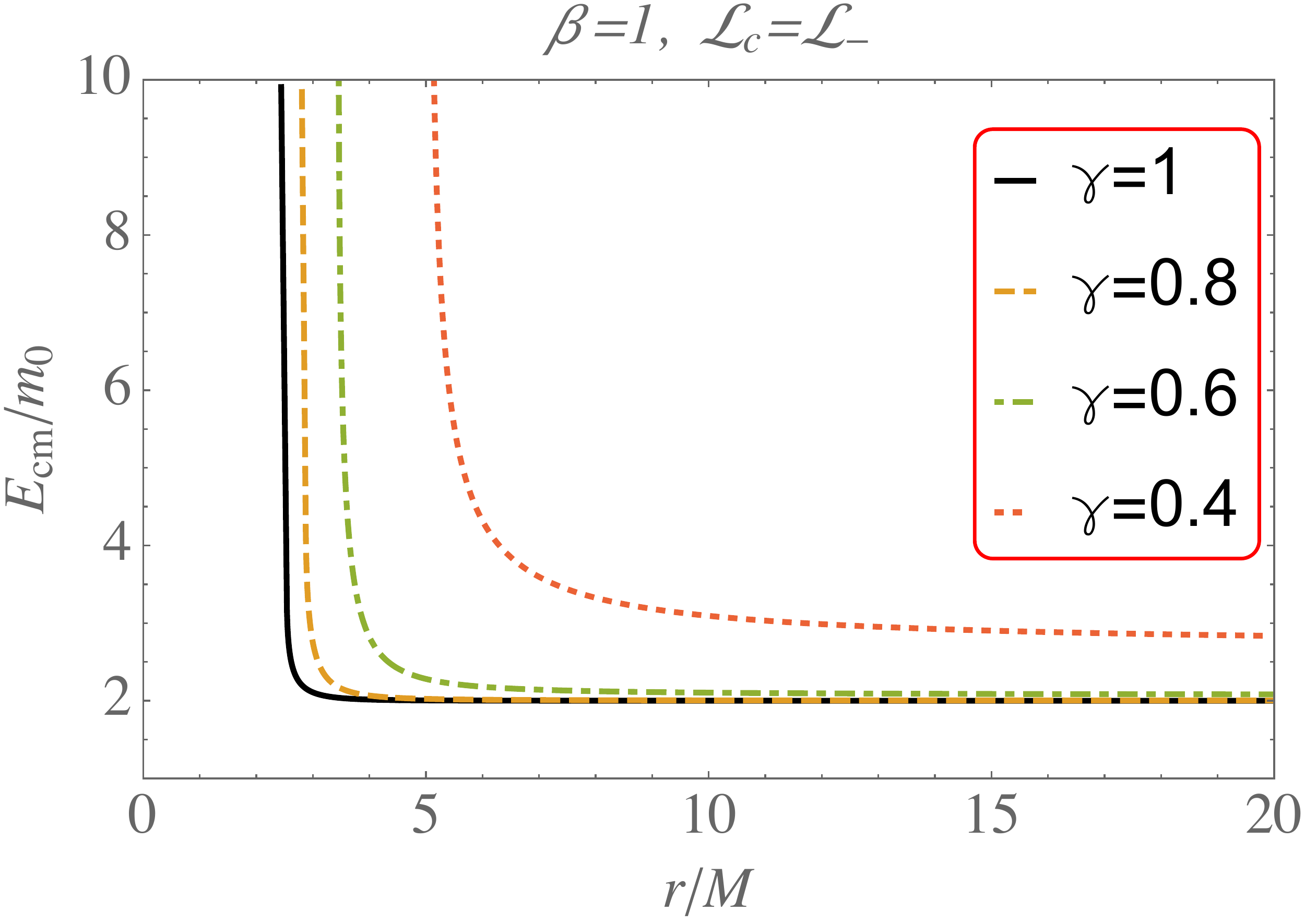}
	
	\caption{Radial dependence of the center of mass energy of two charged particles moving on circular orbits for the different values of $\gamma$ parameter and magnetic parameter $\beta$. Here we considered both '+' and '-' type of circular orbits for charged particles corresponding to angular momentum ${\cal L}_+$ and ${\cal L}_-$, respectively. \label{ecmch-ch}}
\end{figure*}

In Fig.~\ref{ecmch-ch} the radial dependence of the center of mass energy of two colliding charged particles is shown for different values of $\gamma$ and $\beta$. The plots have been made for two classes of circular orbits: with plus and minus signs corresponding to the different values of momenta of the charged particles (see Eq.~(\ref{lcc})). From Fig.~\ref{ecmch-ch} one can see that by increasing the value of the magnetic field the center of mass energy of the system also increases. The magnetic field, in principle, plays the role of an accelerator for the particles. Moreover the effect of the $\gamma$ parameter is also very important. First of all it is worth noting that the distance at which the particle's energy diverges changes with the value of $\gamma$: when $\gamma>1$ it becomes closer to the center, whereas for $\gamma<1$ the radius of this position increases. This behavior is related to the radius of the infinitely red-shifted surface, which, in terms of the total gravitational mass $M$ is given by $r=2M/\gamma$.
On the other hand, we observe that the center of mass energy of the system increases for both $\gamma<1$ and $\gamma>1$ with respect to Schwarzschild case ($\gamma=1$)
for any fixed value of the radial coordinate. This can be explained with the decrease the value of $r_{\rm isco}/M$ for both $\gamma<1$ and $\gamma>1$ cases. The ISCO radius for the Schwarzschild black hole is the maximum in terms of total gravitational mass $M$ ($r_{\rm isco} = 6 M$) and for $\gamma\neq1$ we get $r_{\rm isco}<6M$.

\subsection{Collision of charged particles on circular orbit with neutral particles infalling from infinity}

\begin{figure*}
	\includegraphics[width=0.45 \textwidth]{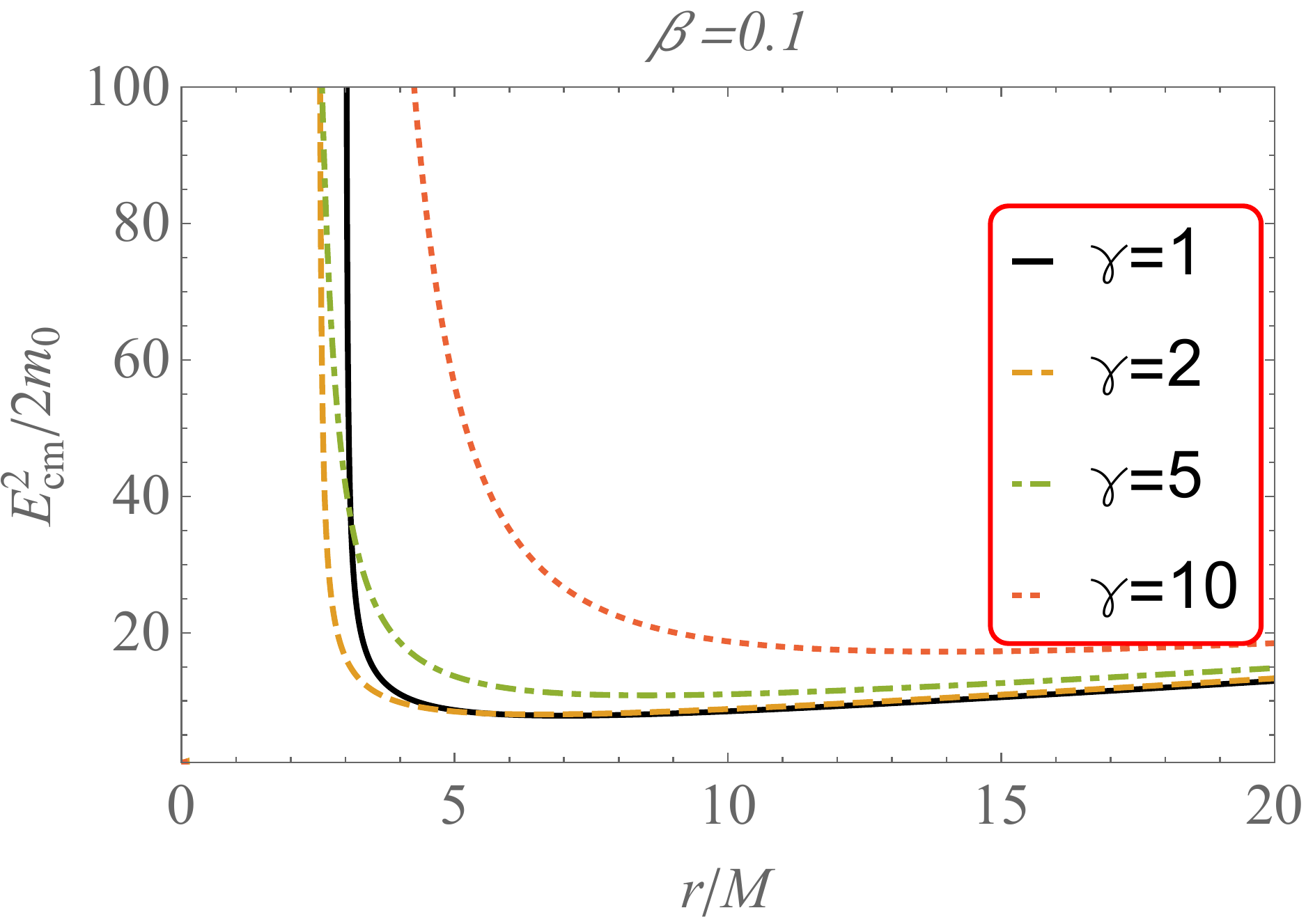}
	\includegraphics[width=0.45 \textwidth]{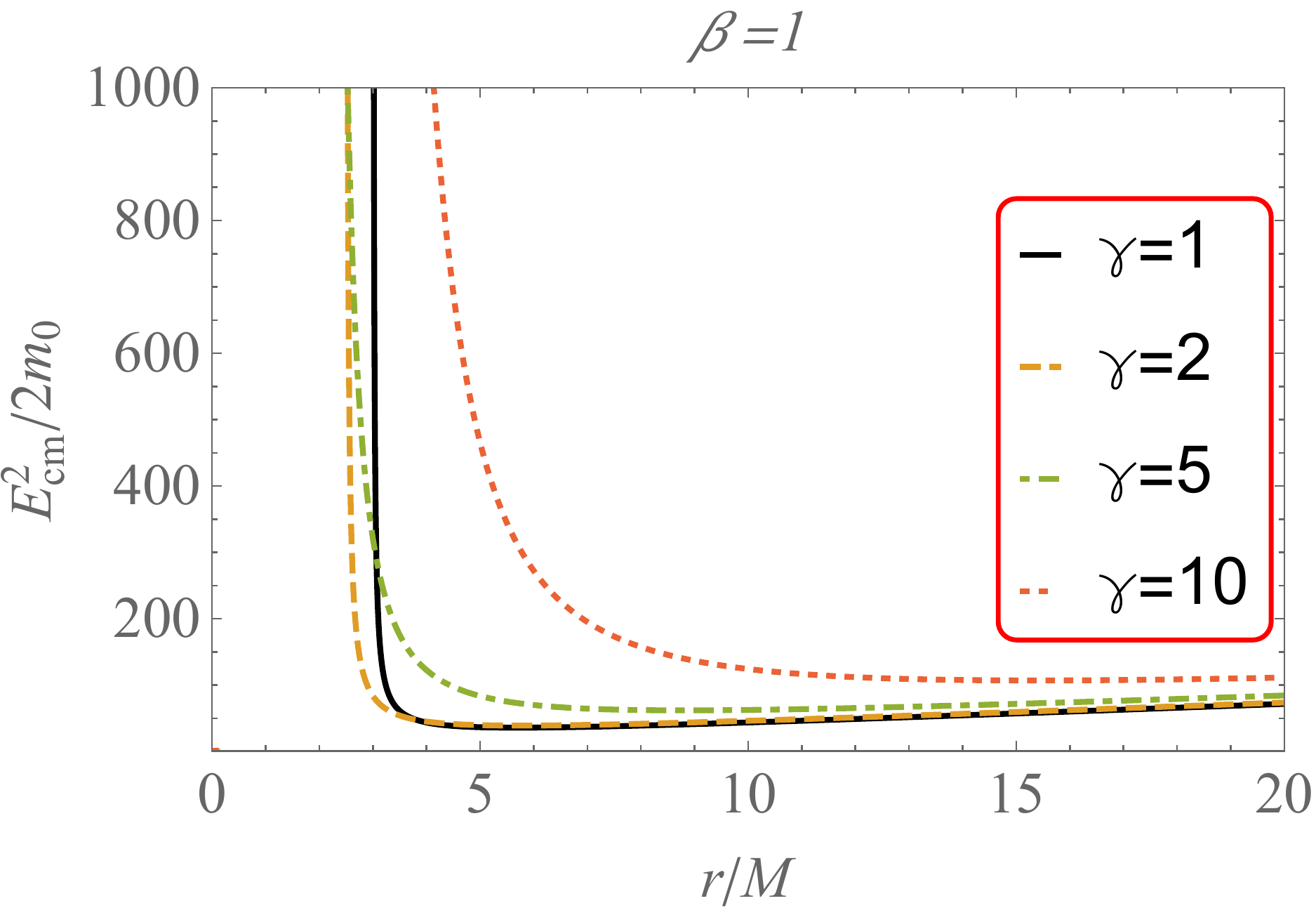}

	\includegraphics[width=0.45 \textwidth]{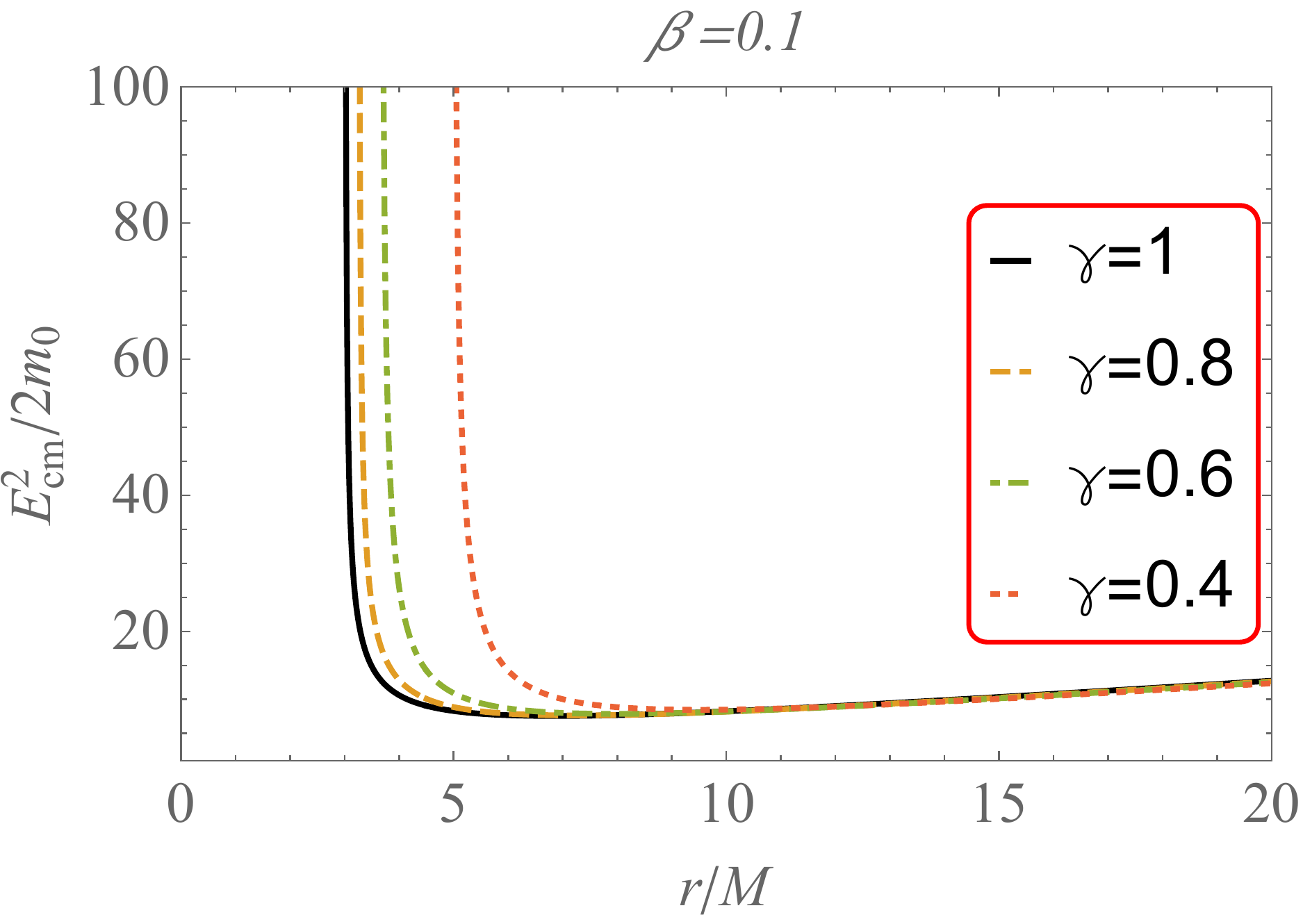}
	\includegraphics[width=0.45 \textwidth]{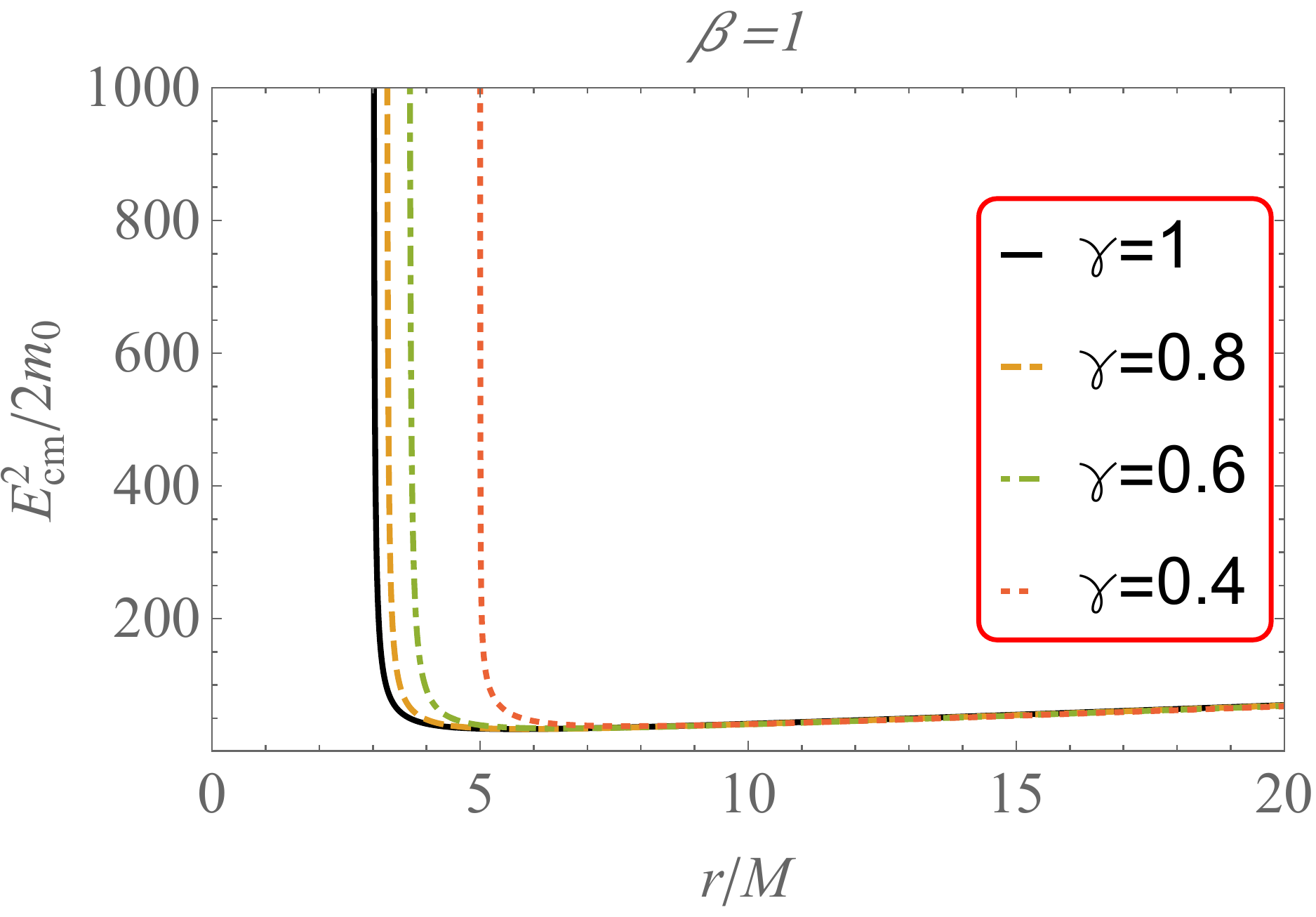}
		
	\caption{Radial dependence of the center of mass energy for the collision of a charged particle moving on circular orbit with a neutral particle infalling from infinity for the different values of $\gamma$ and $\beta$ parameters. \label{ecmch-neu}}
\end{figure*}

Finally we shall consider the case of the collision between a charged particle and a neutral particle, where the charged particle is moving on a circular orbit and the neutral particle is infalling from infinity.
The four velocities of the colliding neutral particle infalling from the infinity and charged particle in the circular orbit  can be expressed with equations (\ref{npvel}) and (\ref{cpvel}), respectively.
%
%\begin{eqnarray}
%u_1^{\alpha}&=& \left(\Gamma F^{-1/2},0,0, \Gamma {v}/{r}\right)\ , \\
%u_2^{\alpha} &=&  \bigg(\frac{1}{F},  \left(\frac{r^2 -2 m r }{r^2-2m r+m^2}\right)^{\frac{1-\gamma ^2}{2}} \nonumber \\ &&\times \sqrt{1 -F-\frac{F^2 {\cal L}^2_n}{r^2-2 mr } }, 0,\frac{F{\cal L}_n}{r^2-2 m r}  \bigg)\ .
%\end{eqnarray}
%
The center of mass energy then takes the form:
\begin{eqnarray}
\frac{E_{\rm cm}^2}{2m_0^2} = 1- {\cal L}_n {\cal B} +\sqrt{\frac{1}{F}+\frac{r^2-2 mr }{F^2}{\cal B}}\ ,
\end{eqnarray}
where ${\cal L}_n$ is the angular momentum of the neutral particle infalling from infinity. 
%{\bf (DM: Again, double check who is infalling from infinity and who is on circular orbit.)}. 
%

Fig.~\ref{ecmch-neu} shows the radial dependence of the above center of mass energy for different values of the magnetic parameter $\beta$ and the deformation parameter $\gamma$. One can see that both the magnetic field and the presence of deformation increase the center of mass energy of colliding particles acting as particle accelerators. Physically, this is related to the decrease in value of innermost stable circular orbits as the magnetic field and $\gamma$ parameter increase. 
%In the case of $\gamma<1$ the increase of the infinite redshift surface causes the increase of the energy. 

\section{Conclusion\label{Summary}}

In the present work we have studied the structure and properties of asymptotically uniform magnetic field in the $\gamma$ spacetime. In particular, in the vicinity of the singular surface we studied the motion of charged particles and the center of mass energy for particles collisions.  The obtained results can be summarized as follows:

The electromagnetic field structure in $\gamma$ spacetime has been considered in the presence of an asymptotically uniform magnetic field. The asymptotic values of the components of the magnetic field for different values of the $\gamma$ parameter tend to the Newtonian limit. The difference in the magnetic field strength becomes very important with decreasing radius. Also, with increasing value of $\gamma$ one can observe an increase of the absolute value of the magnetic field components. For $\gamma<1$ one can see that the azimuthal component of the magnetic field does not cross the surface of infinite redshift $r=2m$, which is similar to Meissner-like effect. 

We studied circular orbits of charged particles in the equatorial plane in the presence of an external magnetic field. The analysis in the absence of magnetic fields had shown that for $\gamma\geq 1/2$ there is only one region above the $r=2m$ singular surface where stable circular orbits are allowed. The location of the ISCO increases with increasing value of $\gamma$. For $\gamma \in (1/\sqrt{5},\  1/2)$ we observe two regions for stable circular orbits, suggesting the appearance of repulsive effects in the vicinity of the singular surface. For $\gamma<1/\sqrt{5}$ there is no limit for the existence of circular orbits. The presence of a test magnetic field increases the size of the region where no ISCO is present and decreases the size of the region where two ranges for circular orbits are present. In other words, the special value of $\gamma$ separating the two regions (which without magnetic field is $\gamma=1/\sqrt{5}$) increases with increase of $\beta$ tending to $1/2$, whereas the value of $\gamma$ for which only one ISCO exists ($\gamma=1/2$) remains unchanged.   
The magnetic field also affects the location of the ISCO by decreasing the value of its radius for fixed values of $\gamma>1/2$. This may have important implications for astrophysics, because the measurements of ISCO have been used to estimate the spin of some astrophysical black hole candidates~\cite{Shafee06,Shafee08,Steiner09,Steiner10,McClintock14,Gou14,Steiner11}. The main idea relies on the possibility of comparing observations coming from the light spectrum of accretion disks with the theoretical results on the ISCO's dependence upon the spin of the black hole~\cite{Shafee06,Shafee08}. However, as we have seen, the location of the ISCO radius depends on several factors. For example, in different theories of gravity the ISCO location can be used to get constraints on other parameters of the central object~\cite{Joshi14,Broderick2007,Abdujabbarov10,Abdujabbarov11a,Abdujabbarov13b,Kong14}. Similarly here we have shown that both the $\gamma$ parameter and the presence of a magnetic field affect the location of the ISCO radius. Therefore, in principle, observations of black holes with a fixed value of the angular momentum can be mimicked by the $\gamma$ metric with corresponding values of $\gamma$ and $\beta$. On the other hand, measurements obtained with different methods can be used to constrain the values of $\gamma$ and $\beta$, thus testing the astrophysical validity of these models.  

 The authors of \cite{Banados09} have shown that two infalling particles in the Kerr spacetime can be accelerated to extremely high (infinite) energies for fine-tuned values of the spin parameter of the black hole and angular momentum of the particles. Here we have shown that the acceleration of the particles to extremely high energies, in principle, is possible without any fine tuning in the case of a prolate geometry corresponding to $\gamma<1$. In this case the center of mass energy of particles diverges near the $r=2m$ singular surface for any value of the particles' angular momentum. However, the situation dramatically changes for $\gamma\geq1$. For, $\gamma=1$ (Schwarzschild case) we get the same result as in~\cite{Banados09}: the center of mass energy of the particles increases but does not diverge near the event horizon $(r=2M)$. For $\gamma>1$ the center of mass energy of the particles remains finite and becomes small for $r$ approaching the singular surface. The situations with the particle acceleration near the singular surface are totally different in the cases of prolate and oblate spacetimes: in the first case we observe the divergence of the center of mass energy without fine-tuning of angular momentum, whereas in the second case the center of mass energy is decreasing with decreasing $r$. 

The analysis of charged and neutral particle acceleration near the singular surface $r=2m$ in the $\gamma$ spacetime showed that the role of deformations and external magnetic fields is important in these processes. Energetically the most interesting case is the collision of neutral particles coming from infinity with charged particles in circular orbits. In general, with the increase of the magnetic field, the center of mass energy of the system is also increasing. The magnetic field, in principle, plays the role of a particle accelerator. Moreover the effect of the deformation parameter is also very important. First, it is natural that the distance where the particle's energy diverges changes with the change of the parameter $\gamma$. When $\gamma>1$ it is close to the center, whereas for $\gamma<1$ the radius of this position increases. This is related to the radius of the infinite redshift surface of the spacetime with a corresponding value of the $\gamma$ parameter. 

Observations of the near horizon regime are now becoming experimentally possible. 
For example, the gravitational redshift in the orbit of the star S2 passing at the point of closest approach (i.e. 1400 Schwarzschild radii) from the supermassive black hole located at the center of the Milky Way galaxy (kmown as SgrA*) was measured~\cite{Gravity18a}. However in order to probe the true nature of the geometry one needs to observe closer to the supposed location of the horizon. 
Recently, the GRAVITY-Very Large Telescope Interferometer (VLTI) observed
the motion and polarization variability of bright flares (hot spots)
in the near-infrared range around SgrA*. 
These observations of features located at the distance of approximately
six to ten times the gravitational radius are the closest observational evidences of motion in the vicinity of the ISCO.
They confirm the existence of clockwise looped orbital motion possibly due to the presence of a strong poloidal magnetic field, on scales of typically 150~$\mu as$ over a few tens of arcminutes,
corresponding to velocities of about 30\% the speed of light~\cite{Gravity18}. Therefore it is possible that in the near future such observations could be used to constrain the values of magnetic fields in the geometry surrounding SgrA*. Therefore the comparison of the motion in the black hole spacetime with the corresponding motion in the $\gamma$ metric could allow one to test the robustness of the conclusion that such an object must be a black hole.
%
%\end{itemize}

\section*{acknowledgements}

This research is supported by by Grants No.~VA-FA-F-2-008 and No. YFA-Ftech-2018-8 of the Uzbekistan Ministry for Innovation Development, and by the
Abdus Salam International Centre for Theoretical
Physics through Grant No.~OEA-NT-01.
This research is partially supported by an
Erasmus+ exchange grant between SU and NUUz.
C.B. acknowledges support from the National Natural Science Foundation of China (Grant No.~U1531117), Fudan University (Grant No.~IDH1512060), and the Alexander von Humboldt Foundation. 
A.A. and B.A. would like to acknowledge Nazarbayev University, Astana,
Kazakhstan for the warm hospitality through support from ORAU
grant SST~2015021. D.M. acknowledges support by Nazarbayev University Faculty Development Competitive Research Grant No. 090118FD5348 and by the Ministry of Education of Kazakhstan's target program IRN: BR05236454. D.M. wishes to thank Prof. Naresh Dadhich for valuable comments and discussions.

\bibliographystyle{apsrev4-1}  %% BibTeX style
\bibliography{gravreferences}

\begin{thebibliography}{84}%
\makeatletter
\providecommand \@ifxundefined [1]{%
 \@ifx{#1\undefined}
}%
\providecommand \@ifnum [1]{%
 \ifnum #1\expandafter \@firstoftwo
 \else \expandafter \@secondoftwo
 \fi
}%
\providecommand \@ifx [1]{%
 \ifx #1\expandafter \@firstoftwo
 \else \expandafter \@secondoftwo
 \fi
}%
\providecommand \natexlab [1]{#1}%
\providecommand \enquote  [1]{``#1''}%
\providecommand \bibnamefont  [1]{#1}%
\providecommand \bibfnamefont [1]{#1}%
\providecommand \citenamefont [1]{#1}%
\providecommand \href@noop [0]{\@secondoftwo}%
\providecommand \href [0]{\begingroup \@sanitize@url \@href}%
\providecommand \@href[1]{\@@startlink{#1}\@@href}%
\providecommand \@@href[1]{\endgroup#1\@@endlink}%
\providecommand \@sanitize@url [0]{\catcode `\\12\catcode `\$12\catcode
  `\&12\catcode `\#12\catcode `\^12\catcode `\_12\catcode `\%12\relax}%
\providecommand \@@startlink[1]{}%
\providecommand \@@endlink[0]{}%
\providecommand \url  [0]{\begingroup\@sanitize@url \@url }%
\providecommand \@url [1]{\endgroup\@href {#1}{\urlprefix }}%
\providecommand \urlprefix  [0]{URL }%
\providecommand \Eprint [0]{\href }%
\providecommand \doibase [0]{http://dx.doi.org/}%
\providecommand \selectlanguage [0]{\@gobble}%
\providecommand \bibinfo  [0]{\@secondoftwo}%
\providecommand \bibfield  [0]{\@secondoftwo}%
\providecommand \translation [1]{[#1]}%
\providecommand \BibitemOpen [0]{}%
\providecommand \bibitemStop [0]{}%
\providecommand \bibitemNoStop [0]{.\EOS\space}%
\providecommand \EOS [0]{\spacefactor3000\relax}%
\providecommand \BibitemShut  [1]{\csname bibitem#1\endcsname}%
\let\auto@bib@innerbib\@empty
%</preamble>
\bibitem [{\citenamefont {{Wald}}(1974)}]{Wald74}%
  \BibitemOpen
  \bibfield  {author} {\bibinfo {author} {\bibfnamefont {R.~M.}\ \bibnamefont
  {{Wald}}},\ }\href {\doibase 10.1103/PhysRevD.10.1680} {\bibfield  {journal}
  {\bibinfo  {journal} {Phys. Rev. D.}\ }\textbf {\bibinfo {volume} {10}},\
  \bibinfo {pages} {1680} (\bibinfo {year} {1974})}\BibitemShut {NoStop}%
\bibitem [{\citenamefont {{Aliev}}\ and\ \citenamefont
  {{{\"O}zdemir}}(2002)}]{Aliev02}%
  \BibitemOpen
  \bibfield  {author} {\bibinfo {author} {\bibfnamefont {A.~N.}\ \bibnamefont
  {{Aliev}}}\ and\ \bibinfo {author} {\bibfnamefont {N.}~\bibnamefont
  {{{\"O}zdemir}}},\ }\href {\doibase 10.1046/j.1365-8711.2002.05727.x}
  {\bibfield  {journal} {\bibinfo  {journal} {Mon. Not. R. Astron. Soc.}\
  }\textbf {\bibinfo {volume} {336}},\ \bibinfo {pages} {241} (\bibinfo {year}
  {2002})},\ \Eprint {http://arxiv.org/abs/gr-qc/0208025} {gr-qc/0208025}
  \BibitemShut {NoStop}%
\bibitem [{\citenamefont {{Aliev}}\ and\ \citenamefont
  {{Frolov}}(2004)}]{Aliev2004}%
  \BibitemOpen
  \bibfield  {author} {\bibinfo {author} {\bibfnamefont {A.~N.}\ \bibnamefont
  {{Aliev}}}\ and\ \bibinfo {author} {\bibfnamefont {V.~P.}\ \bibnamefont
  {{Frolov}}},\ }\href {\doibase 10.1103/PhysRevD.69.084022} {\bibfield
  {journal} {\bibinfo  {journal} {Phys. Rev. D}\ }\textbf {\bibinfo {volume}
  {69}},\ \bibinfo {eid} {084022} (\bibinfo {year} {2004})},\ \Eprint
  {http://arxiv.org/abs/arXiv:hep-th/0401095} {arXiv:hep-th/0401095}
  \BibitemShut {NoStop}%
\bibitem [{\citenamefont {{Aliev}}\ and\ \citenamefont {{G{\"u}mr{\"u}k{\c
  c}{\"u}o{\v g}lu}}(2005)}]{Aliev05}%
  \BibitemOpen
  \bibfield  {author} {\bibinfo {author} {\bibfnamefont {A.~N.}\ \bibnamefont
  {{Aliev}}}\ and\ \bibinfo {author} {\bibfnamefont {A.~E.}\ \bibnamefont
  {{G{\"u}mr{\"u}k{\c c}{\"u}o{\v g}lu}}},\ }\href {\doibase
  10.1103/PhysRevD.71.104027} {\bibfield  {journal} {\bibinfo  {journal} {Phys.
  Rev. D}\ }\textbf {\bibinfo {volume} {71}},\ \bibinfo {eid} {104027}
  (\bibinfo {year} {2005})},\ \Eprint {http://arxiv.org/abs/hep-th/0502223}
  {hep-th/0502223} \BibitemShut {NoStop}%
\bibitem [{\citenamefont {{Aliev}}(2006)}]{Aliev06}%
  \BibitemOpen
  \bibfield  {author} {\bibinfo {author} {\bibfnamefont {A.~N.}\ \bibnamefont
  {{Aliev}}},\ }\href {\doibase 10.1103/PhysRevD.74.024011} {\bibfield
  {journal} {\bibinfo  {journal} {Phys. Rev. D}\ }\textbf {\bibinfo {volume}
  {74}},\ \bibinfo {eid} {024011} (\bibinfo {year} {2006})},\ \Eprint
  {http://arxiv.org/abs/hep-th/0604207} {hep-th/0604207} \BibitemShut {NoStop}%
\bibitem [{\citenamefont {{Frolov}}\ and\ \citenamefont
  {{Shoom}}(2010)}]{Frolov10}%
  \BibitemOpen
  \bibfield  {author} {\bibinfo {author} {\bibfnamefont {V.~P.}\ \bibnamefont
  {{Frolov}}}\ and\ \bibinfo {author} {\bibfnamefont {A.~A.}\ \bibnamefont
  {{Shoom}}},\ }\href {\doibase 10.1103/PhysRevD.82.084034} {\bibfield
  {journal} {\bibinfo  {journal} {Phys. Rev. D.}\ }\textbf {\bibinfo {volume}
  {82}},\ \bibinfo {eid} {084034} (\bibinfo {year} {2010})},\ \Eprint
  {http://arxiv.org/abs/1008.2985} {arXiv:1008.2985 [gr-qc]} \BibitemShut
  {NoStop}%
\bibitem [{\citenamefont {{Abdujabbarov}}\ and\ \citenamefont
  {{Ahmedov}}(2010)}]{Abdujabbarov10}%
  \BibitemOpen
  \bibfield  {author} {\bibinfo {author} {\bibfnamefont {A.}~\bibnamefont
  {{Abdujabbarov}}}\ and\ \bibinfo {author} {\bibfnamefont {B.}~\bibnamefont
  {{Ahmedov}}},\ }\href {\doibase 10.1103/PhysRevD.81.044022} {\bibfield
  {journal} {\bibinfo  {journal} {Phys. Rev. D}\ }\textbf {\bibinfo {volume}
  {81}},\ \bibinfo {eid} {044022} (\bibinfo {year} {2010})},\ \Eprint
  {http://arxiv.org/abs/0905.2730} {arXiv:0905.2730 [gr-qc]} \BibitemShut
  {NoStop}%
\bibitem [{\citenamefont {{Abdujabbarov}}\ \emph
  {et~al.}(2011{\natexlab{a}})\citenamefont {{Abdujabbarov}}, \citenamefont
  {{Ahmedov}},\ and\ \citenamefont {{Hakimov}}}]{Abdujabbarov11a}%
  \BibitemOpen
  \bibfield  {author} {\bibinfo {author} {\bibfnamefont {A.}~\bibnamefont
  {{Abdujabbarov}}}, \bibinfo {author} {\bibfnamefont {B.}~\bibnamefont
  {{Ahmedov}}}, \ and\ \bibinfo {author} {\bibfnamefont {A.}~\bibnamefont
  {{Hakimov}}},\ }\href {\doibase 10.1103/PhysRevD.83.044053} {\bibfield
  {journal} {\bibinfo  {journal} {Phys.Rev. D}\ }\textbf {\bibinfo {volume}
  {83}},\ \bibinfo {eid} {044053} (\bibinfo {year} {2011}{\natexlab{a}})},\
  \Eprint {http://arxiv.org/abs/1101.4741} {arXiv:1101.4741 [gr-qc]}
  \BibitemShut {NoStop}%
\bibitem [{\citenamefont {{Frolov}}(2012)}]{Frolov12}%
  \BibitemOpen
  \bibfield  {author} {\bibinfo {author} {\bibfnamefont {V.~P.}\ \bibnamefont
  {{Frolov}}},\ }\href {\doibase 10.1103/PhysRevD.85.024020} {\bibfield
  {journal} {\bibinfo  {journal} {Phys. Rev. D.}\ }\textbf {\bibinfo {volume}
  {85}},\ \bibinfo {eid} {024020} (\bibinfo {year} {2012})},\ \Eprint
  {http://arxiv.org/abs/1110.6274} {arXiv:1110.6274 [gr-qc]} \BibitemShut
  {NoStop}%
\bibitem [{\citenamefont {{Karas}}\ \emph {et~al.}(2012)\citenamefont
  {{Karas}}, \citenamefont {{Kovar}}, \citenamefont {{Kopacek}}, \citenamefont
  {{Kojima}}, \citenamefont {{Slany}},\ and\ \citenamefont
  {{Stuchlik}}}]{Karas12a}%
  \BibitemOpen
  \bibfield  {author} {\bibinfo {author} {\bibfnamefont {V.}~\bibnamefont
  {{Karas}}}, \bibinfo {author} {\bibfnamefont {J.}~\bibnamefont {{Kovar}}},
  \bibinfo {author} {\bibfnamefont {O.}~\bibnamefont {{Kopacek}}}, \bibinfo
  {author} {\bibfnamefont {Y.}~\bibnamefont {{Kojima}}}, \bibinfo {author}
  {\bibfnamefont {P.}~\bibnamefont {{Slany}}}, \ and\ \bibinfo {author}
  {\bibfnamefont {Z.}~\bibnamefont {{Stuchlik}}},\ }in\ \href@noop {} {\emph
  {\bibinfo {booktitle} {American Astronomical Society Meeting Abstracts
  \#220}}},\ \bibinfo {series} {American Astronomical Society Meeting
  Abstracts}, Vol.\ \bibinfo {volume} {220}\ (\bibinfo {year} {2012})\ p.\
  \bibinfo {pages} {430.07}\BibitemShut {NoStop}%
\bibitem [{\citenamefont {{Hakimov}}\ \emph {et~al.}(2013)\citenamefont
  {{Hakimov}}, \citenamefont {{Abdujabbarov}},\ and\ \citenamefont
  {{Ahmedov}}}]{Hakimov13}%
  \BibitemOpen
  \bibfield  {author} {\bibinfo {author} {\bibfnamefont {A.}~\bibnamefont
  {{Hakimov}}}, \bibinfo {author} {\bibfnamefont {A.}~\bibnamefont
  {{Abdujabbarov}}}, \ and\ \bibinfo {author} {\bibfnamefont {B.}~\bibnamefont
  {{Ahmedov}}},\ }\href {\doibase 10.1103/PhysRevD.88.024008} {\bibfield
  {journal} {\bibinfo  {journal} {Phys. Rev. D}\ }\textbf {\bibinfo {volume}
  {88}},\ \bibinfo {eid} {024008} (\bibinfo {year} {2013})}\BibitemShut
  {NoStop}%
\bibitem [{\citenamefont {{Stuchl{\'{\i}}k}}\ \emph {et~al.}(2014)\citenamefont
  {{Stuchl{\'{\i}}k}}, \citenamefont {{Schee}},\ and\ \citenamefont
  {{Abdujabbarov}}}]{Stuchlik14a}%
  \BibitemOpen
  \bibfield  {author} {\bibinfo {author} {\bibfnamefont {Z.}~\bibnamefont
  {{Stuchl{\'{\i}}k}}}, \bibinfo {author} {\bibfnamefont {J.}~\bibnamefont
  {{Schee}}}, \ and\ \bibinfo {author} {\bibfnamefont {A.}~\bibnamefont
  {{Abdujabbarov}}},\ }\href {\doibase 10.1103/PhysRevD.89.104048} {\bibfield
  {journal} {\bibinfo  {journal} {Phys. Rev. D}\ }\textbf {\bibinfo {volume}
  {89}},\ \bibinfo {eid} {104048} (\bibinfo {year} {2014})}\BibitemShut
  {NoStop}%
\bibitem [{\citenamefont {{Stuchl{\'{\i}}k}}\ and\ \citenamefont {{Kolo{\v
  s}}}(2016)}]{Stuchlik16}%
  \BibitemOpen
  \bibfield  {author} {\bibinfo {author} {\bibfnamefont {Z.}~\bibnamefont
  {{Stuchl{\'{\i}}k}}}\ and\ \bibinfo {author} {\bibfnamefont {M.}~\bibnamefont
  {{Kolo{\v s}}}},\ }\href {\doibase 10.1140/epjc/s10052-015-3862-2} {\bibfield
   {journal} {\bibinfo  {journal} {European Physical Journal C}\ }\textbf
  {\bibinfo {volume} {76}},\ \bibinfo {eid} {32} (\bibinfo {year} {2016})},\
  \Eprint {http://arxiv.org/abs/1511.02936} {arXiv:1511.02936 [gr-qc]}
  \BibitemShut {NoStop}%
\bibitem [{\citenamefont {{Ba{\~n}ados}}\ \emph {et~al.}(2009)\citenamefont
  {{Ba{\~n}ados}}, \citenamefont {{Silk}},\ and\ \citenamefont
  {{West}}}]{Banados09}%
  \BibitemOpen
  \bibfield  {author} {\bibinfo {author} {\bibfnamefont {M.}~\bibnamefont
  {{Ba{\~n}ados}}}, \bibinfo {author} {\bibfnamefont {J.}~\bibnamefont
  {{Silk}}}, \ and\ \bibinfo {author} {\bibfnamefont {S.~M.}\ \bibnamefont
  {{West}}},\ }\href {\doibase 10.1103/PhysRevLett.103.111102} {\bibfield
  {journal} {\bibinfo  {journal} {Physical Review Letters}\ }\textbf {\bibinfo
  {volume} {103}},\ \bibinfo {eid} {111102} (\bibinfo {year}
  {2009})}\BibitemShut {NoStop}%
\bibitem [{\citenamefont {{Abdujabbarov}}\ \emph
  {et~al.}(2013{\natexlab{a}})\citenamefont {{Abdujabbarov}}, \citenamefont
  {{Tursunov}}, \citenamefont {{Ahmedov}},\ and\ \citenamefont
  {{Kuvatov}}}]{Abdujabbarov13a}%
  \BibitemOpen
  \bibfield  {author} {\bibinfo {author} {\bibfnamefont {A.~A.}\ \bibnamefont
  {{Abdujabbarov}}}, \bibinfo {author} {\bibfnamefont {A.~A.}\ \bibnamefont
  {{Tursunov}}}, \bibinfo {author} {\bibfnamefont {B.~J.}\ \bibnamefont
  {{Ahmedov}}}, \ and\ \bibinfo {author} {\bibfnamefont {A.}~\bibnamefont
  {{Kuvatov}}},\ }\href {\doibase 10.1007/s10509-012-1251-y} {\bibfield
  {journal} {\bibinfo  {journal} {Astrophys Space Sci}\ }\textbf {\bibinfo
  {volume} {343}},\ \bibinfo {pages} {173} (\bibinfo {year}
  {2013}{\natexlab{a}})},\ \Eprint {http://arxiv.org/abs/1209.2680}
  {arXiv:1209.2680 [gr-qc]} \BibitemShut {NoStop}%
\bibitem [{\citenamefont {{Tursunov}}\ \emph {et~al.}(2013)\citenamefont
  {{Tursunov}}, \citenamefont {{Kolo{\v s}}}, \citenamefont {{Abdujabbarov}},
  \citenamefont {{Ahmedov}},\ and\ \citenamefont
  {{Stuchl{\'{\i}}k}}}]{Tursunov13}%
  \BibitemOpen
  \bibfield  {author} {\bibinfo {author} {\bibfnamefont {A.}~\bibnamefont
  {{Tursunov}}}, \bibinfo {author} {\bibfnamefont {M.}~\bibnamefont {{Kolo{\v
  s}}}}, \bibinfo {author} {\bibfnamefont {A.}~\bibnamefont {{Abdujabbarov}}},
  \bibinfo {author} {\bibfnamefont {B.}~\bibnamefont {{Ahmedov}}}, \ and\
  \bibinfo {author} {\bibfnamefont {Z.}~\bibnamefont {{Stuchl{\'{\i}}k}}},\
  }\href {\doibase 10.1103/PhysRevD.88.124001} {\bibfield  {journal} {\bibinfo
  {journal} {Phys. Rev. D}\ }\textbf {\bibinfo {volume} {88}},\ \bibinfo {eid}
  {124001} (\bibinfo {year} {2013})}\BibitemShut {NoStop}%
\bibitem [{\citenamefont {{Tursunov}}\ \emph {et~al.}(2016)\citenamefont
  {{Tursunov}}, \citenamefont {{Stuchl{\'{\i}}k}},\ and\ \citenamefont
  {{Kolo{\v s}}}}]{Tursunov16}%
  \BibitemOpen
  \bibfield  {author} {\bibinfo {author} {\bibfnamefont {A.}~\bibnamefont
  {{Tursunov}}}, \bibinfo {author} {\bibfnamefont {Z.}~\bibnamefont
  {{Stuchl{\'{\i}}k}}}, \ and\ \bibinfo {author} {\bibfnamefont
  {M.}~\bibnamefont {{Kolo{\v s}}}},\ }\href {\doibase
  10.1103/PhysRevD.93.084012} {\bibfield  {journal} {\bibinfo  {journal} {Phys.
  Rev. D}\ }\textbf {\bibinfo {volume} {93}},\ \bibinfo {eid} {084012}
  (\bibinfo {year} {2016})},\ \Eprint {http://arxiv.org/abs/1603.07264}
  {arXiv:1603.07264 [gr-qc]} \BibitemShut {NoStop}%
\bibitem [{\citenamefont {{Kolo{\v s}}}\ \emph {et~al.}(2015)\citenamefont
  {{Kolo{\v s}}}, \citenamefont {{Stuchl{\'{\i}}k}},\ and\ \citenamefont
  {{Tursunov}}}]{Kolos15}%
  \BibitemOpen
  \bibfield  {author} {\bibinfo {author} {\bibfnamefont {M.}~\bibnamefont
  {{Kolo{\v s}}}}, \bibinfo {author} {\bibfnamefont {Z.}~\bibnamefont
  {{Stuchl{\'{\i}}k}}}, \ and\ \bibinfo {author} {\bibfnamefont
  {A.}~\bibnamefont {{Tursunov}}},\ }\href {\doibase
  10.1088/0264-9381/32/16/165009} {\bibfield  {journal} {\bibinfo  {journal}
  {Classical and Quantum Gravity}\ }\textbf {\bibinfo {volume} {32}},\ \bibinfo
  {eid} {165009} (\bibinfo {year} {2015})},\ \Eprint
  {http://arxiv.org/abs/1506.06799} {arXiv:1506.06799 [gr-qc]} \BibitemShut
  {NoStop}%
\bibitem [{\citenamefont {{Kolo{\v s}}}\ \emph {et~al.}(2017)\citenamefont
  {{Kolo{\v s}}}, \citenamefont {{Tursunov}},\ and\ \citenamefont
  {{Stuchl{\'{\i}}k}}}]{Kolos17}%
  \BibitemOpen
  \bibfield  {author} {\bibinfo {author} {\bibfnamefont {M.}~\bibnamefont
  {{Kolo{\v s}}}}, \bibinfo {author} {\bibfnamefont {A.}~\bibnamefont
  {{Tursunov}}}, \ and\ \bibinfo {author} {\bibfnamefont {Z.}~\bibnamefont
  {{Stuchl{\'{\i}}k}}},\ }\href@noop {} {\bibfield  {journal} {\bibinfo
  {journal} {Eur. Phys. J. C.}\ }\textbf {\bibinfo {volume} {77}},\ \bibinfo
  {pages} {860} (\bibinfo {year} {2017})},\ \Eprint
  {http://arxiv.org/abs/1707.02224} {arXiv:1707.02224 [astro-ph.HE]}
  \BibitemShut {NoStop}%
\bibitem [{\citenamefont {{Tursunov}}\ and\ \citenamefont
  {{Kolo{\v{s}}}}(2018)}]{Tursunov18a}%
  \BibitemOpen
  \bibfield  {author} {\bibinfo {author} {\bibfnamefont {A.~A.}\ \bibnamefont
  {{Tursunov}}}\ and\ \bibinfo {author} {\bibfnamefont {M.}~\bibnamefont
  {{Kolo{\v{s}}}}},\ }\href {\doibase 10.1134/S1063778818020187} {\bibfield
  {journal} {\bibinfo  {journal} {Physics of Atomic Nuclei}\ }\textbf {\bibinfo
  {volume} {81}},\ \bibinfo {pages} {279} (\bibinfo {year} {2018})},\ \Eprint
  {http://arxiv.org/abs/1803.08144} {arXiv:1803.08144 [astro-ph.HE]}
  \BibitemShut {NoStop}%
\bibitem [{\citenamefont {{Tursunov}}\ \emph {et~al.}(2018)\citenamefont
  {{Tursunov}}, \citenamefont {{Kolo{\v{s}}}}, \citenamefont
  {{Stuchl{\'\i}k}},\ and\ \citenamefont {{Gal'tsov}}}]{Tursunov18}%
  \BibitemOpen
  \bibfield  {author} {\bibinfo {author} {\bibfnamefont {A.}~\bibnamefont
  {{Tursunov}}}, \bibinfo {author} {\bibfnamefont {M.}~\bibnamefont
  {{Kolo{\v{s}}}}}, \bibinfo {author} {\bibfnamefont {Z.}~\bibnamefont
  {{Stuchl{\'\i}k}}}, \ and\ \bibinfo {author} {\bibfnamefont {D.~V.}\
  \bibnamefont {{Gal'tsov}}},\ }\href {\doibase 10.3847/1538-4357/aac7c5}
  {\bibfield  {journal} {\bibinfo  {journal} {Astrophys. J.}\ }\textbf
  {\bibinfo {volume} {861}},\ \bibinfo {eid} {2} (\bibinfo {year} {2018})},\
  \Eprint {http://arxiv.org/abs/1803.09682} {arXiv:1803.09682 [gr-qc]}
  \BibitemShut {NoStop}%
\bibitem [{\citenamefont {{Shaymatov}}\ \emph {et~al.}(2018)\citenamefont
  {{Shaymatov}}, \citenamefont {{Ahmedov}}, \citenamefont {{Stuchl{\'{\i}}k}},\
  and\ \citenamefont {{Abdujabbarov}}}]{Shaymatov18}%
  \BibitemOpen
  \bibfield  {author} {\bibinfo {author} {\bibfnamefont {S.}~\bibnamefont
  {{Shaymatov}}}, \bibinfo {author} {\bibfnamefont {B.}~\bibnamefont
  {{Ahmedov}}}, \bibinfo {author} {\bibfnamefont {Z.}~\bibnamefont
  {{Stuchl{\'{\i}}k}}}, \ and\ \bibinfo {author} {\bibfnamefont
  {A.}~\bibnamefont {{Abdujabbarov}}},\ }\href {\doibase
  10.1142/S0218271818500888} {\bibfield  {journal} {\bibinfo  {journal}
  {International Journal of Modern Physics D}\ }\textbf {\bibinfo {volume}
  {27}},\ \bibinfo {eid} {1850088} (\bibinfo {year} {2018})}\BibitemShut
  {NoStop}%
\bibitem [{\citenamefont {{Kimura}}\ \emph {et~al.}(2011)\citenamefont
  {{Kimura}}, \citenamefont {{Nakao}},\ and\ \citenamefont
  {{Tagoshi}}}]{Kimura11}%
  \BibitemOpen
  \bibfield  {author} {\bibinfo {author} {\bibfnamefont {M.}~\bibnamefont
  {{Kimura}}}, \bibinfo {author} {\bibfnamefont {K.-I.}\ \bibnamefont
  {{Nakao}}}, \ and\ \bibinfo {author} {\bibfnamefont {H.}~\bibnamefont
  {{Tagoshi}}},\ }\href {\doibase 10.1103/PhysRevD.83.044013} {\bibfield
  {journal} {\bibinfo  {journal} {Phys. Rev. D}\ }\textbf {\bibinfo {volume}
  {83}},\ \bibinfo {eid} {044013} (\bibinfo {year} {2011})},\ \Eprint
  {http://arxiv.org/abs/1010.5438} {arXiv:1010.5438 [gr-qc]} \BibitemShut
  {NoStop}%
\bibitem [{\citenamefont {{Zaslavskii}}(2012{\natexlab{a}})}]{Zaslavskii12a}%
  \BibitemOpen
  \bibfield  {author} {\bibinfo {author} {\bibfnamefont {O.~B.}\ \bibnamefont
  {{Zaslavskii}}},\ }\href {\doibase 10.1103/PhysRevD.86.124039} {\bibfield
  {journal} {\bibinfo  {journal} {Phys. Rev. D}\ }\textbf {\bibinfo {volume}
  {86}},\ \bibinfo {eid} {124039} (\bibinfo {year} {2012}{\natexlab{a}})},\
  \Eprint {http://arxiv.org/abs/1207.5209} {arXiv:1207.5209 [gr-qc]}
  \BibitemShut {NoStop}%
\bibitem [{\citenamefont {{Ba{\~n}ados}}\ \emph {et~al.}(2011)\citenamefont
  {{Ba{\~n}ados}}, \citenamefont {{Hassanain}}, \citenamefont {{Silk}},\ and\
  \citenamefont {{West}}}]{Banados11}%
  \BibitemOpen
  \bibfield  {author} {\bibinfo {author} {\bibfnamefont {M.}~\bibnamefont
  {{Ba{\~n}ados}}}, \bibinfo {author} {\bibfnamefont {B.}~\bibnamefont
  {{Hassanain}}}, \bibinfo {author} {\bibfnamefont {J.}~\bibnamefont {{Silk}}},
  \ and\ \bibinfo {author} {\bibfnamefont {S.~M.}\ \bibnamefont {{West}}},\
  }\href {\doibase 10.1103/PhysRevD.83.023004} {\bibfield  {journal} {\bibinfo
  {journal} {Physical Review D}\ }\textbf {\bibinfo {volume} {83}},\ \bibinfo
  {eid} {023004} (\bibinfo {year} {2011})},\ \Eprint
  {http://arxiv.org/abs/1010.2724} {arXiv:1010.2724 [astro-ph.CO]} \BibitemShut
  {NoStop}%
\bibitem [{\citenamefont {{Oteev}}\ \emph {et~al.}(2016)\citenamefont
  {{Oteev}}, \citenamefont {{Abdujabbarov}}, \citenamefont
  {{Stuchl{\'{\i}}k}},\ and\ \citenamefont {{Ahmedov}}}]{Oteev16}%
  \BibitemOpen
  \bibfield  {author} {\bibinfo {author} {\bibfnamefont {T.}~\bibnamefont
  {{Oteev}}}, \bibinfo {author} {\bibfnamefont {A.}~\bibnamefont
  {{Abdujabbarov}}}, \bibinfo {author} {\bibfnamefont {Z.}~\bibnamefont
  {{Stuchl{\'{\i}}k}}}, \ and\ \bibinfo {author} {\bibfnamefont
  {B.}~\bibnamefont {{Ahmedov}}},\ }\href {\doibase 10.1007/s10509-016-2850-9}
  {\bibfield  {journal} {\bibinfo  {journal} {Astrophys. Space Sci.}\ }\textbf
  {\bibinfo {volume} {361}},\ \bibinfo {eid} {269} (\bibinfo {year}
  {2016})}\BibitemShut {NoStop}%
\bibitem [{\citenamefont {A.A.~Hakimov}(2014)}]{Hakimov14a}%
  \BibitemOpen
  \bibfield  {author} {\bibinfo {author} {\bibfnamefont {M.~P.}\ \bibnamefont
  {A.A.~Hakimov}, \bibfnamefont {S.R.~Shaymatov}},\ }\href@noop {} {\bibfield
  {journal} {\bibinfo  {journal} {NEWS of the National Academy of Sciences of
  the Republic of Kazakhstan}\ }\textbf {\bibinfo {volume} {294}},\ \bibinfo
  {pages} {33} (\bibinfo {year} {2014})}\BibitemShut {NoStop}%
\bibitem [{\citenamefont {{Igata}}\ \emph {et~al.}(2012)\citenamefont
  {{Igata}}, \citenamefont {{Harada}},\ and\ \citenamefont
  {{Kimura}}}]{Igata12}%
  \BibitemOpen
  \bibfield  {author} {\bibinfo {author} {\bibfnamefont {T.}~\bibnamefont
  {{Igata}}}, \bibinfo {author} {\bibfnamefont {T.}~\bibnamefont {{Harada}}}, \
  and\ \bibinfo {author} {\bibfnamefont {M.}~\bibnamefont {{Kimura}}},\ }\href
  {\doibase 10.1103/PhysRevD.85.104028} {\bibfield  {journal} {\bibinfo
  {journal} {Phys. Rev. D}\ }\textbf {\bibinfo {volume} {85}},\ \bibinfo {eid}
  {104028} (\bibinfo {year} {2012})},\ \Eprint {http://arxiv.org/abs/1202.4859}
  {arXiv:1202.4859 [gr-qc]} \BibitemShut {NoStop}%
\bibitem [{\citenamefont {{Toshmatov}}\ \emph {et~al.}(2015)\citenamefont
  {{Toshmatov}}, \citenamefont {{Abdujabbarov}}, \citenamefont {{Ahmedov}},\
  and\ \citenamefont {{Stuchl{\'{\i}}k}}}]{Toshmatov15d}%
  \BibitemOpen
  \bibfield  {author} {\bibinfo {author} {\bibfnamefont {B.}~\bibnamefont
  {{Toshmatov}}}, \bibinfo {author} {\bibfnamefont {A.}~\bibnamefont
  {{Abdujabbarov}}}, \bibinfo {author} {\bibfnamefont {B.}~\bibnamefont
  {{Ahmedov}}}, \ and\ \bibinfo {author} {\bibfnamefont {Z.}~\bibnamefont
  {{Stuchl{\'{\i}}k}}},\ }\href {\doibase 10.1007/s10509-015-2533-y} {\bibfield
   {journal} {\bibinfo  {journal} {Astrophys Space Sci}\ }\textbf {\bibinfo
  {volume} {360}},\ \bibinfo {eid} {19} (\bibinfo {year} {2015})}\BibitemShut
  {NoStop}%
\bibitem [{\citenamefont {{Gao}}\ and\ \citenamefont {{Zhong}}(2011)}]{Gao11}%
  \BibitemOpen
  \bibfield  {author} {\bibinfo {author} {\bibfnamefont {S.}~\bibnamefont
  {{Gao}}}\ and\ \bibinfo {author} {\bibfnamefont {C.}~\bibnamefont
  {{Zhong}}},\ }\href {\doibase 10.1103/PhysRevD.84.044006} {\bibfield
  {journal} {\bibinfo  {journal} {Phys. Rev. D}\ }\textbf {\bibinfo {volume}
  {84}},\ \bibinfo {eid} {044006} (\bibinfo {year} {2011})},\ \Eprint
  {http://arxiv.org/abs/1106.2852} {arXiv:1106.2852 [gr-qc]} \BibitemShut
  {NoStop}%
\bibitem [{\citenamefont {{Liu}}\ \emph {et~al.}(2011)\citenamefont {{Liu}},
  \citenamefont {{Chen}}, \citenamefont {{Ding}},\ and\ \citenamefont
  {{Jing}}}]{Liu11}%
  \BibitemOpen
  \bibfield  {author} {\bibinfo {author} {\bibfnamefont {C.}~\bibnamefont
  {{Liu}}}, \bibinfo {author} {\bibfnamefont {S.}~\bibnamefont {{Chen}}},
  \bibinfo {author} {\bibfnamefont {C.}~\bibnamefont {{Ding}}}, \ and\ \bibinfo
  {author} {\bibfnamefont {J.}~\bibnamefont {{Jing}}},\ }\href {\doibase
  10.1016/j.physletb.2011.05.070} {\bibfield  {journal} {\bibinfo  {journal}
  {Physics Letters B}\ }\textbf {\bibinfo {volume} {701}},\ \bibinfo {pages}
  {285} (\bibinfo {year} {2011})},\ \Eprint {http://arxiv.org/abs/1012.5126}
  {arXiv:1012.5126 [gr-qc]} \BibitemShut {NoStop}%
\bibitem [{\citenamefont {{Patil}}\ \emph {et~al.}(2011)\citenamefont
  {{Patil}}, \citenamefont {{Joshi}},\ and\ \citenamefont
  {{Malafarina}}}]{Patil11b}%
  \BibitemOpen
  \bibfield  {author} {\bibinfo {author} {\bibfnamefont {M.}~\bibnamefont
  {{Patil}}}, \bibinfo {author} {\bibfnamefont {P.~S.}\ \bibnamefont
  {{Joshi}}}, \ and\ \bibinfo {author} {\bibfnamefont {D.}~\bibnamefont
  {{Malafarina}}},\ }\href {\doibase 10.1103/PhysRevD.83.064007} {\bibfield
  {journal} {\bibinfo  {journal} {Phys. Rev. D}\ }\textbf {\bibinfo {volume}
  {83}},\ \bibinfo {eid} {064007} (\bibinfo {year} {2011})},\ \Eprint
  {http://arxiv.org/abs/1102.2030} {arXiv:1102.2030 [gr-qc]} \BibitemShut
  {NoStop}%
\bibitem [{\citenamefont {{Patil}}\ \emph {et~al.}(2012)\citenamefont
  {{Patil}}, \citenamefont {{Joshi}}, \citenamefont {{Kimura}},\ and\
  \citenamefont {{Nakao}}}]{Patil12}%
  \BibitemOpen
  \bibfield  {author} {\bibinfo {author} {\bibfnamefont {M.}~\bibnamefont
  {{Patil}}}, \bibinfo {author} {\bibfnamefont {P.~S.}\ \bibnamefont
  {{Joshi}}}, \bibinfo {author} {\bibfnamefont {M.}~\bibnamefont {{Kimura}}}, \
  and\ \bibinfo {author} {\bibfnamefont {K.-i.}\ \bibnamefont {{Nakao}}},\
  }\href {\doibase 10.1103/PhysRevD.86.084023} {\bibfield  {journal} {\bibinfo
  {journal} {Phys. Rev. D}\ }\textbf {\bibinfo {volume} {86}},\ \bibinfo {eid}
  {084023} (\bibinfo {year} {2012})},\ \Eprint {http://arxiv.org/abs/1108.0288}
  {arXiv:1108.0288 [gr-qc]} \BibitemShut {NoStop}%
\bibitem [{\citenamefont {{Patil}}\ and\ \citenamefont
  {{Joshi}}(2010)}]{Patil10}%
  \BibitemOpen
  \bibfield  {author} {\bibinfo {author} {\bibfnamefont {M.}~\bibnamefont
  {{Patil}}}\ and\ \bibinfo {author} {\bibfnamefont {P.~S.}\ \bibnamefont
  {{Joshi}}},\ }\href {\doibase 10.1103/PhysRevD.82.104049} {\bibfield
  {journal} {\bibinfo  {journal} {Phys. Rev. D}\ }\textbf {\bibinfo {volume}
  {82}},\ \bibinfo {eid} {104049} (\bibinfo {year} {2010})},\ \Eprint
  {http://arxiv.org/abs/1011.5550} {arXiv:1011.5550 [gr-qc]} \BibitemShut
  {NoStop}%
\bibitem [{\citenamefont {{Zaslavskii}}(2011{\natexlab{a}})}]{Zaslavskii11}%
  \BibitemOpen
  \bibfield  {author} {\bibinfo {author} {\bibfnamefont {O.~B.}\ \bibnamefont
  {{Zaslavskii}}},\ }\href {\doibase 10.1103/PhysRevD.84.024007} {\bibfield
  {journal} {\bibinfo  {journal} {Phys. Rev. D}\ }\textbf {\bibinfo {volume}
  {84}},\ \bibinfo {eid} {024007} (\bibinfo {year} {2011}{\natexlab{a}})},\
  \Eprint {http://arxiv.org/abs/1104.4802} {arXiv:1104.4802 [gr-qc]}
  \BibitemShut {NoStop}%
\bibitem [{\citenamefont {{Zaslavskii}}(2010)}]{Zaslavskii10b}%
  \BibitemOpen
  \bibfield  {author} {\bibinfo {author} {\bibfnamefont {O.~B.}\ \bibnamefont
  {{Zaslavskii}}},\ }\href {\doibase 10.1134/S0021364010210010} {\bibfield
  {journal} {\bibinfo  {journal} {Soviet Journal of Experimental and
  Theoretical Physics Letters}\ }\textbf {\bibinfo {volume} {92}},\ \bibinfo
  {pages} {571} (\bibinfo {year} {2010})}\BibitemShut {NoStop}%
\bibitem [{\citenamefont {{Zaslavskii}}(2011{\natexlab{b}})}]{Zaslavskii11c}%
  \BibitemOpen
  \bibfield  {author} {\bibinfo {author} {\bibfnamefont {O.~B.}\ \bibnamefont
  {{Zaslavskii}}},\ }\href {\doibase 10.1088/0264-9381/28/10/105010} {\bibfield
   {journal} {\bibinfo  {journal} {Classical and Quantum Gravity}\ }\textbf
  {\bibinfo {volume} {28}},\ \bibinfo {eid} {105010} (\bibinfo {year}
  {2011}{\natexlab{b}})},\ \Eprint {http://arxiv.org/abs/1011.0167}
  {arXiv:1011.0167 [gr-qc]} \BibitemShut {NoStop}%
\bibitem [{\citenamefont {{Ghosh}}\ \emph {et~al.}(2014)\citenamefont
  {{Ghosh}}, \citenamefont {{Sheoran}},\ and\ \citenamefont
  {{Amir}}}]{Ghosh14}%
  \BibitemOpen
  \bibfield  {author} {\bibinfo {author} {\bibfnamefont {S.~G.}\ \bibnamefont
  {{Ghosh}}}, \bibinfo {author} {\bibfnamefont {P.}~\bibnamefont {{Sheoran}}},
  \ and\ \bibinfo {author} {\bibfnamefont {M.}~\bibnamefont {{Amir}}},\ }\href
  {\doibase 10.1103/PhysRevD.90.103006} {\bibfield  {journal} {\bibinfo
  {journal} {Phys. Rev. D}\ }\textbf {\bibinfo {volume} {90}},\ \bibinfo {eid}
  {103006} (\bibinfo {year} {2014})},\ \Eprint {http://arxiv.org/abs/1410.5588}
  {arXiv:1410.5588 [gr-qc]} \BibitemShut {NoStop}%
\bibitem [{\citenamefont {{Abdujabbarov}}\ \emph {et~al.}(2014)\citenamefont
  {{Abdujabbarov}}, \citenamefont {{Ahmedov}}, \citenamefont {{Rahimov}},\ and\
  \citenamefont {{Salikhbaev}}}]{Abdujabbarov14}%
  \BibitemOpen
  \bibfield  {author} {\bibinfo {author} {\bibfnamefont {A.}~\bibnamefont
  {{Abdujabbarov}}}, \bibinfo {author} {\bibfnamefont {B.}~\bibnamefont
  {{Ahmedov}}}, \bibinfo {author} {\bibfnamefont {O.}~\bibnamefont
  {{Rahimov}}}, \ and\ \bibinfo {author} {\bibfnamefont {U.}~\bibnamefont
  {{Salikhbaev}}},\ }\href {\doibase 10.1088/0031-8949/89/8/084008} {\bibfield
  {journal} {\bibinfo  {journal} {Physica Scripta}\ }\textbf {\bibinfo {volume}
  {89}},\ \bibinfo {eid} {084008} (\bibinfo {year} {2014})}\BibitemShut
  {NoStop}%
\bibitem [{\citenamefont {{Shaymatov}}\ \emph {et~al.}(2013)\citenamefont
  {{Shaymatov}}, \citenamefont {{Ahmedov}},\ and\ \citenamefont
  {{Abdujabbarov}}}]{Shaymatov13}%
  \BibitemOpen
  \bibfield  {author} {\bibinfo {author} {\bibfnamefont {S.~R.}\ \bibnamefont
  {{Shaymatov}}}, \bibinfo {author} {\bibfnamefont {B.~J.}\ \bibnamefont
  {{Ahmedov}}}, \ and\ \bibinfo {author} {\bibfnamefont {A.~A.}\ \bibnamefont
  {{Abdujabbarov}}},\ }\href {\doibase 10.1103/PhysRevD.88.024016} {\bibfield
  {journal} {\bibinfo  {journal} {Phys. Rev. D}\ }\textbf {\bibinfo {volume}
  {88}},\ \bibinfo {eid} {024016} (\bibinfo {year} {2013})}\BibitemShut
  {NoStop}%
\bibitem [{\citenamefont {{Dadhich}}\ \emph {et~al.}(2018)\citenamefont
  {{Dadhich}}, \citenamefont {{Tursunov}}, \citenamefont {{Ahmedov}},\ and\
  \citenamefont {{Stuchl{\'{\i}}k}}}]{Dadhich18}%
  \BibitemOpen
  \bibfield  {author} {\bibinfo {author} {\bibfnamefont {N.}~\bibnamefont
  {{Dadhich}}}, \bibinfo {author} {\bibfnamefont {A.}~\bibnamefont
  {{Tursunov}}}, \bibinfo {author} {\bibfnamefont {B.}~\bibnamefont
  {{Ahmedov}}}, \ and\ \bibinfo {author} {\bibfnamefont {Z.}~\bibnamefont
  {{Stuchl{\'{\i}}k}}},\ }\href {\doibase 10.1093/mnrasl/sly073} {\bibfield
  {journal} {\bibinfo  {journal} {Mon. Not. R. Astron. Soc}\ }\textbf {\bibinfo
  {volume} {478}},\ \bibinfo {pages} {L89} (\bibinfo {year} {2018})},\ \Eprint
  {http://arxiv.org/abs/1804.09679} {arXiv:1804.09679 [astro-ph.HE]}
  \BibitemShut {NoStop}%
\bibitem [{\citenamefont {{Zaslavskii}}(2012{\natexlab{b}})}]{Zaslavskii12b}%
  \BibitemOpen
  \bibfield  {author} {\bibinfo {author} {\bibfnamefont {O.~B.}\ \bibnamefont
  {{Zaslavskii}}},\ }\href {\doibase 10.1103/PhysRevD.86.084030} {\bibfield
  {journal} {\bibinfo  {journal} {Phys. Rev. D}\ }\textbf {\bibinfo {volume}
  {86}},\ \bibinfo {eid} {084030} (\bibinfo {year} {2012}{\natexlab{b}})},\
  \Eprint {http://arxiv.org/abs/1205.4410} {arXiv:1205.4410 [gr-qc]}
  \BibitemShut {NoStop}%
\bibitem [{\citenamefont {{Dadhich}}(2012)}]{dadhich12b}%
  \BibitemOpen
  \bibfield  {author} {\bibinfo {author} {\bibfnamefont {N.}~\bibnamefont
  {{Dadhich}}},\ }\href@noop {} {\bibfield  {journal} {\bibinfo  {journal}
  {ArXiv e-prints}\ } (\bibinfo {year} {2012})},\ \Eprint
  {http://arxiv.org/abs/1210.1041} {arXiv:1210.1041 [astro-ph.HE]} \BibitemShut
  {NoStop}%
\bibitem [{\citenamefont {{Wagh}}\ \emph {et~al.}(1985)\citenamefont {{Wagh}},
  \citenamefont {{Dhurandhar}},\ and\ \citenamefont {{Dadhich}}}]{wagh85}%
  \BibitemOpen
  \bibfield  {author} {\bibinfo {author} {\bibfnamefont {S.~M.}\ \bibnamefont
  {{Wagh}}}, \bibinfo {author} {\bibfnamefont {S.~V.}\ \bibnamefont
  {{Dhurandhar}}}, \ and\ \bibinfo {author} {\bibfnamefont {N.}~\bibnamefont
  {{Dadhich}}},\ }\href {\doibase 10.1086/162952} {\bibfield  {journal}
  {\bibinfo  {journal} {Astrophys J.}\ }\textbf {\bibinfo {volume} {290}},\
  \bibinfo {pages} {12} (\bibinfo {year} {1985})}\BibitemShut {NoStop}%
\bibitem [{\citenamefont {{Dhurandhar}}\ and\ \citenamefont
  {{Dadhich}}(1983)}]{Dhurandhar83}%
  \BibitemOpen
  \bibfield  {author} {\bibinfo {author} {\bibfnamefont {S.~V.}\ \bibnamefont
  {{Dhurandhar}}}\ and\ \bibinfo {author} {\bibfnamefont {N.}~\bibnamefont
  {{Dadhich}}},\ }\href@noop {} {\bibfield  {journal} {\bibinfo  {journal}
  {Bulletin of the Astronomical Society of India}\ }\textbf {\bibinfo {volume}
  {11}},\ \bibinfo {pages} {85} (\bibinfo {year} {1983})}\BibitemShut {NoStop}%
\bibitem [{\citenamefont {{Dhurandhar}}\ and\ \citenamefont
  {{Dadhich}}(1984{\natexlab{a}})}]{Dhurandhar84b}%
  \BibitemOpen
  \bibfield  {author} {\bibinfo {author} {\bibfnamefont {S.~V.}\ \bibnamefont
  {{Dhurandhar}}}\ and\ \bibinfo {author} {\bibfnamefont {N.}~\bibnamefont
  {{Dadhich}}},\ }\href {\doibase 10.1103/PhysRevD.30.1625} {\bibfield
  {journal} {\bibinfo  {journal} {Phys. Rev. D}\ }\textbf {\bibinfo {volume}
  {30}},\ \bibinfo {pages} {1625} (\bibinfo {year}
  {1984}{\natexlab{a}})}\BibitemShut {NoStop}%
\bibitem [{\citenamefont {{Dhurandhar}}\ and\ \citenamefont
  {{Dadhich}}(1984{\natexlab{b}})}]{Dhurandhar84}%
  \BibitemOpen
  \bibfield  {author} {\bibinfo {author} {\bibfnamefont {S.~V.}\ \bibnamefont
  {{Dhurandhar}}}\ and\ \bibinfo {author} {\bibfnamefont {N.}~\bibnamefont
  {{Dadhich}}},\ }\href {\doibase 10.1103/PhysRevD.29.2712} {\bibfield
  {journal} {\bibinfo  {journal} {Phys. Rev. D}\ }\textbf {\bibinfo {volume}
  {29}},\ \bibinfo {pages} {2712} (\bibinfo {year}
  {1984}{\natexlab{b}})}\BibitemShut {NoStop}%
\bibitem [{\citenamefont {{Abdujabbarov}}\ \emph
  {et~al.}(2011{\natexlab{b}})\citenamefont {{Abdujabbarov}}, \citenamefont
  {{Ahmedov}},\ and\ \citenamefont {{Ahmedov}}}]{Abdujabbarov11b}%
  \BibitemOpen
  \bibfield  {author} {\bibinfo {author} {\bibfnamefont {A.}~\bibnamefont
  {{Abdujabbarov}}}, \bibinfo {author} {\bibfnamefont {B.}~\bibnamefont
  {{Ahmedov}}}, \ and\ \bibinfo {author} {\bibfnamefont {B.}~\bibnamefont
  {{Ahmedov}}},\ }\href {\doibase 10.1103/PhysRevD.84.044044} {\bibfield
  {journal} {\bibinfo  {journal} {Phys. Rev. D}\ }\textbf {\bibinfo {volume}
  {84}},\ \bibinfo {eid} {044044} (\bibinfo {year} {2011}{\natexlab{b}})},\
  \Eprint {http://arxiv.org/abs/1107.5389} {arXiv:1107.5389 [astro-ph.SR]}
  \BibitemShut {NoStop}%
\bibitem [{\citenamefont {{Zipoy}}(1966)}]{Zipoy66}%
  \BibitemOpen
  \bibfield  {author} {\bibinfo {author} {\bibfnamefont {D.~M.}\ \bibnamefont
  {{Zipoy}}},\ }\href {\doibase 10.1063/1.1705005} {\bibfield  {journal}
  {\bibinfo  {journal} {Journal of Mathematical Physics}\ }\textbf {\bibinfo
  {volume} {7}},\ \bibinfo {pages} {1137} (\bibinfo {year} {1966})}\BibitemShut
  {NoStop}%
\bibitem [{\citenamefont {{Voorhees}}(1970)}]{Voorhees70}%
  \BibitemOpen
  \bibfield  {author} {\bibinfo {author} {\bibfnamefont {B.~H.}\ \bibnamefont
  {{Voorhees}}},\ }\href {\doibase 10.1103/PhysRevD.2.2119} {\bibfield
  {journal} {\bibinfo  {journal} {Phys. Rev. D}\ }\textbf {\bibinfo {volume}
  {2}},\ \bibinfo {pages} {2119} (\bibinfo {year} {1970})}\BibitemShut
  {NoStop}%
\bibitem [{\citenamefont {{Bambi}}(2017)}]{Bambi17c}%
  \BibitemOpen
  \bibfield  {author} {\bibinfo {author} {\bibfnamefont {C.}~\bibnamefont
  {{Bambi}}},\ }\href {\doibase 10.1103/RevModPhys.89.025001} {\bibfield
  {journal} {\bibinfo  {journal} {Reviews of Modern Physics}\ }\textbf
  {\bibinfo {volume} {89}},\ \bibinfo {eid} {025001} (\bibinfo {year}
  {2017})}\BibitemShut {NoStop}%
\bibitem [{\citenamefont {{Mazur}}\ and\ \citenamefont
  {{Mottola}}(2004)}]{Mazur2004}%
  \BibitemOpen
  \bibfield  {author} {\bibinfo {author} {\bibfnamefont {P.~O.}\ \bibnamefont
  {{Mazur}}}\ and\ \bibinfo {author} {\bibfnamefont {E.}~\bibnamefont
  {{Mottola}}},\ }\href {\doibase 10.1073/pnas.0402717101} {\bibfield
  {journal} {\bibinfo  {journal} {Proceedings of the National Academy of
  Science}\ }\textbf {\bibinfo {volume} {101}},\ \bibinfo {pages} {9545}
  (\bibinfo {year} {2004})},\ \Eprint
  {http://arxiv.org/abs/arXiv:gr-qc/0407075} {arXiv:gr-qc/0407075} \BibitemShut
  {NoStop}%
\bibitem [{\citenamefont {{Chirenti}}\ and\ \citenamefont
  {{Rezzolla}}(2007)}]{Chirenti07}%
  \BibitemOpen
  \bibfield  {author} {\bibinfo {author} {\bibfnamefont {C.~B.~M.~H.}\
  \bibnamefont {{Chirenti}}}\ and\ \bibinfo {author} {\bibfnamefont
  {L.}~\bibnamefont {{Rezzolla}}},\ }\href {\doibase
  10.1088/0264-9381/24/16/013} {\bibfield  {journal} {\bibinfo  {journal}
  {Class. Quantum Grav.}\ }\textbf {\bibinfo {volume} {24}},\ \bibinfo {pages}
  {4191} (\bibinfo {year} {2007})},\ \Eprint {http://arxiv.org/abs/0706.1513}
  {arXiv:0706.1513} \BibitemShut {NoStop}%
\bibitem [{\citenamefont {{Chirenti}}\ and\ \citenamefont
  {{Rezzolla}}(2016)}]{Chirenti16}%
  \BibitemOpen
  \bibfield  {author} {\bibinfo {author} {\bibfnamefont {C.}~\bibnamefont
  {{Chirenti}}}\ and\ \bibinfo {author} {\bibfnamefont {L.}~\bibnamefont
  {{Rezzolla}}},\ }\href {\doibase 10.1103/PhysRevD.94.084016} {\bibfield
  {journal} {\bibinfo  {journal} {Phys. Rev. D}\ }\textbf {\bibinfo {volume}
  {94}},\ \bibinfo {eid} {084016} (\bibinfo {year} {2016})},\ \Eprint
  {http://arxiv.org/abs/1602.08759} {arXiv:1602.08759 [gr-qc]} \BibitemShut
  {NoStop}%
\bibitem [{\citenamefont {{Cardoso}}\ \emph
  {et~al.}(2016{\natexlab{a}})\citenamefont {{Cardoso}}, \citenamefont
  {{Franzin}},\ and\ \citenamefont {{Pani}}}]{Cardoso16}%
  \BibitemOpen
  \bibfield  {author} {\bibinfo {author} {\bibfnamefont {V.}~\bibnamefont
  {{Cardoso}}}, \bibinfo {author} {\bibfnamefont {E.}~\bibnamefont
  {{Franzin}}}, \ and\ \bibinfo {author} {\bibfnamefont {P.}~\bibnamefont
  {{Pani}}},\ }\href {\doibase 10.1103/PhysRevLett.116.171101} {\bibfield
  {journal} {\bibinfo  {journal} {Physical Review Letters}\ }\textbf {\bibinfo
  {volume} {116}},\ \bibinfo {eid} {171101} (\bibinfo {year}
  {2016}{\natexlab{a}})},\ \Eprint {http://arxiv.org/abs/1602.07309}
  {arXiv:1602.07309 [gr-qc]} \BibitemShut {NoStop}%
\bibitem [{\citenamefont {{Cardoso}}\ \emph
  {et~al.}(2016{\natexlab{b}})\citenamefont {{Cardoso}}, \citenamefont
  {{Franzin}},\ and\ \citenamefont {{Pani}}}]{Cardoso16erratum}%
  \BibitemOpen
  \bibfield  {author} {\bibinfo {author} {\bibfnamefont {V.}~\bibnamefont
  {{Cardoso}}}, \bibinfo {author} {\bibfnamefont {E.}~\bibnamefont
  {{Franzin}}}, \ and\ \bibinfo {author} {\bibfnamefont {P.}~\bibnamefont
  {{Pani}}},\ }\href {\doibase 10.1103/PhysRevLett.117.089902} {\bibfield
  {journal} {\bibinfo  {journal} {Physical Review Letters}\ }\textbf {\bibinfo
  {volume} {117}},\ \bibinfo {eid} {089902} (\bibinfo {year}
  {2016}{\natexlab{b}})}\BibitemShut {NoStop}%
\bibitem [{\citenamefont {{Carballo-Rubio}}\ \emph {et~al.}(2018)\citenamefont
  {{Carballo-Rubio}}, \citenamefont {{Di Filippo}}, \citenamefont
  {{Liberati}},\ and\ \citenamefont {{Visser}}}]{Carballo-Rubio18}%
  \BibitemOpen
  \bibfield  {author} {\bibinfo {author} {\bibfnamefont {R.}~\bibnamefont
  {{Carballo-Rubio}}}, \bibinfo {author} {\bibfnamefont {F.}~\bibnamefont {{Di
  Filippo}}}, \bibinfo {author} {\bibfnamefont {S.}~\bibnamefont {{Liberati}}},
  \ and\ \bibinfo {author} {\bibfnamefont {M.}~\bibnamefont {{Visser}}},\
  }\href@noop {} {\bibfield  {journal} {\bibinfo  {journal} {ArXiv e-prints}\ }
  (\bibinfo {year} {2018})},\ \Eprint {http://arxiv.org/abs/1809.08238}
  {arXiv:1809.08238 [gr-qc]} \BibitemShut {NoStop}%
\bibitem [{\citenamefont {{Johannsen}}\ and\ \citenamefont
  {{Psaltis}}(2011)}]{Johannsen11}%
  \BibitemOpen
  \bibfield  {author} {\bibinfo {author} {\bibfnamefont {T.}~\bibnamefont
  {{Johannsen}}}\ and\ \bibinfo {author} {\bibfnamefont {D.}~\bibnamefont
  {{Psaltis}}},\ }\href {\doibase 10.1103/PhysRevD.83.124015} {\bibfield
  {journal} {\bibinfo  {journal} {Phys. Rev. D}\ }\textbf {\bibinfo {volume}
  {83}},\ \bibinfo {eid} {124015} (\bibinfo {year} {2011})},\ \Eprint
  {http://arxiv.org/abs/1105.3191} {arXiv:1105.3191 [gr-qc]} \BibitemShut
  {NoStop}%
\bibitem [{\citenamefont {{Yagi}}\ and\ \citenamefont
  {{Stein}}(2016)}]{Yagi16}%
  \BibitemOpen
  \bibfield  {author} {\bibinfo {author} {\bibfnamefont {K.}~\bibnamefont
  {{Yagi}}}\ and\ \bibinfo {author} {\bibfnamefont {L.~C.}\ \bibnamefont
  {{Stein}}},\ }\href {\doibase 10.1088/0264-9381/33/5/054001} {\bibfield
  {journal} {\bibinfo  {journal} {Classical and Quantum Gravity}\ }\textbf
  {\bibinfo {volume} {33}},\ \bibinfo {eid} {054001} (\bibinfo {year}
  {2016})},\ \Eprint {http://arxiv.org/abs/1602.02413} {arXiv:1602.02413
  [gr-qc]} \BibitemShut {NoStop}%
\bibitem [{\citenamefont {{Erez}}\ and\ \citenamefont {{Rosen}}()}]{Erez59}%
  \BibitemOpen
  \bibfield  {author} {\bibinfo {author} {\bibfnamefont {G.}~\bibnamefont
  {{Erez}}}\ and\ \bibinfo {author} {\bibfnamefont {N.}~\bibnamefont
  {{Rosen}}},\ }\href@noop {} {\bibfield  {journal} {\bibinfo  {journal} {Bull.
  Res. Council Israel}\ }\textbf {\bibinfo {volume} {8}},\ \bibinfo {eid}
  {47}}\BibitemShut {NoStop}%
\bibitem [{\citenamefont {{Herrera}}\ and\ \citenamefont
  {{Pastora}}(2000)}]{Herrera00}%
  \BibitemOpen
  \bibfield  {author} {\bibinfo {author} {\bibfnamefont {L.}~\bibnamefont
  {{Herrera}}}\ and\ \bibinfo {author} {\bibfnamefont {J.~L.~H.}\ \bibnamefont
  {{Pastora}}},\ }\href {\doibase 10.1063/1.1319517} {\bibfield  {journal}
  {\bibinfo  {journal} {Journal of Mathematical Physics}\ }\textbf {\bibinfo
  {volume} {41}},\ \bibinfo {pages} {7544} (\bibinfo {year} {2000})},\ \Eprint
  {http://arxiv.org/abs/gr-qc/0010003} {gr-qc/0010003} \BibitemShut {NoStop}%
\bibitem [{\citenamefont {{Hern{\'a}ndez-Pastora}}\ and\ \citenamefont
  {{Mart{\'{\i}}n}}(1994)}]{Hernandez-Pastora94}%
  \BibitemOpen
  \bibfield  {author} {\bibinfo {author} {\bibfnamefont {J.~L.}\ \bibnamefont
  {{Hern{\'a}ndez-Pastora}}}\ and\ \bibinfo {author} {\bibfnamefont
  {J.}~\bibnamefont {{Mart{\'{\i}}n}}},\ }\href {\doibase 10.1007/BF02107146}
  {\bibfield  {journal} {\bibinfo  {journal} {General Relativity and
  Gravitation}\ }\textbf {\bibinfo {volume} {26}},\ \bibinfo {pages} {877}
  (\bibinfo {year} {1994})}\BibitemShut {NoStop}%
\bibitem [{\citenamefont {{Hernandez}}(1967)}]{Hernandez67}%
  \BibitemOpen
  \bibfield  {author} {\bibinfo {author} {\bibfnamefont {W.~C.}\ \bibnamefont
  {{Hernandez}}},\ }\href {\doibase 10.1103/PhysRev.153.1359} {\bibfield
  {journal} {\bibinfo  {journal} {Physical Review}\ }\textbf {\bibinfo {volume}
  {153}},\ \bibinfo {pages} {1359} (\bibinfo {year} {1967})}\BibitemShut
  {NoStop}%
\bibitem [{\citenamefont {{Stewart}}\ \emph {et~al.}(1982)\citenamefont
  {{Stewart}}, \citenamefont {{Papadopoulos}}, \citenamefont {{Witten}},
  \citenamefont {{Berezdivin}},\ and\ \citenamefont {{Herrera}}}]{Stewart82}%
  \BibitemOpen
  \bibfield  {author} {\bibinfo {author} {\bibfnamefont {B.~W.}\ \bibnamefont
  {{Stewart}}}, \bibinfo {author} {\bibfnamefont {D.}~\bibnamefont
  {{Papadopoulos}}}, \bibinfo {author} {\bibfnamefont {L.}~\bibnamefont
  {{Witten}}}, \bibinfo {author} {\bibfnamefont {R.}~\bibnamefont
  {{Berezdivin}}}, \ and\ \bibinfo {author} {\bibfnamefont {L.}~\bibnamefont
  {{Herrera}}},\ }\href {\doibase 10.1007/BF00756201} {\bibfield  {journal}
  {\bibinfo  {journal} {General Relativity and Gravitation}\ }\textbf {\bibinfo
  {volume} {14}},\ \bibinfo {pages} {97} (\bibinfo {year} {1982})}\BibitemShut
  {NoStop}%
\bibitem [{\citenamefont {{Herrera}}\ \emph {et~al.}(2005)\citenamefont
  {{Herrera}}, \citenamefont {{Magli}},\ and\ \citenamefont
  {{Malafarina}}}]{Herrera05}%
  \BibitemOpen
  \bibfield  {author} {\bibinfo {author} {\bibfnamefont {L.}~\bibnamefont
  {{Herrera}}}, \bibinfo {author} {\bibfnamefont {G.}~\bibnamefont {{Magli}}},
  \ and\ \bibinfo {author} {\bibfnamefont {D.}~\bibnamefont {{Malafarina}}},\
  }\href {\doibase 10.1007/s10714-005-0120-1} {\bibfield  {journal} {\bibinfo
  {journal} {General Relativity and Gravitation}\ }\textbf {\bibinfo {volume}
  {37}},\ \bibinfo {pages} {1371} (\bibinfo {year} {2005})},\ \Eprint
  {http://arxiv.org/abs/gr-qc/0407037} {gr-qc/0407037} \BibitemShut {NoStop}%
\bibitem [{\citenamefont {{Virbhadra}}(1996)}]{Virbhadra96}%
  \BibitemOpen
  \bibfield  {author} {\bibinfo {author} {\bibfnamefont {K.~S.}\ \bibnamefont
  {{Virbhadra}}},\ }\href@noop {} {\bibfield  {journal} {\bibinfo  {journal}
  {ArXiv General Relativity and Quantum Cosmology e-prints}\ } (\bibinfo {year}
  {1996})},\ \Eprint {http://arxiv.org/abs/gr-qc/9606004} {gr-qc/9606004}
  \BibitemShut {NoStop}%
\bibitem [{\citenamefont {{Herrera}}\ \emph {et~al.}(1999)\citenamefont
  {{Herrera}}, \citenamefont {{Paiva}},\ and\ \citenamefont
  {{Santos}}}]{Herrera99}%
  \BibitemOpen
  \bibfield  {author} {\bibinfo {author} {\bibfnamefont {L.}~\bibnamefont
  {{Herrera}}}, \bibinfo {author} {\bibfnamefont {F.~M.}\ \bibnamefont
  {{Paiva}}}, \ and\ \bibinfo {author} {\bibfnamefont {N.~O.}\ \bibnamefont
  {{Santos}}},\ }\href {\doibase 10.1063/1.532943} {\bibfield  {journal}
  {\bibinfo  {journal} {Journal of Mathematical Physics}\ }\textbf {\bibinfo
  {volume} {40}},\ \bibinfo {pages} {4064} (\bibinfo {year} {1999})},\ \Eprint
  {http://arxiv.org/abs/gr-qc/9810079} {gr-qc/9810079} \BibitemShut {NoStop}%
\bibitem [{\citenamefont {{Liu}}\ \emph {et~al.}(2018)\citenamefont {{Liu}},
  \citenamefont {{Zhou}},\ and\ \citenamefont {{Bambi}}}]{Liu18}%
  \BibitemOpen
  \bibfield  {author} {\bibinfo {author} {\bibfnamefont {H.}~\bibnamefont
  {{Liu}}}, \bibinfo {author} {\bibfnamefont {M.}~\bibnamefont {{Zhou}}}, \
  and\ \bibinfo {author} {\bibfnamefont {C.}~\bibnamefont {{Bambi}}},\ }\href
  {\doibase 10.1088/1475-7516/2018/08/044} {\bibfield  {journal} {\bibinfo
  {journal} {JCAP}\ }\textbf {\bibinfo {volume} {8}},\ \bibinfo {eid} {044}
  (\bibinfo {year} {2018})},\ \Eprint {http://arxiv.org/abs/1801.00867}
  {arXiv:1801.00867 [gr-qc]} \BibitemShut {NoStop}%
\bibitem [{\citenamefont {{Chowdhury}}\ \emph {et~al.}(2012)\citenamefont
  {{Chowdhury}}, \citenamefont {{Patil}}, \citenamefont {{Malafarina}},\ and\
  \citenamefont {{Joshi}}}]{Chowdhury12}%
  \BibitemOpen
  \bibfield  {author} {\bibinfo {author} {\bibfnamefont {A.~N.}\ \bibnamefont
  {{Chowdhury}}}, \bibinfo {author} {\bibfnamefont {M.}~\bibnamefont
  {{Patil}}}, \bibinfo {author} {\bibfnamefont {D.}~\bibnamefont
  {{Malafarina}}}, \ and\ \bibinfo {author} {\bibfnamefont {P.~S.}\
  \bibnamefont {{Joshi}}},\ }\href {\doibase 10.1103/PhysRevD.85.104031}
  {\bibfield  {journal} {\bibinfo  {journal} {Phys. Rev. D.}\ }\textbf
  {\bibinfo {volume} {85}},\ \bibinfo {eid} {104031} (\bibinfo {year}
  {2012})},\ \Eprint {http://arxiv.org/abs/1112.2522} {arXiv:1112.2522 [gr-qc]}
  \BibitemShut {NoStop}%
\bibitem [{\citenamefont {{Joshi}}\ \emph {et~al.}(2011)\citenamefont
  {{Joshi}}, \citenamefont {{Malafarina}},\ and\ \citenamefont
  {{Narayan}}}]{Joshi11}%
  \BibitemOpen
  \bibfield  {author} {\bibinfo {author} {\bibfnamefont {P.~S.}\ \bibnamefont
  {{Joshi}}}, \bibinfo {author} {\bibfnamefont {D.}~\bibnamefont
  {{Malafarina}}}, \ and\ \bibinfo {author} {\bibfnamefont {R.}~\bibnamefont
  {{Narayan}}},\ }\href {\doibase 10.1088/0264-9381/28/23/235018} {\bibfield
  {journal} {\bibinfo  {journal} {Classical and Quantum Gravity}\ }\textbf
  {\bibinfo {volume} {28}},\ \bibinfo {eid} {235018} (\bibinfo {year}
  {2011})},\ \Eprint {http://arxiv.org/abs/1106.5438} {arXiv:1106.5438 [gr-qc]}
  \BibitemShut {NoStop}%
\bibitem [{\citenamefont {{Joshi}}\ \emph {et~al.}(2014)\citenamefont
  {{Joshi}}, \citenamefont {{Malafarina}},\ and\ \citenamefont
  {{Narayan}}}]{Joshi14}%
  \BibitemOpen
  \bibfield  {author} {\bibinfo {author} {\bibfnamefont {P.~S.}\ \bibnamefont
  {{Joshi}}}, \bibinfo {author} {\bibfnamefont {D.}~\bibnamefont
  {{Malafarina}}}, \ and\ \bibinfo {author} {\bibfnamefont {R.}~\bibnamefont
  {{Narayan}}},\ }\href {\doibase 10.1088/0264-9381/31/1/015002} {\bibfield
  {journal} {\bibinfo  {journal} {Classical and Quantum Gravity}\ }\textbf
  {\bibinfo {volume} {31}},\ \bibinfo {eid} {015002} (\bibinfo {year}
  {2014})},\ \Eprint {http://arxiv.org/abs/1304.7331} {arXiv:1304.7331 [gr-qc]}
  \BibitemShut {NoStop}%
\bibitem [{\citenamefont {{Bambi}}\ and\ \citenamefont
  {{Malafarina}}(2013)}]{Bambi13d}%
  \BibitemOpen
  \bibfield  {author} {\bibinfo {author} {\bibfnamefont {C.}~\bibnamefont
  {{Bambi}}}\ and\ \bibinfo {author} {\bibfnamefont {D.}~\bibnamefont
  {{Malafarina}}},\ }\href {\doibase 10.1103/PhysRevD.88.064022} {\bibfield
  {journal} {\bibinfo  {journal} {Phys. Rev. D}\ }\textbf {\bibinfo {volume}
  {88}},\ \bibinfo {eid} {064022} (\bibinfo {year} {2013})},\ \Eprint
  {http://arxiv.org/abs/1307.2106} {arXiv:1307.2106 [gr-qc]} \BibitemShut
  {NoStop}%
\bibitem [{\citenamefont {{Shafee}}\ \emph {et~al.}(2006)\citenamefont
  {{Shafee}}, \citenamefont {{McClintock}}, \citenamefont {{Narayan}},
  \citenamefont {{Davis}}, \citenamefont {{Li}},\ and\ \citenamefont
  {{Remillard}}}]{Shafee06}%
  \BibitemOpen
  \bibfield  {author} {\bibinfo {author} {\bibfnamefont {R.}~\bibnamefont
  {{Shafee}}}, \bibinfo {author} {\bibfnamefont {J.~E.}\ \bibnamefont
  {{McClintock}}}, \bibinfo {author} {\bibfnamefont {R.}~\bibnamefont
  {{Narayan}}}, \bibinfo {author} {\bibfnamefont {S.~W.}\ \bibnamefont
  {{Davis}}}, \bibinfo {author} {\bibfnamefont {L.-X.}\ \bibnamefont {{Li}}}, \
  and\ \bibinfo {author} {\bibfnamefont {R.~A.}\ \bibnamefont {{Remillard}}},\
  }\href {\doibase 10.1086/498938} {\bibfield  {journal} {\bibinfo  {journal}
  {Astrophys. J. Lett}\ }\textbf {\bibinfo {volume} {636}},\ \bibinfo {pages}
  {L113} (\bibinfo {year} {2006})},\ \Eprint
  {http://arxiv.org/abs/astro-ph/0508302} {astro-ph/0508302} \BibitemShut
  {NoStop}%
\bibitem [{\citenamefont {{Shafee}}\ \emph {et~al.}(2008)\citenamefont
  {{Shafee}}, \citenamefont {{Narayan}},\ and\ \citenamefont
  {{McClintock}}}]{Shafee08}%
  \BibitemOpen
  \bibfield  {author} {\bibinfo {author} {\bibfnamefont {R.}~\bibnamefont
  {{Shafee}}}, \bibinfo {author} {\bibfnamefont {R.}~\bibnamefont {{Narayan}}},
  \ and\ \bibinfo {author} {\bibfnamefont {J.~E.}\ \bibnamefont
  {{McClintock}}},\ }\href {\doibase 10.1086/527346} {\bibfield  {journal}
  {\bibinfo  {journal} {Astrophys J.}\ }\textbf {\bibinfo {volume} {676}},\
  \bibinfo {pages} {549} (\bibinfo {year} {2008})},\ \Eprint
  {http://arxiv.org/abs/0705.2244} {arXiv:0705.2244} \BibitemShut {NoStop}%
\bibitem [{\citenamefont {{Steiner}}\ \emph {et~al.}(2009)\citenamefont
  {{Steiner}}, \citenamefont {{McClintock}}, \citenamefont {{Remillard}},
  \citenamefont {{Narayan}},\ and\ \citenamefont {{Gou}}}]{Steiner09}%
  \BibitemOpen
  \bibfield  {author} {\bibinfo {author} {\bibfnamefont {J.~F.}\ \bibnamefont
  {{Steiner}}}, \bibinfo {author} {\bibfnamefont {J.~E.}\ \bibnamefont
  {{McClintock}}}, \bibinfo {author} {\bibfnamefont {R.~A.}\ \bibnamefont
  {{Remillard}}}, \bibinfo {author} {\bibfnamefont {R.}~\bibnamefont
  {{Narayan}}}, \ and\ \bibinfo {author} {\bibfnamefont {L.}~\bibnamefont
  {{Gou}}},\ }\href {\doibase 10.1088/0004-637X/701/2/L83} {\bibfield
  {journal} {\bibinfo  {journal} {Astrophys. J. Lett}\ }\textbf {\bibinfo
  {volume} {701}},\ \bibinfo {pages} {L83} (\bibinfo {year} {2009})},\ \Eprint
  {http://arxiv.org/abs/0907.2920} {arXiv:0907.2920 [astro-ph.HE]} \BibitemShut
  {NoStop}%
\bibitem [{\citenamefont {{Steiner}}\ \emph {et~al.}(2010)\citenamefont
  {{Steiner}}, \citenamefont {{McClintock}}, \citenamefont {{Remillard}},
  \citenamefont {{Gou}}, \citenamefont {{Yamada}},\ and\ \citenamefont
  {{Narayan}}}]{Steiner10}%
  \BibitemOpen
  \bibfield  {author} {\bibinfo {author} {\bibfnamefont {J.~F.}\ \bibnamefont
  {{Steiner}}}, \bibinfo {author} {\bibfnamefont {J.~E.}\ \bibnamefont
  {{McClintock}}}, \bibinfo {author} {\bibfnamefont {R.~A.}\ \bibnamefont
  {{Remillard}}}, \bibinfo {author} {\bibfnamefont {L.}~\bibnamefont {{Gou}}},
  \bibinfo {author} {\bibfnamefont {S.}~\bibnamefont {{Yamada}}}, \ and\
  \bibinfo {author} {\bibfnamefont {R.}~\bibnamefont {{Narayan}}},\ }\href
  {\doibase 10.1088/2041-8205/718/2/L117} {\bibfield  {journal} {\bibinfo
  {journal} {Astrophys. J. Lett.}\ }\textbf {\bibinfo {volume} {718}},\
  \bibinfo {pages} {L117} (\bibinfo {year} {2010})},\ \Eprint
  {http://arxiv.org/abs/1006.5729} {arXiv:1006.5729 [astro-ph.HE]} \BibitemShut
  {NoStop}%
\bibitem [{\citenamefont {{McClintock}}\ \emph {et~al.}(2014)\citenamefont
  {{McClintock}}, \citenamefont {{Narayan}},\ and\ \citenamefont
  {{Steiner}}}]{McClintock14}%
  \BibitemOpen
  \bibfield  {author} {\bibinfo {author} {\bibfnamefont {J.~E.}\ \bibnamefont
  {{McClintock}}}, \bibinfo {author} {\bibfnamefont {R.}~\bibnamefont
  {{Narayan}}}, \ and\ \bibinfo {author} {\bibfnamefont {J.~F.}\ \bibnamefont
  {{Steiner}}},\ }\href {\doibase 10.1007/s11214-013-0003-9} {\bibfield
  {journal} {\bibinfo  {journal} {Space Science Reviews}\ }\textbf {\bibinfo
  {volume} {183}},\ \bibinfo {pages} {295} (\bibinfo {year} {2014})},\ \Eprint
  {http://arxiv.org/abs/1303.1583} {arXiv:1303.1583 [astro-ph.HE]} \BibitemShut
  {NoStop}%
\bibitem [{\citenamefont {{Gou}}\ \emph {et~al.}(2014)\citenamefont {{Gou}},
  \citenamefont {{McClintock}}, \citenamefont {{Remillard}}, \citenamefont
  {{Steiner}}, \citenamefont {{Reid}}, \citenamefont {{Orosz}}, \citenamefont
  {{Narayan}}, \citenamefont {{Hanke}},\ and\ \citenamefont
  {{Garc{\'{\i}}a}}}]{Gou14}%
  \BibitemOpen
  \bibfield  {author} {\bibinfo {author} {\bibfnamefont {L.}~\bibnamefont
  {{Gou}}}, \bibinfo {author} {\bibfnamefont {J.~E.}\ \bibnamefont
  {{McClintock}}}, \bibinfo {author} {\bibfnamefont {R.~A.}\ \bibnamefont
  {{Remillard}}}, \bibinfo {author} {\bibfnamefont {J.~F.}\ \bibnamefont
  {{Steiner}}}, \bibinfo {author} {\bibfnamefont {M.~J.}\ \bibnamefont
  {{Reid}}}, \bibinfo {author} {\bibfnamefont {J.~A.}\ \bibnamefont {{Orosz}}},
  \bibinfo {author} {\bibfnamefont {R.}~\bibnamefont {{Narayan}}}, \bibinfo
  {author} {\bibfnamefont {M.}~\bibnamefont {{Hanke}}}, \ and\ \bibinfo
  {author} {\bibfnamefont {J.}~\bibnamefont {{Garc{\'{\i}}a}}},\ }\href
  {\doibase 10.1088/0004-637X/790/1/29} {\bibfield  {journal} {\bibinfo
  {journal} {Astrophys. J.}\ }\textbf {\bibinfo {volume} {790}},\ \bibinfo
  {eid} {29} (\bibinfo {year} {2014})},\ \Eprint
  {http://arxiv.org/abs/1308.4760} {arXiv:1308.4760 [astro-ph.HE]} \BibitemShut
  {NoStop}%
\bibitem [{\citenamefont {{Steiner}}\ \emph {et~al.}(2011)\citenamefont
  {{Steiner}}, \citenamefont {{Reis}}, \citenamefont {{McClintock}},
  \citenamefont {{Narayan}}, \citenamefont {{Remillard}}, \citenamefont
  {{Orosz}}, \citenamefont {{Gou}}, \citenamefont {{Fabian}},\ and\
  \citenamefont {{Torres}}}]{Steiner11}%
  \BibitemOpen
  \bibfield  {author} {\bibinfo {author} {\bibfnamefont {J.~F.}\ \bibnamefont
  {{Steiner}}}, \bibinfo {author} {\bibfnamefont {R.~C.}\ \bibnamefont
  {{Reis}}}, \bibinfo {author} {\bibfnamefont {J.~E.}\ \bibnamefont
  {{McClintock}}}, \bibinfo {author} {\bibfnamefont {R.}~\bibnamefont
  {{Narayan}}}, \bibinfo {author} {\bibfnamefont {R.~A.}\ \bibnamefont
  {{Remillard}}}, \bibinfo {author} {\bibfnamefont {J.~A.}\ \bibnamefont
  {{Orosz}}}, \bibinfo {author} {\bibfnamefont {L.}~\bibnamefont {{Gou}}},
  \bibinfo {author} {\bibfnamefont {A.~C.}\ \bibnamefont {{Fabian}}}, \ and\
  \bibinfo {author} {\bibfnamefont {M.~A.~P.}\ \bibnamefont {{Torres}}},\
  }\href {\doibase 10.1111/j.1365-2966.2011.19089.x} {\bibfield  {journal}
  {\bibinfo  {journal} {Mon. Not. R. Astron. Soc}\ }\textbf {\bibinfo {volume}
  {416}},\ \bibinfo {pages} {941} (\bibinfo {year} {2011})},\ \Eprint
  {http://arxiv.org/abs/1010.1013} {arXiv:1010.1013 [astro-ph.HE]} \BibitemShut
  {NoStop}%
\bibitem [{\citenamefont {{Broderick}}\ and\ \citenamefont
  {{Narayan}}(2007)}]{Broderick2007}%
  \BibitemOpen
  \bibfield  {author} {\bibinfo {author} {\bibfnamefont {A.~E.}\ \bibnamefont
  {{Broderick}}}\ and\ \bibinfo {author} {\bibfnamefont {R.}~\bibnamefont
  {{Narayan}}},\ }\href {\doibase 10.1088/0264-9381/24/3/009} {\bibfield
  {journal} {\bibinfo  {journal} {Class. Quantum Grav.}\ }\textbf {\bibinfo
  {volume} {24}},\ \bibinfo {pages} {659} (\bibinfo {year} {2007})},\ \Eprint
  {http://arxiv.org/abs/arXiv:gr-qc/0701154} {arXiv:gr-qc/0701154} \BibitemShut
  {NoStop}%
\bibitem [{\citenamefont {{Abdujabbarov}}\ \emph
  {et~al.}(2013{\natexlab{b}})\citenamefont {{Abdujabbarov}}, \citenamefont
  {{Ahmedov}},\ and\ \citenamefont {{Jurayeva}}}]{Abdujabbarov13b}%
  \BibitemOpen
  \bibfield  {author} {\bibinfo {author} {\bibfnamefont {A.~A.}\ \bibnamefont
  {{Abdujabbarov}}}, \bibinfo {author} {\bibfnamefont {B.~J.}\ \bibnamefont
  {{Ahmedov}}}, \ and\ \bibinfo {author} {\bibfnamefont {N.~B.}\ \bibnamefont
  {{Jurayeva}}},\ }\href {\doibase 10.1103/PhysRevD.87.064042} {\bibfield
  {journal} {\bibinfo  {journal} {Phys. Rev. D.}\ }\textbf {\bibinfo {volume}
  {87}},\ \bibinfo {eid} {064042} (\bibinfo {year}
  {2013}{\natexlab{b}})}\BibitemShut {NoStop}%
\bibitem [{\citenamefont {{Kong}}\ \emph {et~al.}(2014)\citenamefont {{Kong}},
  \citenamefont {{Li}},\ and\ \citenamefont {{Bambi}}}]{Kong14}%
  \BibitemOpen
  \bibfield  {author} {\bibinfo {author} {\bibfnamefont {L.}~\bibnamefont
  {{Kong}}}, \bibinfo {author} {\bibfnamefont {Z.}~\bibnamefont {{Li}}}, \ and\
  \bibinfo {author} {\bibfnamefont {C.}~\bibnamefont {{Bambi}}},\ }\href
  {\doibase 10.1088/0004-637X/797/2/78} {\bibfield  {journal} {\bibinfo
  {journal} {Astrophys. J.}\ }\textbf {\bibinfo {volume} {797}},\ \bibinfo
  {eid} {78} (\bibinfo {year} {2014})},\ \Eprint
  {http://arxiv.org/abs/1405.1508} {arXiv:1405.1508 [gr-qc]} \BibitemShut
  {NoStop}%
\bibitem [{\citenamefont {{Gravity Collaboration}}\ \emph
  {et~al.}(2018{\natexlab{a}})\citenamefont {{Gravity Collaboration}},
  \citenamefont {{Abuter}}, \citenamefont {{Amorim}}, \citenamefont {{Anugu}},
  \citenamefont {{Baub{\"o}ck}}, \citenamefont {{Benisty}}, \citenamefont
  {{Berger}}, \citenamefont {{Blind}}, \citenamefont {{Bonnet}}, \citenamefont
  {{Brandner}}, \citenamefont {{Buron}}, \citenamefont {{Collin}},
  \citenamefont {{Chapron}}, \citenamefont {{Cl{\'e}net}}, \citenamefont
  {{Coud{\'e} Du Foresto}}, \citenamefont {{de Zeeuw}}, \citenamefont {{Deen}},
  \citenamefont {{Delplancke-Str{\"o}bele}}, \citenamefont {{Dembet}},
  \citenamefont {{Dexter}}, \citenamefont {{Duvert}}, \citenamefont {{Eckart}},
  \citenamefont {{Eisenhauer}}, \citenamefont {{Finger}}, \citenamefont
  {{F{\"o}rster Schreiber}}, \citenamefont {{F{\'e}dou}}, \citenamefont
  {{Garcia}}, \citenamefont {{Garcia Lopez}}, \citenamefont {{Gao}},
  \citenamefont {{Gendron}}, \citenamefont {{Genzel}}, \citenamefont
  {{Gillessen}}, \citenamefont {{Gordo}}, \citenamefont {{Habibi}},
  \citenamefont {{Haubois}}, \citenamefont {{Haug}}, \citenamefont
  {{Hau{\ss}mann}}, \citenamefont {{Henning}}, \citenamefont {{Hippler}},
  \citenamefont {{Horrobin}}, \citenamefont {{Hubert}}, \citenamefont
  {{Hubin}}, \citenamefont {{Jimenez Rosales}}, \citenamefont {{Jochum}},
  \citenamefont {{Jocou}}, \citenamefont {{Kaufer}}, \citenamefont {{Kellner}},
  \citenamefont {{Kendrew}}, \citenamefont {{Kervella}}, \citenamefont {{Kok}},
  \citenamefont {{Kulas}}, \citenamefont {{Lacour}}, \citenamefont
  {{Lapeyr{\`e}re}}, \citenamefont {{Lazareff}}, \citenamefont {{Le Bouquin}},
  \citenamefont {{L{\'e}na}}, \citenamefont {{Lippa}}, \citenamefont
  {{Lenzen}}, \citenamefont {{M{\'e}rand}}, \citenamefont {{M{\"u}ler}},
  \citenamefont {{Neumann}}, \citenamefont {{Ott}}, \citenamefont {{Palanca}},
  \citenamefont {{Paumard}}, \citenamefont {{Pasquini}}, \citenamefont
  {{Perraut}}, \citenamefont {{Perrin}}, \citenamefont {{Pfuhl}}, \citenamefont
  {{Plewa}}, \citenamefont {{Rabien}}, \citenamefont {{Ram{\'{\i}}rez}},
  \citenamefont {{Ramos}}, \citenamefont {{Rau}}, \citenamefont
  {{Rodr{\'{\i}}guez-Coira}}, \citenamefont {{Rohloff}}, \citenamefont
  {{Rousset}}, \citenamefont {{Sanchez-Bermudez}}, \citenamefont
  {{Scheithauer}}, \citenamefont {{Sch{\"o}ller}}, \citenamefont {{Schuler}},
  \citenamefont {{Spyromilio}}, \citenamefont {{Straub}}, \citenamefont
  {{Straubmeier}}, \citenamefont {{Sturm}}, \citenamefont {{Tacconi}},
  \citenamefont {{Tristram}}, \citenamefont {{Vincent}}, \citenamefont {{von
  Fellenberg}}, \citenamefont {{Wank}}, \citenamefont {{Waisberg}},
  \citenamefont {{Widmann}}, \citenamefont {{Wieprecht}}, \citenamefont
  {{Wiest}}, \citenamefont {{Wiezorrek}}, \citenamefont {{Woillez}},
  \citenamefont {{Yazici}}, \citenamefont {{Ziegler}},\ and\ \citenamefont
  {{Zins}}}]{Gravity18a}%
  \BibitemOpen
  \bibfield  {author} {\bibinfo {author} {\bibnamefont {{Gravity
  Collaboration}}}, \bibinfo {author} {\bibfnamefont {R.}~\bibnamefont
  {{Abuter}}}, \bibinfo {author} {\bibfnamefont {A.}~\bibnamefont {{Amorim}}},
  \bibinfo {author} {\bibfnamefont {N.}~\bibnamefont {{Anugu}}}, \bibinfo
  {author} {\bibfnamefont {M.}~\bibnamefont {{Baub{\"o}ck}}}, \bibinfo {author}
  {\bibfnamefont {M.}~\bibnamefont {{Benisty}}}, \bibinfo {author}
  {\bibfnamefont {J.~P.}\ \bibnamefont {{Berger}}}, \bibinfo {author}
  {\bibfnamefont {N.}~\bibnamefont {{Blind}}}, \bibinfo {author} {\bibfnamefont
  {H.}~\bibnamefont {{Bonnet}}}, \bibinfo {author} {\bibfnamefont
  {W.}~\bibnamefont {{Brandner}}}, \bibinfo {author} {\bibfnamefont
  {A.}~\bibnamefont {{Buron}}}, \bibinfo {author} {\bibfnamefont
  {C.}~\bibnamefont {{Collin}}}, \bibinfo {author} {\bibfnamefont
  {F.}~\bibnamefont {{Chapron}}}, \bibinfo {author} {\bibfnamefont
  {Y.}~\bibnamefont {{Cl{\'e}net}}}, \bibinfo {author} {\bibfnamefont
  {V.}~\bibnamefont {{Coud{\'e} Du Foresto}}}, \bibinfo {author} {\bibfnamefont
  {P.~T.}\ \bibnamefont {{de Zeeuw}}}, \bibinfo {author} {\bibfnamefont
  {C.}~\bibnamefont {{Deen}}}, \bibinfo {author} {\bibfnamefont
  {F.}~\bibnamefont {{Delplancke-Str{\"o}bele}}}, \bibinfo {author}
  {\bibfnamefont {R.}~\bibnamefont {{Dembet}}}, \bibinfo {author}
  {\bibfnamefont {J.}~\bibnamefont {{Dexter}}}, \bibinfo {author}
  {\bibfnamefont {G.}~\bibnamefont {{Duvert}}}, \bibinfo {author}
  {\bibfnamefont {A.}~\bibnamefont {{Eckart}}}, \bibinfo {author}
  {\bibfnamefont {F.}~\bibnamefont {{Eisenhauer}}}, \bibinfo {author}
  {\bibfnamefont {G.}~\bibnamefont {{Finger}}}, \bibinfo {author}
  {\bibfnamefont {N.~M.}\ \bibnamefont {{F{\"o}rster Schreiber}}}, \bibinfo
  {author} {\bibfnamefont {P.}~\bibnamefont {{F{\'e}dou}}}, \bibinfo {author}
  {\bibfnamefont {P.}~\bibnamefont {{Garcia}}}, \bibinfo {author}
  {\bibfnamefont {R.}~\bibnamefont {{Garcia Lopez}}}, \bibinfo {author}
  {\bibfnamefont {F.}~\bibnamefont {{Gao}}}, \bibinfo {author} {\bibfnamefont
  {E.}~\bibnamefont {{Gendron}}}, \bibinfo {author} {\bibfnamefont
  {R.}~\bibnamefont {{Genzel}}}, \bibinfo {author} {\bibfnamefont
  {S.}~\bibnamefont {{Gillessen}}}, \bibinfo {author} {\bibfnamefont
  {P.}~\bibnamefont {{Gordo}}}, \bibinfo {author} {\bibfnamefont
  {M.}~\bibnamefont {{Habibi}}}, \bibinfo {author} {\bibfnamefont
  {X.}~\bibnamefont {{Haubois}}}, \bibinfo {author} {\bibfnamefont
  {M.}~\bibnamefont {{Haug}}}, \bibinfo {author} {\bibfnamefont
  {F.}~\bibnamefont {{Hau{\ss}mann}}}, \bibinfo {author} {\bibfnamefont
  {T.}~\bibnamefont {{Henning}}}, \bibinfo {author} {\bibfnamefont
  {S.}~\bibnamefont {{Hippler}}}, \bibinfo {author} {\bibfnamefont
  {M.}~\bibnamefont {{Horrobin}}}, \bibinfo {author} {\bibfnamefont
  {Z.}~\bibnamefont {{Hubert}}}, \bibinfo {author} {\bibfnamefont
  {N.}~\bibnamefont {{Hubin}}}, \bibinfo {author} {\bibfnamefont
  {A.}~\bibnamefont {{Jimenez Rosales}}}, \bibinfo {author} {\bibfnamefont
  {L.}~\bibnamefont {{Jochum}}}, \bibinfo {author} {\bibfnamefont
  {K.}~\bibnamefont {{Jocou}}}, \bibinfo {author} {\bibfnamefont
  {A.}~\bibnamefont {{Kaufer}}}, \bibinfo {author} {\bibfnamefont
  {S.}~\bibnamefont {{Kellner}}}, \bibinfo {author} {\bibfnamefont
  {S.}~\bibnamefont {{Kendrew}}}, \bibinfo {author} {\bibfnamefont
  {P.}~\bibnamefont {{Kervella}}}, \bibinfo {author} {\bibfnamefont
  {Y.}~\bibnamefont {{Kok}}}, \bibinfo {author} {\bibfnamefont
  {M.}~\bibnamefont {{Kulas}}}, \bibinfo {author} {\bibfnamefont
  {S.}~\bibnamefont {{Lacour}}}, \bibinfo {author} {\bibfnamefont
  {V.}~\bibnamefont {{Lapeyr{\`e}re}}}, \bibinfo {author} {\bibfnamefont
  {B.}~\bibnamefont {{Lazareff}}}, \bibinfo {author} {\bibfnamefont {J.-B.}\
  \bibnamefont {{Le Bouquin}}}, \bibinfo {author} {\bibfnamefont
  {P.}~\bibnamefont {{L{\'e}na}}}, \bibinfo {author} {\bibfnamefont
  {M.}~\bibnamefont {{Lippa}}}, \bibinfo {author} {\bibfnamefont
  {R.}~\bibnamefont {{Lenzen}}}, \bibinfo {author} {\bibfnamefont
  {A.}~\bibnamefont {{M{\'e}rand}}}, \bibinfo {author} {\bibfnamefont
  {E.}~\bibnamefont {{M{\"u}ler}}}, \bibinfo {author} {\bibfnamefont
  {U.}~\bibnamefont {{Neumann}}}, \bibinfo {author} {\bibfnamefont
  {T.}~\bibnamefont {{Ott}}}, \bibinfo {author} {\bibfnamefont
  {L.}~\bibnamefont {{Palanca}}}, \bibinfo {author} {\bibfnamefont
  {T.}~\bibnamefont {{Paumard}}}, \bibinfo {author} {\bibfnamefont
  {L.}~\bibnamefont {{Pasquini}}}, \bibinfo {author} {\bibfnamefont
  {K.}~\bibnamefont {{Perraut}}}, \bibinfo {author} {\bibfnamefont
  {G.}~\bibnamefont {{Perrin}}}, \bibinfo {author} {\bibfnamefont
  {O.}~\bibnamefont {{Pfuhl}}}, \bibinfo {author} {\bibfnamefont {P.~M.}\
  \bibnamefont {{Plewa}}}, \bibinfo {author} {\bibfnamefont {S.}~\bibnamefont
  {{Rabien}}}, \bibinfo {author} {\bibfnamefont {A.}~\bibnamefont
  {{Ram{\'{\i}}rez}}}, \bibinfo {author} {\bibfnamefont {J.}~\bibnamefont
  {{Ramos}}}, \bibinfo {author} {\bibfnamefont {C.}~\bibnamefont {{Rau}}},
  \bibinfo {author} {\bibfnamefont {G.}~\bibnamefont
  {{Rodr{\'{\i}}guez-Coira}}}, \bibinfo {author} {\bibfnamefont {R.-R.}\
  \bibnamefont {{Rohloff}}}, \bibinfo {author} {\bibfnamefont {G.}~\bibnamefont
  {{Rousset}}}, \bibinfo {author} {\bibfnamefont {J.}~\bibnamefont
  {{Sanchez-Bermudez}}}, \bibinfo {author} {\bibfnamefont {S.}~\bibnamefont
  {{Scheithauer}}}, \bibinfo {author} {\bibfnamefont {M.}~\bibnamefont
  {{Sch{\"o}ller}}}, \bibinfo {author} {\bibfnamefont {N.}~\bibnamefont
  {{Schuler}}}, \bibinfo {author} {\bibfnamefont {J.}~\bibnamefont
  {{Spyromilio}}}, \bibinfo {author} {\bibfnamefont {O.}~\bibnamefont
  {{Straub}}}, \bibinfo {author} {\bibfnamefont {C.}~\bibnamefont
  {{Straubmeier}}}, \bibinfo {author} {\bibfnamefont {E.}~\bibnamefont
  {{Sturm}}}, \bibinfo {author} {\bibfnamefont {L.~J.}\ \bibnamefont
  {{Tacconi}}}, \bibinfo {author} {\bibfnamefont {K.~R.~W.}\ \bibnamefont
  {{Tristram}}}, \bibinfo {author} {\bibfnamefont {F.}~\bibnamefont
  {{Vincent}}}, \bibinfo {author} {\bibfnamefont {S.}~\bibnamefont {{von
  Fellenberg}}}, \bibinfo {author} {\bibfnamefont {I.}~\bibnamefont {{Wank}}},
  \bibinfo {author} {\bibfnamefont {I.}~\bibnamefont {{Waisberg}}}, \bibinfo
  {author} {\bibfnamefont {F.}~\bibnamefont {{Widmann}}}, \bibinfo {author}
  {\bibfnamefont {E.}~\bibnamefont {{Wieprecht}}}, \bibinfo {author}
  {\bibfnamefont {M.}~\bibnamefont {{Wiest}}}, \bibinfo {author} {\bibfnamefont
  {E.}~\bibnamefont {{Wiezorrek}}}, \bibinfo {author} {\bibfnamefont
  {J.}~\bibnamefont {{Woillez}}}, \bibinfo {author} {\bibfnamefont
  {S.}~\bibnamefont {{Yazici}}}, \bibinfo {author} {\bibfnamefont
  {D.}~\bibnamefont {{Ziegler}}}, \ and\ \bibinfo {author} {\bibfnamefont
  {G.}~\bibnamefont {{Zins}}},\ }\href {\doibase 10.1051/0004-6361/201833718}
  {\bibfield  {journal} {\bibinfo  {journal} {Astronomy \& Astrophysics}\
  }\textbf {\bibinfo {volume} {615}},\ \bibinfo {eid} {L15} (\bibinfo {year}
  {2018}{\natexlab{a}})},\ \Eprint {http://arxiv.org/abs/1807.09409}
  {arXiv:1807.09409} \BibitemShut {NoStop}%
\bibitem [{\citenamefont {{Gravity Collaboration}}\ \emph
  {et~al.}(2018{\natexlab{b}})\citenamefont {{Gravity Collaboration}},
  \citenamefont {{Abuter}}, \citenamefont {{Amorim}}, \citenamefont
  {{Baub{\"o}ck}}, \citenamefont {{Berger}}, \citenamefont {{Bonnet}},
  \citenamefont {{Brandner}}, \citenamefont {{Cl{\'e}net}}, \citenamefont
  {{Coud{\'e} Du Foresto}}, \citenamefont {{de Zeeuw}}, \citenamefont {{Deen}},
  \citenamefont {{Dexter}}, \citenamefont {{Duvert}}, \citenamefont {{Eckart}},
  \citenamefont {{Eisenhauer}}, \citenamefont {{F{\"o}rster Schreiber}},
  \citenamefont {{Garcia}}, \citenamefont {{Gao}}, \citenamefont {{Gendron}},
  \citenamefont {{Genzel}}, \citenamefont {{Gillessen}}, \citenamefont
  {{Guajardo}}, \citenamefont {{Habibi}}, \citenamefont {{Haubois}},
  \citenamefont {{Henning}}, \citenamefont {{Hippler}}, \citenamefont
  {{Horrobin}}, \citenamefont {{Huber}}, \citenamefont {{Jim{\'e}nez-Rosales}},
  \citenamefont {{Jocou}}, \citenamefont {{Kervella}}, \citenamefont
  {{Lacour}}, \citenamefont {{Lapeyr{\`e}re}}, \citenamefont {{Lazareff}},
  \citenamefont {{Le Bouquin}}, \citenamefont {{L{\'e}na}}, \citenamefont
  {{Lippa}}, \citenamefont {{Ott}}, \citenamefont {{Panduro}}, \citenamefont
  {{Paumard}}, \citenamefont {{Perraut}}, \citenamefont {{Perrin}},
  \citenamefont {{Pfuhl}}, \citenamefont {{Plewa}}, \citenamefont {{Rabien}},
  \citenamefont {{Rodr{\'{\i}}guez-Coira}}, \citenamefont {{Rousset}},
  \citenamefont {{Sternberg}}, \citenamefont {{Straub}}, \citenamefont
  {{Straubmeier}}, \citenamefont {{Sturm}}, \citenamefont {{Tacconi}},
  \citenamefont {{Vincent}}, \citenamefont {{von Fellenberg}}, \citenamefont
  {{Waisberg}}, \citenamefont {{Widmann}}, \citenamefont {{Wieprecht}},
  \citenamefont {{Wiezorrek}}, \citenamefont {{Woillez}},\ and\ \citenamefont
  {{Yazici}}}]{Gravity18}%
  \BibitemOpen
  \bibfield  {author} {\bibinfo {author} {\bibnamefont {{Gravity
  Collaboration}}}, \bibinfo {author} {\bibfnamefont {R.}~\bibnamefont
  {{Abuter}}}, \bibinfo {author} {\bibfnamefont {A.}~\bibnamefont {{Amorim}}},
  \bibinfo {author} {\bibfnamefont {M.}~\bibnamefont {{Baub{\"o}ck}}}, \bibinfo
  {author} {\bibfnamefont {J.~P.}\ \bibnamefont {{Berger}}}, \bibinfo {author}
  {\bibfnamefont {H.}~\bibnamefont {{Bonnet}}}, \bibinfo {author}
  {\bibfnamefont {W.}~\bibnamefont {{Brandner}}}, \bibinfo {author}
  {\bibfnamefont {Y.}~\bibnamefont {{Cl{\'e}net}}}, \bibinfo {author}
  {\bibfnamefont {V.}~\bibnamefont {{Coud{\'e} Du Foresto}}}, \bibinfo {author}
  {\bibfnamefont {P.~T.}\ \bibnamefont {{de Zeeuw}}}, \bibinfo {author}
  {\bibfnamefont {C.}~\bibnamefont {{Deen}}}, \bibinfo {author} {\bibfnamefont
  {J.}~\bibnamefont {{Dexter}}}, \bibinfo {author} {\bibfnamefont
  {G.}~\bibnamefont {{Duvert}}}, \bibinfo {author} {\bibfnamefont
  {A.}~\bibnamefont {{Eckart}}}, \bibinfo {author} {\bibfnamefont
  {F.}~\bibnamefont {{Eisenhauer}}}, \bibinfo {author} {\bibfnamefont {N.~M.}\
  \bibnamefont {{F{\"o}rster Schreiber}}}, \bibinfo {author} {\bibfnamefont
  {P.}~\bibnamefont {{Garcia}}}, \bibinfo {author} {\bibfnamefont
  {F.}~\bibnamefont {{Gao}}}, \bibinfo {author} {\bibfnamefont
  {E.}~\bibnamefont {{Gendron}}}, \bibinfo {author} {\bibfnamefont
  {R.}~\bibnamefont {{Genzel}}}, \bibinfo {author} {\bibfnamefont
  {S.}~\bibnamefont {{Gillessen}}}, \bibinfo {author} {\bibfnamefont
  {P.}~\bibnamefont {{Guajardo}}}, \bibinfo {author} {\bibfnamefont
  {M.}~\bibnamefont {{Habibi}}}, \bibinfo {author} {\bibfnamefont
  {X.}~\bibnamefont {{Haubois}}}, \bibinfo {author} {\bibfnamefont
  {T.}~\bibnamefont {{Henning}}}, \bibinfo {author} {\bibfnamefont
  {S.}~\bibnamefont {{Hippler}}}, \bibinfo {author} {\bibfnamefont
  {M.}~\bibnamefont {{Horrobin}}}, \bibinfo {author} {\bibfnamefont
  {A.}~\bibnamefont {{Huber}}}, \bibinfo {author} {\bibfnamefont
  {A.}~\bibnamefont {{Jim{\'e}nez-Rosales}}}, \bibinfo {author} {\bibfnamefont
  {L.}~\bibnamefont {{Jocou}}}, \bibinfo {author} {\bibfnamefont
  {P.}~\bibnamefont {{Kervella}}}, \bibinfo {author} {\bibfnamefont
  {S.}~\bibnamefont {{Lacour}}}, \bibinfo {author} {\bibfnamefont
  {V.}~\bibnamefont {{Lapeyr{\`e}re}}}, \bibinfo {author} {\bibfnamefont
  {B.}~\bibnamefont {{Lazareff}}}, \bibinfo {author} {\bibfnamefont {J.-B.}\
  \bibnamefont {{Le Bouquin}}}, \bibinfo {author} {\bibfnamefont
  {P.}~\bibnamefont {{L{\'e}na}}}, \bibinfo {author} {\bibfnamefont
  {M.}~\bibnamefont {{Lippa}}}, \bibinfo {author} {\bibfnamefont
  {T.}~\bibnamefont {{Ott}}}, \bibinfo {author} {\bibfnamefont
  {J.}~\bibnamefont {{Panduro}}}, \bibinfo {author} {\bibfnamefont
  {T.}~\bibnamefont {{Paumard}}}, \bibinfo {author} {\bibfnamefont
  {K.}~\bibnamefont {{Perraut}}}, \bibinfo {author} {\bibfnamefont
  {G.}~\bibnamefont {{Perrin}}}, \bibinfo {author} {\bibfnamefont
  {O.}~\bibnamefont {{Pfuhl}}}, \bibinfo {author} {\bibfnamefont {P.~M.}\
  \bibnamefont {{Plewa}}}, \bibinfo {author} {\bibfnamefont {S.}~\bibnamefont
  {{Rabien}}}, \bibinfo {author} {\bibfnamefont {G.}~\bibnamefont
  {{Rodr{\'{\i}}guez-Coira}}}, \bibinfo {author} {\bibfnamefont
  {G.}~\bibnamefont {{Rousset}}}, \bibinfo {author} {\bibfnamefont
  {A.}~\bibnamefont {{Sternberg}}}, \bibinfo {author} {\bibfnamefont
  {O.}~\bibnamefont {{Straub}}}, \bibinfo {author} {\bibfnamefont
  {C.}~\bibnamefont {{Straubmeier}}}, \bibinfo {author} {\bibfnamefont
  {E.}~\bibnamefont {{Sturm}}}, \bibinfo {author} {\bibfnamefont {L.~J.}\
  \bibnamefont {{Tacconi}}}, \bibinfo {author} {\bibfnamefont {F.}~\bibnamefont
  {{Vincent}}}, \bibinfo {author} {\bibfnamefont {S.}~\bibnamefont {{von
  Fellenberg}}}, \bibinfo {author} {\bibfnamefont {I.}~\bibnamefont
  {{Waisberg}}}, \bibinfo {author} {\bibfnamefont {F.}~\bibnamefont
  {{Widmann}}}, \bibinfo {author} {\bibfnamefont {E.}~\bibnamefont
  {{Wieprecht}}}, \bibinfo {author} {\bibfnamefont {E.}~\bibnamefont
  {{Wiezorrek}}}, \bibinfo {author} {\bibfnamefont {J.}~\bibnamefont
  {{Woillez}}}, \ and\ \bibinfo {author} {\bibfnamefont {S.}~\bibnamefont
  {{Yazici}}},\ }\href {\doibase 10.1051/0004-6361/201834294} {\bibfield
  {journal} {\bibinfo  {journal} {Astronomy \& Astrophysics}\ }\textbf
  {\bibinfo {volume} {618}},\ \bibinfo {eid} {L10} (\bibinfo {year}
  {2018}{\natexlab{b}})},\ \Eprint {http://arxiv.org/abs/1810.12641}
  {arXiv:1810.12641} \BibitemShut {NoStop}%
\end{thebibliography}%

\end{document}